%% file: main_uqRobustWRChan_arxiv1.tex
\documentclass[preprint,3p]{elsarticle}      

\usepackage{amsmath}
\usepackage{natbib}
\usepackage{amssymb}
\usepackage{graphicx}
\usepackage{dcolumn}
\usepackage{bm}
\usepackage{mathrsfs}
\usepackage{color}
\usepackage{psfrag}
\usepackage{url}
\usepackage{soul}
\usepackage{float}  

\usepackage{cancel}  

\usepackage[normalem]{ulem}
\usepackage{color}
\newcommand{\rev}[1]{{\color{black}#1}}
\newcommand{\revs}[1]{{\color{black}#1}} 

\usepackage{booktabs}
  
\include{commands}

\begin{document}

\title{An Uncertainty-Quantification Framework for Assessing Accuracy, Sensitivity, and Robustness in Computational Fluid Dynamics}

\author[focal1,focal2]{S.~Rezaeiravesh\corref{cor1}}
\ead{salehr@kth.se}
\author[focal1,focal2]{R.~Vinuesa\corref{cor2}}
\ead{rvinuesa@mech.kth.se}
\author[focal1,focal2]{P.~Schlatter\corref{cor2}}
\ead{pschlatt@mech.kth.se}
\cortext[cor1]{Principal Corresponding Author}
\cortext[cor2]{Corresponding Author}
\address[focal1]{SimEx/FLOW, Engineering Mechanics, KTH Royal Institute of Technology, SE-100 44 Stockholm, Sweden}
\address[focal2]{Swedish e-Science Research Centre (SeRC), Stockholm, Sweden}

\date{\today}

\begin{keyword}
Uncertainty quantification,
Validation and verification, 
Computational fluid dynamics,
Polynomial chaos expansion,
Gaussian process regression,
Wall turbulence.
\end{keyword}

\begin{abstract}
A framework is developed based on different uncertainty quantification (UQ) techniques in order to assess validation and verification (V\&V) metrics in computational physics problems, in general, and computational fluid dynamics (CFD), in particular.
The metrics include accuracy, sensitivity and robustness of the simulator's outputs with respect to uncertain inputs and computational parameters. 
These parameters are divided into two groups: based on the variation of the first group, a computer experiment is designed, the data of which may become uncertain due to the parameters of the second group.
To construct a surrogate model based on uncertain data, Gaussian process regression (GPR) with observation-dependent (heteroscedastic) noise structure is used.
To estimate the propagated uncertainties in the simulator's outputs from first and also the combination of first and second groups of parameters, standard and probabilistic polynomial chaos expansions (PCE) are employed, respectively. 
Global sensitivity analysis based on Sobol decomposition is performed in connection with the computer experiment to rank the parameters based on their influence on the simulator's output. 
To illustrate its capabilities, the framework is applied to the scale-resolving simulations of turbulent channel flow using the open-source CFD solver Nek5000. Due to the high-order nature of Nek5000 a thorough assessment of the results' accuracy and reliability is crucial, as the code is aimed at high-fidelity simulations.
The detailed analyses and the resulting conclusions can enhance our insight into the influence of different factors on physics simulations, in particular the simulations of wall-bounded turbulence. 
\end{abstract}

\maketitle

\section{Introduction}\label{sec:intro}
In any scientific and engineering field, the ultimate aim of applying analytical, numerical, and experimental approaches to fluid dynamics problems is to obtain \emph{high-quality} results. 
In a broad classical sense, quality can be exclusively interpreted as \emph{accuracy}. 
But, as explained in the landmark studies by~\citet{roache}, and \citet{oberkampf:02}, \emph{reliability} \rev{and \emph{robustness}} should be also included in the interpretation of the quality. 
In \rfs~\cite{oberkampf:02,oberkampf:10}, a thorough review is given on approaches based on verification and validation (V\&V) which have a key importance to assess accuracy and reliability of computational simulations. 
Following the terminologies introduced by~\citet{schlesinger:79} and extended in \rfs~\cite{roache,oberkampf:02,oberkampf:10}, verification in computational simulations means investigating the errors in the computational model and its solution. 
On the other hand, validation aims at quantifying how accurate the solutions of a computational model are compared to reference experimental~data.

The present study aims at introducing a framework based on uncertainty quantification (UQ) techniques in order to quantify accuracy, robustness and sensitivity of computational simulations.
The three mentioned concepts are closely related to V\&V. 
It is also noteworthy that V\&V and UQ together build the foundations of predictive computational sciences, see~\citet{oden:10:1}.
The framework developed in this study is general in nature and hence applicable to any computational physics problem.
Nevertheless, we provide the examples and supporting discussions with regard to CFD (computational fluid dynamics) problems and with emphasis on the simulation of turbulent \rev{wall-bounded} flows.
These types of flows are complex and appear in many physics and engineering applications.

In order to put the work in the context, let us start by providing the necessary terminology and definitions. 
In an ideal environment, the \emph{true value} of the quantities of interest (QoIs) of a \rev{physical system such as a} flow could be measured in a highly-accurate experiment or a high-fidelity numerical simulation. 
Borrowing the definition from Rabinovich~\cite{rabinovich:17}, the true value is ``the value of a quantity that, were it known, would ideally reflect the property of an object with respect to the purpose of the measurement". 
Clearly, in practice, through repeated realizations in which the possible errors and uncertainties are kept \rev{at a} minimum, only a \emph{reference true value} would be possibly achieved.  
The deviation of the QoIs of a simulation from a reference true value is denoted as the \emph{error} in the QoIs. 
Naturally, a lower error means higher \emph{accuracy} and vice versa. 
According to \rfs~\cite{oberkampf:02,oberkampf:10}, in the verification step the accuracy of a computational model is assessed with respect to analytical and highly-accurate numerical data. 
Such actions require software testing.
Conversely, in the validation step, the results of the computational model are compared to experimental data which within some confidence reflect the true reality. 

Another point of view for the definition of error is as follows:
According to the AIAA (American Institute of Aeronautics and Astronautics) Guide on V\&V~\cite{aiaa:98}, an error is defined as ``a recognizable deficiency in any phase or activity of modeling and simulations that is not due to lack of knowledge".
According to the same guide, two types of errors can be considered. 
The unacknowledged errors are due to  mistakes, bugs in the codes, and compiler errors, for instance. 
In contrast, the acknowledged errors are identified and quantified as the deviation from a reference true value. 
This is aligned with the definition of error given above. 
Referring to the same Guide~\cite{aiaa:98}, \emph{uncertainty} is defined as ``a potential deficiency in any phase or activity of the modeling process that is due to lack of knowledge". 
As compared to the definition of error, the key distinguishing characteristics of uncertainty are that it is, first, a potential deficiency, and second, it is due to the lack of knowledge.
The UQ literature, for instance~\citet{uqHandbook} and \citet{smith}, defines uncertainty as the lack of certainty in a model output which may stem from incomplete knowledge or lack of knowledge, incomplete modeling or data, uncertain or variable factors in the model, etc. 
Hereafter, we use the word \emph{factor} for combination of parameters and inputs. 
In a broad view, the uncertainties can be divided into two main groups: epistemic (bias) and aleatoric (random) uncertainties.
The treatment of each group requires employing appropriate sets of mathematical and statistical tools, see~\eg~\cite{uqHandbook,smith}.
A common strategy in the UQ framework is, however, to consider the uncertainties to be random (even if they are not really so), in order to make probabilistic methods applicable, see \eg~\cite{smith}.

Given the above general definitions, we provide some examples related to CFD and turbulence simulation which help understanding the connection between UQ and V\&V, and also motivate the aim and strategy of the present study. 
The general definitions for the error can also be adopted for numerical and associated computational models developed for solving the continuous form of a system of PDEs (partial differential equations) which for CFD applications are the Navier--Stokes equations. 
What is known as the numerical error includes discretization and projection (truncation) components\rev{, and also numerical parameters such as tolerances for iterative solvers}. 
Computational errors may originate from, for instance, bugs in the computer codes, parallelization algorithms, compilers, and perhaps hardware. 
Clearly, due to unavailability of analytical general solutions for the Navier--Stokes equations, except for very few idealized flows, the reference true solution in the verification step is chosen from the most accurate available numerical simulations \rev{and experimental data}. 
\rev{However, the reference data can also be uncertain, see \eg~\cite{oliver:14,mariotti:17,salehPoF:18,salehHWA:18}}.

The simulation of the turbulent flows is more subtle, due to the additional types of errors and uncertainties, and at the same time, the small margin of tolerance allowed in engineering designs relying on these flows.   
The use of the DNS (direct numerical simulation) approach, which would provide the highest fidelity results, at high Reynolds numbers that are relevant to engineering applications is computationally infeasible. 
The computational cost can be reduced using the LES (large-eddy simulation) approach which directly resolves scales larger than a characteristic length and models the remaining subgrid scales (SGS), see for instance the textbook by \citet{sagaut:06}.
However, the use of SGS models to compensate for the effect of unresolved scales adds more complexity.
In fact, not only the numerical and modeling errors may become intertwined, but also the structure and calibration parameters of the SGS models introduce new sources of uncertainties. 
\revs{
As shown in \rfs~\cite{kravchenko:97,vreman:96}, the contribution and effect of the components of numerical errors vary for different terms in the Navier-Stokes equations and also depend on the order of the numerical scheme used for discretizing the equations. 
For high-order methods, the aliasing error has been found to be dominant, whereas in low-order methods the truncation error may be so high that even exceeds the SGS effects.}
The uncertainties arising from turbulence modeling become more dominant when moving toward lower-fidelity approaches such as hybrid LES/RANS (Reynolds-averaged Navier--Stokes), wall-modelled LES and RANS, see \rf~\cite{sagaut:13}. 
In addition to all these, when using scale-resolving simulations such as DNS and LES, extra uncertainties in the averaged QoIs due to finite time-averaging appear\rev{. These errors could in principle be reduced by running longer, at an obvious computational cost.}

To summarize up to this point, the reported QoIs of any CFD simulation are \emph{uncertain}, at least up to some extent.
The uncertainty in the QoIs is driven by many sources of uncertainties and errors which can, however, be intertwined and interconnected. 
As a result, tracing the uncertainties by analytical approaches is usually very hard, if not impossible, and in any case the outcome could be inaccurate and unreliable. 

A remedy can be found in the field of UQ.
In particular, the UQ forward problem is considered, which aims at estimating the propagation of uncertainties in a \emph{simulator}'s (\ie~the computational code) outputs due to the uncertainty in different factors (inputs and parameters). 
If the factor under investigation is not uncertain or random by nature, the impact of its \emph{variability} on the outputs is evaluated.
Considering the costs involved in running a simulator, a main part of the process is focused on building \rev{\emph{surrogate models}} to express approximate relationship between the simulator's inputs and outputs. 
Higher uncertainties estimated in the model outputs indicate low \emph{robustness} of the model to a particular set of uncertain factors. 
The forward problem is complemented by \emph{sensitivity analysis} through which the contribution of each of the uncertain factors in the resulting uncertainty in the model output is evaluated. 
The robustness and sensitivity can be estimated for primary QoIs as well as their errors.
Therefore, depending on the reference data used to evaluate the error, both verification and validation of the computational models can be conducted in a probabilistic framework.

The \emph{non-intrusive} implementation of the UQ forward problem can be carried out using \emph{computer experiments}, see \rfs~\cite{sacks:89, santner:03}.
As will be detailed in \sect~\ref{sec:uq}, the implementation includes the following steps: 
(i) parameterization of the factors, (ii) assuming the parameters to be random variables with known distributions over given spaces, and (iii) running the simulator at a limited number of samples taken for the uncertain parameters.
In the rest of this \rev{paper}, the factors which are considered in the described methodology are referred to as \emph{\catI} parameters.

Several studies have investigated the impact of variation of \catI~parameters on the outputs of numerical simulations of turbulent flows. 
A short survey of the relevant factors and corresponding example studies is given below. 
Our focus is on scale-resolving approaches such as DNS and LES which are computationally expensive but capable of revealing more physics. 
For RANS simulations, the reader is referred to the review paper~\cite{xiao:19} and the references therein.
The impact of combined numerical errors which include spatial discretization and time integration on scale-resolving simulations have been investigated for different flows. 
Meyers and Sagaut~\cite{meyers:07} and Rezaeiravesh and Liefvendahl~\cite{salehPoF:18} studied turbulent channel flow, Meldi~\et~\cite{meldi:12} considered spatially-evolving mixing layers, and Mariotti \et~\cite{mariotti:17} investigated the flow over a rectangular cylinder. 
The influence of the coefficients appearing in the SGS models in LES have been studied by Lucor~\et\cite{lucor:07} and Meldi \et~\cite{meldi:12}.
Congedo \et~\cite{congedo:13} investigated the sensitivity of LES of turbulent pipe flow with respect to the inflow condition. 
For the wall-modeled LES, the influence of different factors on the approximate wall boundary condition and consequently on the whole flow field is studied by Rezaeiravesh~\et~\cite{salehCaF:19}.

A computer experiment relies on several realizations of the simulator's outputs or QoIs, where there exists a one-to-one correspondence between a realization and a sample taken from the space of \catI~parameters.
The QoIs in each of such realizations can be uncertain due to what is hereafter referred to as \emph{\catII}~uncertainties. 
An example is the uncertainty in the time-averaged QoIs of turbulent flows as a result of using finite number of time samples.
The way of incorporating the \emph{known} \catII~uncertainties in the design of experiments for \catI~uncertainties constitutes another focus of the present study. 
It is stressed that the evaluation of \catII~uncertainties is beyond the scope of the present  paper. 
In particular, for assessment of uncertainty due to lack of time samples, the reader is referred to Oliver \et~\cite{oliver:14} and Russo and Luchini~\cite{russo:17}.

The unified framework of the present study is developed based on different UQ techniques. 
To estimate uncertainties in, and equivalently the robustness of, the outputs of a simulator with respect to the factors falling under \catI, the non-intrusive polynomial chaos expansion (PCE) \cite{xiu:05,xiu:07,ghanem:91,pettersson:15} technique is used. 
Different types of construction of PCE as well as the impact of the number of samples on the accuracy of the estimations are discussed.
A novel aspect of this study is to show how to incorporate the already-known uncertainties of \catII~when aiming for quantifying the uncertainties from the \catI~factors.
This is obtained through employing a more general form of observation noise in the Gaussian process regression (GPR),~\cite{rasmussen:05,gramacy:20}. 
To estimate uncertainties in the QoIs in these situations, probabilistic polynomial chaos expansion (PPCE) is an appropriate tool. 
The formulation and use of this method within the present unified framework is another novelty of this study. 
In addition to these, to perform global sensitivity analysis~\cite{uqHandbook,smith,santner:03}, Sobol indices relying on the ANOVA (analysis of variance) technique~\cite{sobol:90,sobol:01} are estimated. 

The remainder of this paper is organized as follow.
\sect~\ref{sec:uq} provides the details of the UQ techniques which form the framework of the present study. 
Moreover, a short description of the numerical implementation of these approaches is given.
The connection between the UQ techniques and the evaluation of accuracy, robustness, and sensitivity in CFD simulations is explained in \sect~\ref{sec:uqCFD}. 
These are followed by \sect~\ref{sec:example}, where it is shown how the framework can be employed for a practical scale-resolving simulation of a canonical wall-bounded turbulent flow. 
Finally, conclusions and plans for extending the current study are~provided in \sect~\ref{sec:conclusions}.

\section{UQ Techniques}\label{sec:uq}
In a UQ forward problem, the uncertainties in a model inputs and parameters are propagated into the model outputs, and the resulting response surface and statistical moments of the outputs are constructed.
In \sect~\ref{sec:uqFWD}, theoretical aspects of a UQ forward problem are discussed. 
First, the concept of a \emph{surrogate} which yields a computationally inexpensive relationship between inputs and outputs of a computational model, is reviewed.
Then the standard \rev{polynomial chaos expansion (PCE)} is discussed, which is a powerful technique to estimate statistical moments of QoIs due to the variability of \catI~parameters in a computer experiment. 
The treatment of the known uncertainties of the factors belonging to \catII~comes next via reviewing \rev{Gaussian process regression (GPR)}. 
We then investigate the properties of probabilistic PCE which is a combination of non-intrusive PCE and GPR. 
As a complement to the UQ forward problem, sensitivity analysis is performed, which is the focus of \sect~\ref{sec:sobol}. 
Finally, in \sect~\ref{sec:uqImplmnt} it is briefly explained how the described UQ techniques are implemented.

The UQ techniques employed here are chosen to comply with two basic requirements: 
first, reliable UQ estimates must be achievable through a limited number of realizations. 
This is essential considering the large computational cost of running the simulator in many computational physics problems, including CFD. 
Second, the resulting UQ framework should be non-intrusively linked to the simulator.  
This provides a great flexibility in practice\rev{, as not specialised codes, potentially with involved numerical algorithms and parallelization approaches, need to be developed, optimized, debugged and maintained}.

\subsection{UQ Forward Problem}\label{sec:uqFWD}

\subsubsection{Surrogate}
The uncertain factors belonging to \catI~are denoted by $\fq$, where $\fq=\{q_1,q_2,\cdots,q_p\}$.
As a convention throughout this paper, all bold-face letters are used for representing tensors \revs{of order greater than zero}. 
The parameters~$\fq$ vary over the $p$-dimensional admissible space $\BQ$, \ie~$\fq\in\BQ\subset \BR^p$. 
Besides $\fq$, a simulator and its outputs can also be dependent on the controlled parameters, denoted by~$\chi$. 
From a single run of the deterministic simulator a realization for the quantities of interest or model outputs $\cR$ are obtained.
Without loss of generality, we only consider univariate outputs~$\cR\in\BR$.
In general, these QoIs can be contaminated by \catII~uncertainties.

In the best of worlds, we could construct a functional~$f(\chi,\fq)$ to describe the relationship between the inputs and the simulations outputs.
Considering the observation or measurement errors~$\feps$ and adopting an additive error model, we could write
\begin{equation}\label{eq:surrGen}
r=f(\chi,\fq) + \feps \,.
\end{equation}
In general, $r=r(\chi,\fq)$, and $\feps$ can be observation-dependent and contains generally both bias and random effects, but here we assume it to be only representing random noise.
Nevertheless, constructing an exact function for the simulator~$f(\chi,\fq)$ is not feasible in practice, considering the prohibitively large number of simulations required.
Instead, a \emph{surrogate}~$\tf(\chi,\fq)$ can be constructed using a limited number of simulations in a computer experiment with outputs~$\mathbf{R}=\{\cR^{(i)}\}_{i=1}^n$, corresponding one-to-one to the samples~$\mathbf{Q}=\{\fq^{(i)}\}_{i=1}^n$ drawn from~$\BQ$. 
We refer to $\cD=\{(\fq^{(i)},r^{(i)})\,|\,i=1,2,\ldots,n\}$ as the \emph{training data} for constructing the surrogate. 
The constructed surrogate can then be used to approximately predict values of~$r$ over the entire $\BQ$ and also be employed for estimating the statistical moments of~$r$ due to the variation of~$\fq$.

There are different methods for constructing surrogates, see~\eg~\cite{uqHandbook,gramacy:20}. A short review is given below on the PCE and GPR methods, which are appropriate for the purpose of the present study. 
The strong point about PCE is that estimating statistical moments of~$\tf(\chi,\fq)$ is a natural outcome of constructing the surrogate. 
However, in the standard use, incorporating the observation uncertainties into the surrogate is not trivial. 
In contrast, such a capability is the main feature of GPR.
Consequently, the predictions of~$r$ by GPR in~$\BQ$ are accompanied by uncertainties.

\subsubsection{Polynomial Chaos Expansion (PCE)}\label{sec:pce}
Consider the factors~$\fq$ to be mutually-independent. 
This property either holds by the problem definition or is achieved through applying a transformation such as Rosenblatt's~\cite{rosenblatt} to the originally dependent parameters. 
Consequently, the joint probability density function (PDF) of these parameters is~$\rho(\fq) = \prod_{i=1}^p \rho_{i}(q_i)$. 
Let each~$q_i$ be mapped to a standard random variable~$\xi_i$, see below, over range~$\Gamma_i$. 
Consequently,~$\fq\in \BQ$ is eventually mapped to~$\fxi\in \Gamma$, where~$\Gamma=\bigotimes_{i=1}^p \Gamma_i$ \rev{and $\bigotimes$ denotes tensor product}.
Using polynomial chaos expansion (PCE), the surrogate~$\tilde{f}(\chi,\fxi)$ is written as a truncated expansion \cite{ghanem:91,xiu:10,pettersson:15},
\begin{equation}\label{eq:pce}
\tilde{f}(\chi,\fxi) \approx 
\sum_{\rk=0}^K \hat{f}_{\rk}(\chi)\Psi_{\rk}(\fxi) \,.
\end{equation}
Here~$\rk$ is a unique re-index corresponding to the multi-index~$\fk=(k_1,k_2,\ldots,k_p)$, and the bases are~$\Psi_\rk(\fxi) = \prod_{i=1}^p \psi_{k_i}(\xi_i)$. 
The PCE surrogates are considered to be of \emph{parametric} type, since through the bases a predefined structure for expressing the uncertain behavior of the model output is considered. 
In the framework of generalized PCE (gPCE), see \rfs~\cite{xiu:02,eldred:09}, for the standard distribution of each univariate random variable~$\xi_i$, where $i=1,2,\cdots,p$, a set of polynomial bases~$\{\psi_{k_i}(\xi_i)\}$ exist which are orthogonal with respect to the PDF~$\rho_i(\xi_i)$. 
This can be expressed as below introducing the weighted inner-product~$\langle \cdot,\cdot \rangle_{\rho_i}$,
\begin{equation}\label{eq:pceIP1}
\langle \psi_{k_i}(\xi_i) \,, \psi_{m_i}(\xi_i) \rangle_{\rho_i}
= \int_{\Gamma_i} \psi_{k_i}(\xi_i) \psi_{m_i}(\xi_i) \rho_i(\xi_i) \dd\xi_i = \gamma_{k_i} \delta_{k_i m_i} \,,
\end{equation}
where,~$\gamma_{k_i}=\langle \psi_{k_i}(\xi_i)\psi_{k_i}(\xi_i)\rangle_{\rho_i}$ and~$\delta_{k_i m_i}$ is the Kronecker delta. 
This also makes the multi-variate bases orthogonal with respect to~$\rho(\fxi)$,
\begin{equation}\label{eq:pceIP2}
\langle \Psi_\fk(\fxi) \,, \Psi_{\mathbf{m}}(\fxi)\rangle_\rho = \gamma_\fk \delta_{\fk \mathbf{m}} \,.
\end{equation}
The inner-products~(\ref{eq:pceIP1}) and (\ref{eq:pceIP2}) can be equivalently interpreted as expectations~$\BE_{q_i}[ \psi_{k_i}(\xi_i) \psi_{m_i}(\xi_i) ]$ and $\BE_{\fq}[\Psi_\fk(\fxi) \Psi_{\mathbf{m}}(\fxi)]$, respectively. 
To construct the PCE (\ref{eq:pce}), we need to: (i) choose a rule to construct the multi-variate bases~$\Psi_{k}(\fxi)$ and accordingly truncate (\ref{eq:pce}) at~$K$, (ii) specify the nodal set~$\{\fxi^{(i)}\}_{i=1}^n$, and (iii) adopt a method to estimate coefficients~$\hf_\rk(\chi)$ given training data $\cD$.

The set of samples~$\{\fxi^{(i)}\}_{i=1}^n$ in $\Gamma$ can be chosen to make a structured grid.
Assume in the~$i$-th dimension of the~$p$-dimensional parameter space~$\Gamma$, we have chosen~$n_i$ sampling nodes. 
Then, the index set~$\Lambda_\fk$ for the tensor product method (TPM) reads as,
\begin{equation}\label{eq:pceTP}
\Lambda^{\rm{TPM}}_\fk = \{\fk\in \BZ^p_{\geq} : \max_{1\leq i\leq p} k_i\leq (n_i-1) \} \,,
\end{equation}
where $\BZ_\geq$ denotes the non-negative integers. 
The TPM rule leads to~$K+1=\prod_{i=1}^p n_i$. 
Therefore, the number of terms in expansion (\ref{eq:pce}) grows exponentially with~$p$, an issue that is referred to as the curse of dimensionality. 
Another choice for the nodal set is to adopt total order method (TOM), for which,
\begin{equation}\label{eq:pceTO}
\Lambda^{\rm{TOM}}_\fk = \{\fk\in\BZ^p_\geq : |\fk|\leq L \} \,,
\end{equation}
where,~$L$ is the maximum polynomial order in each dimension of the parameter space and $|\fk|=\sum_{i=1}^p k_i$  is the magnitude of the multi-index $\fk$.
For TOM, we have~$K+1= {(L+p)!}/{L!p!}$, see \rfs~\cite{eldred:09,smith}.
It is noted that TOM can be considered as a special case of the more general hyperbolic truncation scheme~\cite{blatman:09}.

For obtaining the coefficients~$\hf_\rk(\chi)$ in (\ref{eq:pce}), different techniques can be employed. 
Applying a projection method (\rev{denoted as pseudo-spectral method in}~\cite{xiu:07}) leads to the following integral,
\begin{equation}\label{eq:pceProj}
\hf_\rk(\chi) 
= \frac{\BE_\fq[f(\chi,\fxi) \Psi_\fk(\fxi)]}{\BE_\fq[\Psi_\fk(\fxi) \Psi_\fk(\fxi)]}
=\frac{1}{\gamma_\fk} \int_\Gamma f(\chi,\fxi) \Psi_\fk(\fxi) \rho(\fxi) \dd \fxi \,,
\end{equation}
\revs{where, $\gamma_\fk$ is defined in~(\ref{eq:pceIP2}).}
For the numerical approximation of the integral, an optimal choice is to use the standard Gaussian quadrature rule which results in, 
\begin{equation}\label{eq:gpr_GQrule}
\hf_\rk(\chi) \approx
\frac{1}{\gamma_\fk}
\sum_{j=1}^n r^{(j)} \Psi_\fk(\fxi^{(j)}) \rho(\fxi^{(j)}){\rw}^{(j)} \,.
\end{equation}
In this construction,~$\fxi^{(j)}=(\xi^{(j)}_1,\cdots,\xi^{(j)}_p)$ and~$\rw^{(j)}$ denote the $j$-th quadrature nodes and weight, respectively. 
In the framework of generalized PCE~\cite{xiu:02,eldred:09}, the~$n_i$ nodes in~$i$-th dimension are the zeros of polynomial basis~$\psi_{n_i}(\xi_i)$. 
Moreover, the nodal set in the multi-dimensional parameter space are then constructed using TPM, which may lead to the curse of dimensionality. 
A remedy to this, while still using the projection method to compute the coefficients~$\hf_\rk(\chi)$ from~(\ref{eq:pceProj}), is to use sparse grid quadrature rules (for both nested and non-nested nodes), see the details in \rfs~\cite{smolyak:63,gerstner:98}. 
It is noteworthy that in projection methods, the constructed surrogate at the sampling nodes necessarily possesses the exact training values, \ie~$\tf(\fxi^{(j)})=r^{(j)}$ for $j=1,2,\cdots,n$. 
This clearly proves the connection between Lagrange interpolation based on  stochastic collocation points (which are the parameter samples) and PCE~(\ref{eq:pce}), see \rfs~\cite{eldred:09,xiu:07}.

Another approach for estimating coefficients~$\hf_\rk(\chi)$ in~(\ref{eq:pce}) is to use regression through which a linear set of equations,
\begin{equation}\label{eq:pceSystem}
\mathbf{\mathbf{A}} \hbf = \mathbf{R}^T \,,
\end{equation}
is solved for $\hbf=[\hf_0,\hf_1,\ldots,\hf_K]^T$ via,
\begin{equation}\label{eq:pceRegress}
\|\mathbf{A} \hbf - \mathbf{R}^T\|_{l_2} \leq \text{tol} \,.
\end{equation}
Here, $\text{tol.}$ is a small preset tolerance, $\mathbf{A}$ is a $n\times (K+1)$ matrix with elements~$A_{i (\rk+1)}=\Psi_\rk(\fxi^{(i)})$ and~$\mathbf{R}=[r^{(1)},r^{(2)},\ldots,r^{(n)}]$. 
When using the regression method, the samples~$\{\fxi^{(i)}\}_{i=1}^n$ in the multi-dimensional space are not required to be necessarily structured. 
However, the sampling strategy significantly controls the accuracy of the constructed PCE~(\ref{eq:pce}), see~\eg~\cite{hosder:07}. 
The linear system~(\ref{eq:pceSystem}) becomes under- and over-determined if~$n$ is smaller and bigger than~$(K+1)$, respectively. 
For cases where~$n\ll (K+1)$, which may occur due to the high computational costs involved in making realizations of~$\cR$, an extra constraint on~$\hbf$ is needed.  
In this case, instead of solving (\ref{eq:pceRegress}), we seek for an optimal~$\hbf_{\rm opt}$ where,
\begin{equation}\label{eq:cmprsdSens}
\hbf_{\rm opt}=\min \|\hbf\|_{l_1} \quad \text{such that} \quad 
\|\mathbf{A} \hbf - \mathbf{R}^T\|_{l_2} \leq \text{tol} \,.
\end{equation}
This method, which is called compressed sensing~\cite{candes:06,moore:12}, results in a unique sparse solution for~$\hbf$.
Contrary to the projection method, the PCE~(\ref{eq:pce}) determined by the regression approach does not necessarily match the training data~$\{(\fxi^{(i)},\cR^{(i)})\}_{i=1}^n$. 
However,~$\feps$ in~(\ref{eq:surrGen}) is more a residual of fitting than an observation uncertainty.
Flexibility in sampling from the parameter space and also more numerical stability when handling noisy simulator outputs are advantages of the regression methods over the projection approach.    

The expansion~(\ref{eq:pce}) with determined coefficients as a surrogate interpolates values of~$r$ over the whole parameter space $\BQ$. 
As a strong point of PCE, it can be shown (see~\eg~\cite{uqHandbook,smith}) that the mean and variance of~$\tf(\chi,\fq)$ (and hence approximately $f(\chi,\fq)$) due to the variability of~$\fq$ can be estimated from the following~expressions, 
\begin{eqnarray}
\BE_\fq[\tf(\chi,\fq)] = \hf_0(\chi)  \,, \label{eq:pceE} \\
\BV_\fq[\tf(\chi,\fq)] = \sum_{\rk=1}^K \hf^2_{\rk}(\chi) \gamma_k \,. \label{eq:pceV}
\end{eqnarray}
If the model response $f(\chi,\fq)$ is sufficiently smooth in the parameter space $\BQ$, then the rate of convergence of the above stochastic moments with respect to the number of terms~$K$ in the expansion~(\ref{eq:pce}) can be high. 
In general, it is necessary to assess the quality by which a surrogate constructed in a computer experiment can predict the simulator outputs (QoIs). 
For this purpose, different approaches have been developed, see~\rfs~\cite{owen:17, owenThesis:17,schobi:15} and the references therein, which can be categorized into two major strategies. 
In the first strategy the predictions by a surrogate are validated using either a new set of data or subsets of the training data. 
The second strategy is based on using a set of diagnostic tools to investigate whether the coefficients or norm of the terms included in PCE~(\ref{eq:pce}), for instance, converge (up to a certain threshold) with the number of training samples and truncation~$K$.

\subsubsection{Gaussian Process Regression (GPR)}\label{sec:gpr}
The PCE (\ref{eq:pce}) is a parametric surrogate which can describe the global behavior of~$f(\chi,\fq)$ over~$\BQ$ through estimating moments~(\ref{eq:pceE}) and~(\ref{eq:pceV}).
In contrast, non-parametric surrogates for~$f(\chi,\fq)$ can be considered with more flexibility in prediction over the input space. 
In particular, here we consider Gaussian process regression (GPR)~\cite{rasmussen:05,gramacy:20}.
According to Rasmussen and Williams~\cite{rasmussen:05}, ``A Gaussian process ($\GP$) is a collection of random variables, any finite number of which have a joint Gaussian distribution."
The natural possibility of using Bayesian formalism besides prediction of an output along with associated uncertainty, makes the GPR a suitable tool for UQ~studies.

The GPR assumes the surrogate $\tf(\chi,\fq)$ for the simulator in~(\ref{eq:surrGen}) to be random, over which a prior distribution in form of a Gaussian process is assumed.
A Gaussian process is fully specified by its mean and covariance function~as, see \rfs~\cite{rasmussen:05,gramacy:20},
\begin{equation}\label{eq:gpr}
\tf(\fq)\sim \GP\left(m(\fq),c(\fq,\fq';\fThetaEps,\bm{\beta})\right) \,,
\end{equation}
where, 
\begin{eqnarray}
m(\fq) &=& \BE[\tf(\fq)] \,, \\
c(\fq,\fq';\fThetaEps,\bm{\beta}) &=& \BE[(\tf(\fq)-m(\fq)) (\tf(\fq')-m(\fq'))] \,.
\end{eqnarray}
In this definition,~$\bm{\Theta}_\varepsilon$ denote the parameters specifying the structure of the observation noise and~$\bm{\beta}$ are the hyperparameters in the kernel function constructed in the space of inputs~$\fq$ to represent the covariance of~$\tf(\cdot)$. 
There are different options for such kernel function, see~\eg~\cite{rasmussen:05}.
Note that in this section we drop~$\chi$ from $\tf(\chi,\fq)$ to ease the notation. 
Also, as it is convenient in the context of GPR, we may refer to~$\fq$ as inputs.

For the so-called universal GPR, the input space is first projected into a high-dimensional space via introducing bases~$\bm{\phi}(q)$. 
A particular choice would be using a set of monomials as bases in each of the~$p$ dimensions of the input space, i.e. $\bm{\phi}(q)=\{1,q,q^2,\ldots,q^{d-1}\}$, where~$d\in \BN$.
As a result \rev{of adopting the weight-space view for the GPR} \cite{rasmussen:05} \revs{and using the training data,} \eq~(\ref{eq:surrGen}) is written~as,
\begin{equation}\label{eq:gp_weight}
\mathbf{R}= \bm{\Phi}(\fQ)\bm{\Theta} + \bm{\feps} \,,
\end{equation}
where \rev{weights $\fTheta$ are of size $pd$} and $\bm{\Phi}(\fQ)$ is a~$n\times p d$ matrix whose $i$-th row is,
\begin{equation}
[\bm{\Phi}(\fQ)]_{i:} = 
[
1,q_1^{(i)} ,q_1^{(i)^2},\ldots,q_1^{(i)^{d-1}},\ldots ,
1,q_p^{(i)} ,q_p^{(i)^2},\ldots,q_p^{(i)^{d-1}}
] \,,
\end{equation}
for $i=1,2,\ldots,n$. 
For these particular settings, the GPR would become essentially similar to PCE~(\ref{eq:pce}) with several extra features. 
Given the training data set~$\cD=\{(q^{(i)},\cR^{(i)})\}_{i=1}^n$ with corresponding observation error~$\feps$ obtained through a computer experiment, the posterior distribution for~$\tf(\fq)$, or equivalently for~$\fTheta$ is inferred conditioned on~$\cD$ and $\bm{\Theta}_\varepsilon$.
Simultaneously, the kernel's hyperparameters $\bm{\beta}$ are optimized. 
Employing the posterior distribution, the posterior predictive at test points~$\fQs$ in the input space can be obtained, \cite{rasmussen:05,gramacy:20}.
It is essential to remark that GPR is an interpolant, so it provides higher accuracy within the input admissible space~$\BQ$. 
To evaluate the quality of a GPR constructed in a computer experiment, different approaches exist, see \rfs~\cite{owenThesis:17,owen:17}. 
If only the mean predictions by GPR are of interest, then the techniques mentioned at the end of \sect~\ref{sec:pce} for PCE can be used. 
For a general case, a set of measures are introduced in~\cite{bastos:09}.

A strong point of using GPR surrogates is the possibility of easily incorporating the uncertainties in training data~$\cD$ into the surrogate construction and hence in the predictions. 
Commonly in the formulation for GPR, the noise~$\varepsilon_i$ for all the~$n$ training data samples are assumed to be independent and identically distributed (iid) by Gaussian distribution~$\cN\sim(0,\sigma_d^2)$ where~$\sigma_d^2$ is fixed, see \rfs~\cite{rasmussen:05,gramacy:20}.
This type of noise is referred as homoscedastic. 
However, for the purpose of the present study we need to relax the assumption of having identical noise levels (yet keeping the independence assumption) and consider heteroscedastic noise structures which allows the noise levels to be input-dependent. 
In particular, the focus is on~$\varepsilon_i\sim \cN(0,\sigma_{d_i}^2)$ for $i=1,2,\ldots,n$, where~$\sigma_{d_i}^2$ are known and constant.\footnote{For GPR with more general heteroscedastic noise structure, see \rfs~\cite{conti:10,fricker:13}.}
In the context of the present study, the value of~$\sigma_{d_i}^2$ represents the magnitude of uncertainty of \catII~which exists in the QoIs of the~$i$-th simulation in a computer experiment comprising~$n$ independent simulations.

As proposed by Goldberg et al.~\cite{goldberg:97} for a GPR with input-dependent noise $\varepsilon_i\sim \cN(0,\bm{\Theta}_{\varepsilon_i})$ where $\bm{\Theta}_{\varepsilon_i}=\sigma^2_{d_i}$, the predicted~$\fR^*$ at new input samples (test samples)~$\fQs=\{\fq^{*^{(i)}}\}_{i=1}^{n^*}$ has a multivariate Gaussian distribution $\cN(m(\fR^*|\cD,\fThetaEps,\fQ^*),v(\fR^*|\cD,\fThetaEps,\fQ^*))$, in which,
\begin{eqnarray}
m(\fR^*|\cD,\fThetaEps,\fQ^*) &=&
\fC(\fQs,\fQ)(\fC(\fQ,\fQ)+\fC_N)^{-1}\mathbf{R}^T \,, \label{eq:gpr_m}\\
v(\fR^*|\cD,\fThetaEps,\fQ^*) &=&
\fC(\fQs,\fQs)-\fC(\fQs,\fQ)(\fC(\fQ,\fQ)+\fC_N)^{-1} \fC(\fQ,\fQs)+\mathbf{e}(\fQs) \,. \label{eq:gpr_v}
\end{eqnarray}
Here, \rev{without loss of generality, mean of $\tf(\fq)$ in (\ref{eq:gpr}) is assumed to be zero, }~$\mathbf{R}=[r^{(1)},r^{(2)},\cdots,r^{(n)}]$,~$\fC(\fQ,\fQ')$ is a~$n\times n'$ matrix for which, $[\fC(\fQ,\fQ')]_{ij}=c(\fq^{(i)},\fq'^{(j)})$ where~$c(\cdot,\cdot)$ is a kernel function dependent on hyperparameters $\bm{\beta}$. 
Moreover,~$\fC_N$ is the covariance matrix of the noise in the training data which is equal to $\text{diag}(\mathbf{e}(\fQ))$, where~$[\mathbf{e}(\fQ)]_i=\sigma_{d_i}^2$ for $i=1,2,\ldots,n$. 
The noise level at the test inputs, \ie~$\mathbf{e}(\fQs)$ is unknown. 
Following Goldberg~\et~\cite{goldberg:97} a GP prior is assumed over the log of the noise variances. 
The kernel function for this GP has the same structure as that of the main GP, however with different values of hyperparameters.
As it is discussed in the next section, the mean predictions by Eq.~(\ref{eq:gpr_m}) are more important in the framework of the present study, and in fact, the confidence intervals (CIs) for those predictions are computed from Eq.~(\ref{eq:gpr_v}).

\subsubsection{Probabilistic PCE (PPCE)}\label{sec:ppce}
Here, we are seeking for a combination of GPR and non-intrusive PCE which takes advantage of the characteristics of both methods.
For such a combination, at least two views can be found in the literature.
In the approach of Sch{\"o}bi \et~\cite{schobi:15}, the mean trend in the universal GPR~(\ref{eq:gp_weight}) is constructed by PCE and the error part is modeled by Gaussian processes. 
In contrast, in the \emph{probabilistic PCE} (PPCE) approach, Owen~\cite{owenThesis:17} proposed to compute the coefficients in the PCE (\ref{eq:pce}) using the Bayesian quadrature technique~\cite{ohagan:91}. 
What is described in this section is in essence similar to the latter, but is more general because of handling observation-dependent noise and also considering more standard distributions for~$\fq$ in PCE.

As discussed above, when we have uncertain factors~$\fq$ of \catI~varying over $\BQ$ and our aim is to estimate the resulting statistical moments of a simulator output~$f(\fq)$, the PCE approach is employed.
In the non-intrusive way, associated to each sample drawn from $\BQ$, a CFD simulation is performed and a realization of the flow QoIs is obtained. 
The output of any of these simulations can be uncertain due to the parameters falling under \catII. 
If these uncertainties are known and can be expressed as Gaussian noise levels, then they, by the use of GPR, can propagate and affect the statistical moments (\ref{eq:pceE}) and (\ref{eq:pceV}) estimated by the standard PCE approach.

To this end, first a GPR surrogate is constructed using the noisy data. 
Then, independent samples from the resulting predictive distribution with mean and variance~(\ref{eq:gpr_m}) and~(\ref{eq:gpr_v}) are drawn.
Any of such samples of the GPR surrogate is denoted by $\tf^{(j)}(\chi,\fq^*)=\tf(\chi,\fq^{*^{(j)}})$, for which a PCE can be constructed by~(\ref{eq:pce}) over~$\BQ$. 
To compute the PCE coefficients, projection or regression methods with any of the truncation schemes~(\ref{eq:pceTP}) or~(\ref{eq:pceTO}) can be employed. 
However, to maximize the convergence of the PCE as well as taking advantage of the framework of gPCE~\cite{xiu:02}, it is recommended to use tensor-product truncation~(\ref{eq:pceTP}) along with the Gauss quadrature rule~(\ref{eq:gpr_GQrule}). 
In this case, the samples~$\xi_i^*\in\Gamma_i$ corresponding to $q_i^*\in\BQ_i$, for $i=1,2,\cdots,p$ are taken as the zeros of polynomial bases $\psi_{n_i}(\xi_i)$ chosen based on the Wiener-Askey rule~\cite{xiu:02} in accordance with PDF~$\rho_i(q_i)$.
Since the evaluation of the surrogate $\tf^{(j)}(\chi,\fq^*)$ corresponding to these samples is computationally inexpensive, an arbitrary number of samples~$\xi_i^*$ over the input space~$\BQ_i$ can be considered. 
Having PCE coefficients $\hat{\mathbf{f}}^{(j)}$ estimated for the $j$-th sample from the GPR surrogate, an estimate for statistical moments (\ref{eq:pceE}) and (\ref{eq:pceV}) can be determined. 
Having a sufficient number of samples, say $n_s$, taken from the GPR predictive distribution, estimates for the sample mean and variance of the statistical moments~(\ref{eq:pceE}) and (\ref{eq:pceV}) are obtained. 
For instance for $\BE_\fq[\tf(\chi,\fq)]$ we get, 
\begin{eqnarray}
\hat{\BE}[\BE_\fq[\tf(\chi,\fq)]] &=&
\frac{1}{n_s} \sum_{j=1}^{n_s} \BE_\fq[\tf^{(j)}(\chi,\fq)] = \frac{1}{n_s}\sum_{j=1}^{n_s} \hf_0^{(j)}(\chi) \,, \\
\hat{\BV}[\BE_\fq[\tf(\chi,\fq)]] &=&
\frac{1}{n_s} \sum_{j=1}^{n_s} \left(  \BE_\fq[\tf^{(j)}(\chi,\fq)] - \hat{\BE}[\BE_\fq[\tf(\chi,\fq)]]\right)^2 \,.
\end{eqnarray} 
Similar expression can be written for $\BV_\fq[\tf(\chi,\fq)]$ given by (\ref{eq:pceV}). 
The estimators $\hat{\BE}[\cdot]$ and $n_s\hat{\BV}[\cdot]$ will respectively converge to the population values~$\BE[\cdot]$ and $\BV[\cdot]$ as $n_s \to \infty$, see~\eg~\cite{ramach:09}.

\subsection{Sensitivity Analysis}\label{sec:sobol}
So far, appropriate tools for UQ forward problems have been discussed, so that the uncertainty propagated into the simulator outputs from a $p$-dimensional uncertain~$\fq$ can be estimated. 
As a complement to this, tools from global sensitivity analysis (GSA) are utilized to rank the~$p$ components of~$\fq$ based on their influence on the output~$\cR$. 
Contrary to the local sensitivity analysis, where the sensitivity of~$\cR$ to small perturbations of~$\fq$ is estimated, in GSA all~$\fq$ are let to simultaneously vary over their admissible space.
For conducting GSA, there are different approaches, see~\eg~\cite{uqHandbook,smith}.
Here, a short description for computing Sobol sensitivity indices is~provided. 
Once again, we drop $\chi$ from $f(\chi,\fq)$ for simplifying the notation.

By analysis of variance (ANOVA) or Sobol decomposition~\cite{sobol:90,sobol:01}, a model function is decomposed as,
\begin{equation}\label{eq:anova_f}
f(\fq) = 
f_0 +\sum_{i=1}^p f_i(q_i) + \sum_{1\leq i<j\leq p} f_{ij}(q_i,q_j)+\cdots\,,
\end{equation}
where,~$f_0$ is the mean of $f(\fq)$,~$f_i(q_i)$~specify the contribution of each parameter, $f_{ij}(q_i,q_j)$ denote effects of
interaction between each pair of parameters, and so on for other interactions. 
These contributors are defined~as,
\begin{eqnarray*}
f_0 &=& \BE_\fq[f(\fq)] \,, \\
f_i(q_i) &=& \BE_\fq[f(\fq)|q_i] - f_0 \,, \\
f_{ij}(q_{i},q_j) &=& \BE_\fq[f(\fq)|q_i,q_j] -f_i(q_i) -f_j(q_j) - f_0 \,. 
\end{eqnarray*}
Here, $\BE_\fq[f(\fq)|q_i]$, for instance, denotes the expected value of $f(\fq)$ conditioned on fixed values of $q_i$. 
Similar to \eq~(\ref{eq:anova_f}), the total variance of $f(\fq)$, denoted by $D$, is decomposed as, 
\begin{equation}
\BV_\fq[f(\fq)] = D=\sum_{i=1}^p D_i + \sum_{1\leq i<j\leq p} D_{ij} + \cdots \,,
\end{equation}
where, $D_i=\BV_\fq[f_i(q_i)]$, $D_{ij}=\BV_\fq[f_{ij}(q_i,q_j)]$, and  so on. 
The main Sobol indices are eventually defined as the contribution of
each of~$D_i$, $D_{ij}$, ... in the total variance $D$:
\begin{equation}\label{eq:sobol}
S_i=D_i/D\,,\quad 
S_{ij}=D_{ij}/D \,,\, \ldots \,, \quad i,j=1,2,\cdots,p
\end{equation}
In connection with a computer experiment, a surrogate $\tf(\fq)$ can be employed to compute the Sobol indices~(\ref{eq:sobol}). 
Then, integrals appearing in different expectations and variances are numerically evaluated.
In case of using the probabilistic PCE discussed in \sect~\ref{sec:ppce}, it is possible to construct confidence intervals for the estimated Sobol indices.

\subsection{Implementation of the UQ Techniques}\label{sec:uqImplmnt}
A \texttt{Python} toolbox has been developed within which different UQ techniques including those used in this study are implemented.
The \revs{toolbox, which will be released in open-source,} can be non-intrusively linked to any CFD solver through appropriate interfaces. 
The PCE technique as well as the sensitivity analysis using Sobol indices are implemented using standard libraries such as \texttt{numpy}~\cite{numpy} and \texttt{scipy}~\cite{scipy}. 
For solving the optimization problem (\ref{eq:cmprsdSens}) when using PCE with compressed sensing, the library \texttt{cvxpy}~\cite{cvxpy} is employed.  
For GPR, we use \texttt{GPyTorch}~\cite{gpytorch} in which the optimization of the kernels' hyperparameters are performed using the ADAM algorithm~\cite{adam}.

\section{Accuracy, Robustness, and Sensitivity Analysis}\label{sec:uqCFD}
In this section, it is shortly explained how the UQ tools introduced in \sect~\ref{sec:uq} can be employed when assessing accuracy, robustness and sensitivity of the CFD simulations. 
To measure the robustness of a QoI in a computer model (\ref{eq:surrGen}), we need to estimate the variance~$\BV_\fq[f(\chi,\fq)]$ in a UQ forward problem where the parameters~$\fq$ vary over~$\BQ$ with a given distribution.
This can be achieved by non-intrusive application of the standard or probabilistic PCE methods reviewed in \sects~\ref{sec:pce} and~\ref{sec:ppce}, respectively.
As a result of using standard PCE, $\BE_\fq[f(\chi,\fq)]\pm t_{\alpha} \sqrt{\BV_\fq[f(\chi,\fq)]}$ is obtained, in which~$t_{\alpha}$ is a constant value that is looked up from the t-table~\cite{ramach:09} for a desired confidence level. 
In this study, we take~$\alpha=95\%$ which leads to~$t_{95\%}=1.96$. 
In case of using probabilistic PCE, different combinations of moments due to the variation of~$\fq$ (\catI~parameters) as well as uncertainties of \catII~are achieved. 
Comparable to the result of standard PCE, we get $\hBE[\BE_\fq[f(\chi,\fq)]]\pm t_{\alpha} \sqrt{\hBE[\BV_\fq[f(\chi,\fq)]]}$. 
The uncertainty term (under the square root) in this expression is itself uncertain for which we have, $\hBE[\BV_\fq[f(\chi,\fq)]]\pm t_{\alpha} \sqrt{\hBV[\BV_\fq[f(\chi,\fq)]]}$. 
Besides these, the uncertainty in the mean expected value $\hBE[\BE_\fq[f(\chi,\fq)]]$ is represented by $\pm t_{\alpha} \sqrt{\hBV[\BE_\fq[f(\chi,\fq)]]}$.

As discussed in~\sect~\ref{sec:intro}, the accuracy is defined as the deviation of the QoIs (simulator's outputs) from what is considered as the reference true value. 
Such a value would be obtained from a high-fidelity simulation or experiment for verification and validation purposes, respectively. 
To measure the a-posteriori error as an indicator of the accuracy for a QoI~$r(\chi,\fq)$, an option is, 
\begin{equation}\label{eq:qoiErrNorm_gen}
\epsilon[\cR(\chi,\fq)] = \|\cR(\chi,\fq)-\cR^\circ(\chi)\|/\|\cR^\circ(\chi)\| \,,
\end{equation}
in which, the symbol~$^\circ$ specifies the reference true value. 
In particular, we use the~$l_\infty$ norm, since it is more restrictive than the~$l_1$ and~$l_2$ norms, see \eg~\cite{kreyszig:78}. 
As formulated here, the error in a QoI can vary due to both \catI~and \catII~uncertainties.  
Therefore,~$\epsilon[\cR(\chi,\fq)]$ can be treated as another output to which different UQ techniques can be applied. 
Of particular importance is the construction of error surfaces in the admissible space of $\fq$.
As pointed out in \rfs~\cite{meyers:07,meyers:10,salehPoF:18}, the resulting error portraits are capable of indicating the complexities in the variation of errors in QoIs of turbulent flows. 
This confirms the shortcoming of the classical numerical analysis techniques in predicting the errors in the averaged QoIs obtained by scale-resolving simulations of turbulent flows.

\section{Illustrative Example}\label{sec:example}
\subsection{Overview}
The focus of this section is on illustrating how the tools introduced in the previous sections can be utilized for the purpose of uncertainty quantification of a standard CFD simulation. 
As a showcase, the wall-resolved simulation of fully-developed turbulent channel flow is considered. 
This canonical case has been extensively used in the literature for studying the physics of wall-bounded turbulence and also for developing and testing numerical algorithms.

The channel comprises of two parallel walls at which no-slip \rev{and no-penetration} boundary conditions for velocity are imposed. 
\revs{The channel half-height is denoted by~$\delta$ and the normalized wall-normal coordinate~$y$ is measured from the bottom wall and varies between~$0$ and~$2$.}
The flow is periodic in the streamwise and spanwise directions which are associated with coordinates~$x$ and~$z$, respectively. 
The flow is driven by fixed mass flux, which reduces to constant bulk velocity $U_b$ for constant density flows. Dimensional analysis shows that only one parameter is needed to fully describe turbulent channel flow, which can be chosen as the bulk Reynolds number~$\rey_b=U_b \delta/\nu$, with the kinematic viscosity $\nu$.

The channel flow simulations are carried out using the high-order spectral-element~\cite{deville:02} open-source solver~\nek~\cite{nek}, see~\sect~\ref{sec:cfdSolver}. 
Starting from initial conditions consisting of a laminar velocity profile with random perturbations, the flow undergoes transition and eventually becomes fully turbulent. 
Quantities of interest of the channel flow simulation can be any functional of the simulation output. 
As detailed in \sect~\ref{sec:chanCompExp}, the QoIs are taken to be the flow quantities averaged over time and periodic directions, after the fully-\rev{developed} turbulent state is established.

\subsection{CFD Solver}\label{sec:cfdSolver}
The numerical code \nek~used in this study was developed by Fischer \et~\cite{nek}.
This code is based on the spectral-element method (SEM), which was originally proposed by Patera~\cite{patera}, in which so-called spectral elements are used to decompose the domain. 
The incompressible Navier--Stokes equations are cast in weak form, and the velocity and pressure fields are expressed in terms of Lagrange interpolants of Legendre polynomials, at the Gauss--Lobatto--Legendre (GLL) quadrature points within the elements. 
Note that the method allows flexible distributions of the spectral elements, a fact that enables highly accurate simulations of turbulent flows in moderately complex geometries (see for instance \rfs~\cite{duct_prf,wing_1M}), while the GLL grid-point distributions within the elements are prescribed by the polynomial order~$N$. 
All the simulations discussed here were conducted using the~$\mathbb{P}_{N}-\mathbb{P}_{N}$ formulation, \ie~polynomials of order~$N$ are used both for the velocity and the pressure. 
In \nek, the nonlinear terms are treated explicitly by third-order extrapolation and the viscous terms are treated implicitly by third-order backward differentiation. 
Velocity and pressure are decoupled and solved iteratively, the former through conjugate gradients and the latter by means of the generalized minimal residual (GMRES) method.
Jacobi preconditioning is used for the velocity, and the additive overlapping Schwarz method is used to build a pressure preconditioner~\citep{schwarz}. 
Two methods are considered for the coarse-grid solve from the latter: the first one (more adequate for smaller problems) is based on a Cholesky factorization and it is denoted as XXT~\citep{xxt}, and the second one is based on an algebraic multigrid (AMG) solver~\citep{amg}. 
Additional details on the way in which the governing equations are solved in \nek~can be found in \rf~\cite{offermans}. 
Furthermore, and although in earlier versions of \nek~an explicit-filtering approach was adopted~\cite{fischer_mullen}, in this work we consider the approach proposed by Schlatter \et~\cite{rt_schlatter}, in which the Navier--Stokes equations are supplemented by a dissipative term based on a high-pass spectral filter (additional details are provided in \rfs~\cite{negi:17}).
Note that \nek~is written in Fortran77/C, and that the message-passing interface (MPI) is used for parallelism.

\subsection{Computer Experiment}\label{sec:chanCompExp}
In a computer experiment, the effect of \rev{inputs}~$\fq$ with a joint PDF~$\rho(\fq)$ varying over~$\BQ$ is evaluated on a set of simulation outputs. 
Here, we consider numerical parameters~$\fq=(\dxp,\dzp)$ where~$\dxp$ and~$\dzp$ are the average spacing between the GLL points in the streamwise and spanwise directions of a channel flow, respectively.
The superscript~$^+$ denotes normalization with respect to the turbulent viscous length scale~$\delta_\nu = \nu/u^\circ_\tau$, where~$u^\circ_\tau$ is the true nominal friction velocity. 
From a reference simulation it is possible to know a-priori the friction-based Reynolds number~$\reyt^\circ=u^\circ_\tau \delta/\nu$ corresponding to a~$\reyb$ employed to setup the channel flow. 
In the illustrative example, we consider a channel flow at~$\reyb=5019.55$ with~$\reyt^\circ=279.899$, according to the DNS by Iwamoto~\et~\cite{iwamoto:02}. 
\rev{Note that choosing this particular reference database is arbitrary and depending on the target~$\reyb$, the DNS data from other resources such as \rfs~\cite{hoyas:08,lee:15} could alternatively be used. 
In any case, as it is inferred from its definition in \sect~\ref{sec:intro}, the accuracy of a QoI generally depends on the chosen reference data.}

In a channel flow simulation, the size of the grid elements in both streamwise and spanwise directions are fixed. 
Therefore, 
\begin{equation}
\dxp = \frac{l_x/\delta}{E_x N} \reyt^\circ \,,
\end{equation}
where,~$l_x$ and~$E_x$ specify domain length and number of elements in the streamwise direction, respectively, and\rev{~$N$ is the polynomial order in the spectral-element method which is taken to be~$7$ for the present simulations.}
A similar expression can be written for~$\dzp$. 
\revs{In the computer experiments to obtain different~$\dxp$ and~$\dzp$ for given~$l_x$ and~$l_z$, the number of elements~$E_x$ and~$E_z$ are changed.}
To ensure the insensitivity of the analysis with respect to~$l_x$ and~$l_z$, we consider a large domain with $l_x/\delta=16$ and $l_z/\delta=9$ after trying different lengths. 
Small modifications of the chosen lengths to setup simulations associated with \revs{some} values of $\dxp$-$\dzp$ samples may have been applied. 
In the wall-normal direction, the element size decreases towards the wall, in such a way that the first off-wall GLL point has~$\dyp_w=0.45$ in all simulations. 
This ensures an adequate resolution in the near-wall turbulent field and hence removes the potential uncertainty due to the no-slip boundary condition at the wall.

Two admissible spaces $\BQ=\BQ_\dxp \otimes \BQ_\dzp$ for $\fq$ are considered: 
$\BQ_1=[5,130]\otimes [5,70]$ and $\BQ_2=[5,50]\otimes [5,30]$. 
The CFD simulations are performed for the samples taken from $\BQ_1$.
Then, having a surrogate constructed on $\BQ_1$, the QoIs can be interpolated to any arbitrary number of samples in $\BQ_2\subset\BQ_1$.

The QoIs are taken to be the averaged wall friction velocity~$\langle u_\tau\rangle$, averaged velocity \rev{profile}~$\langle u_i\rangle$ and second-order moments of velocity, $\langle u'_i u'_j \rangle$~\rev{(so called Reynolds stresses)}, where $u'_i=u_i-\langle u_i \rangle$ and~$i,j=1,2,3$.
\rev{Note that indices $1$, $2$, and $3$ refer to the streamwise, wall-normal, and spanwise directions, respectively.
\revs{The general toolbox~\cite{nek-stats} can be used to compute the turbulence statistics in \nek.}
Using the second-order moments of velocity, the root-mean-square (rms) velocity fluctuations are defined as, $u'_{\rms_i}=\sqrt{\langle u'_i u'_i \rangle}$ for $i=1,2,3$.}
The averaged profiles are only dependent on the wall-normal coordinate,~$y$. 
Note that as a property of the spectral-element method, the values of quantities can be interpolated at any arbitrary number of points in an element using Lagrange interpolation constructed from the original GLL points in that element. 
\rev{Using this feature in post-processing the simulation results, the values of the discrete profiles are obtained at~$N_g$ equi-spaced points in the wall-normal direction.}

Assessment of the uncertainties in the above QoIs due to time-averaging using the methods in \rfs~\cite{russo:17,oliver:14,vinuesa:16} is left to be thoroughly studied in a future study. 
In this regard, computing~$\langle\cdot\rangle$ is performed over a long time span so that the associated uncertainties become negligible and do not hinder quantifying uncertainties due to the variation of~$\fq$.
Nevertheless, with the use of GPR and PPCE, it is shown how to combine the uncertainties due to time averaging or other \catII~factors with the \catI ~uncertainties.

\subsubsection{Accuracy \& Sensitivity Analysis}
Following the discussion in \sect~\ref{sec:uqCFD}, the error in the QoIs of channel flow simulations are defined as a normalized deviation from a reference true data (specified by~$^\circ$), which are here taken to be the DNS data of Iwamoto~\et~\cite{iwamoto:02}. 
For a scalar QoI, such as~$\lut$, the error is defined as,
\begin{equation}
\epsilon[\lut] = (\lut - \lut^\circ)/\lut^\circ \,.
\end{equation}
\rev{Note that $u_{\tau}$ is a very relevant quantity in the context of wall-bounded turbulence, and that errors in $u_{\tau}$ may lead to inadequate interpretations of the data~\citep{noise_eif}.} For a QoI that is averaged over both time and periodic directions, the profile depends on the wall-normal coordinate~$y$ and the definition~(\ref{eq:qoiErrNorm_gen}) is adopted. \rev{This means in practise that the wall-normal maximum of the deviation is considered (maximum norm)}.

\begin{figure}
\centering   
   \begin{tabular}{cc}
      \includegraphics[scale=0.4]{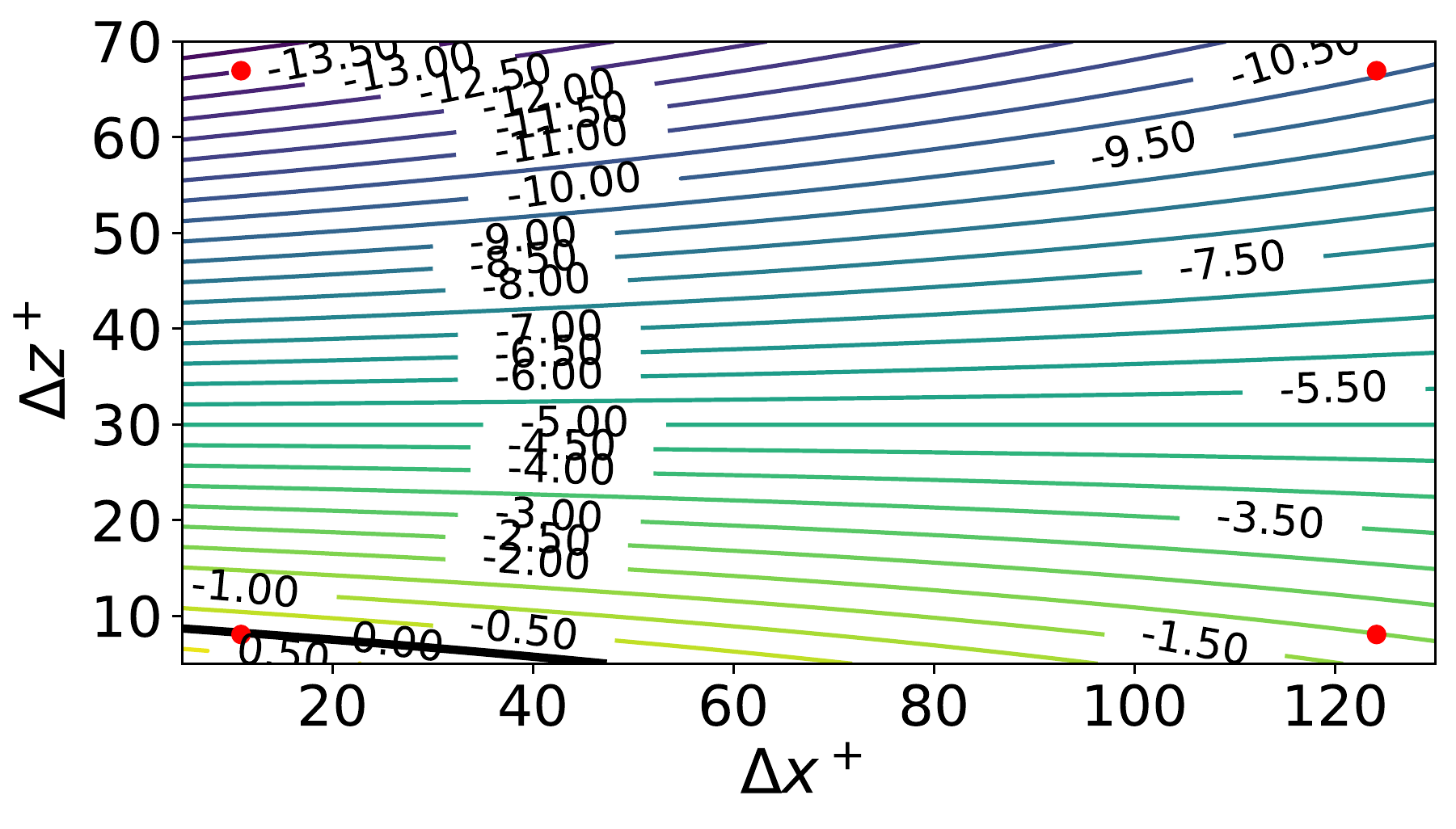} &
      \includegraphics[scale=0.4]{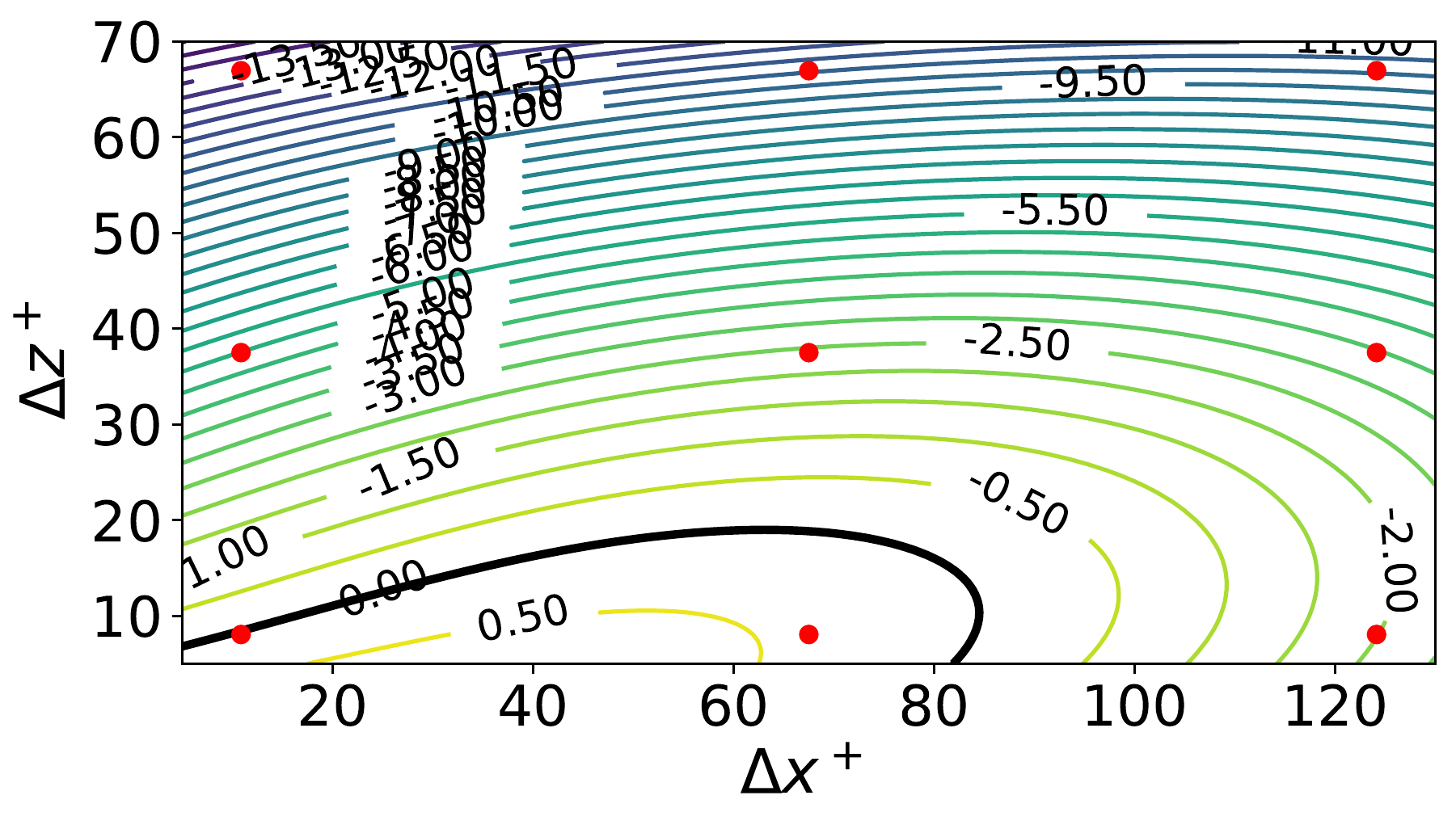} \\
      {\small{(a)}} &    {\small{(b)}} \\      
      \includegraphics[scale=0.4]{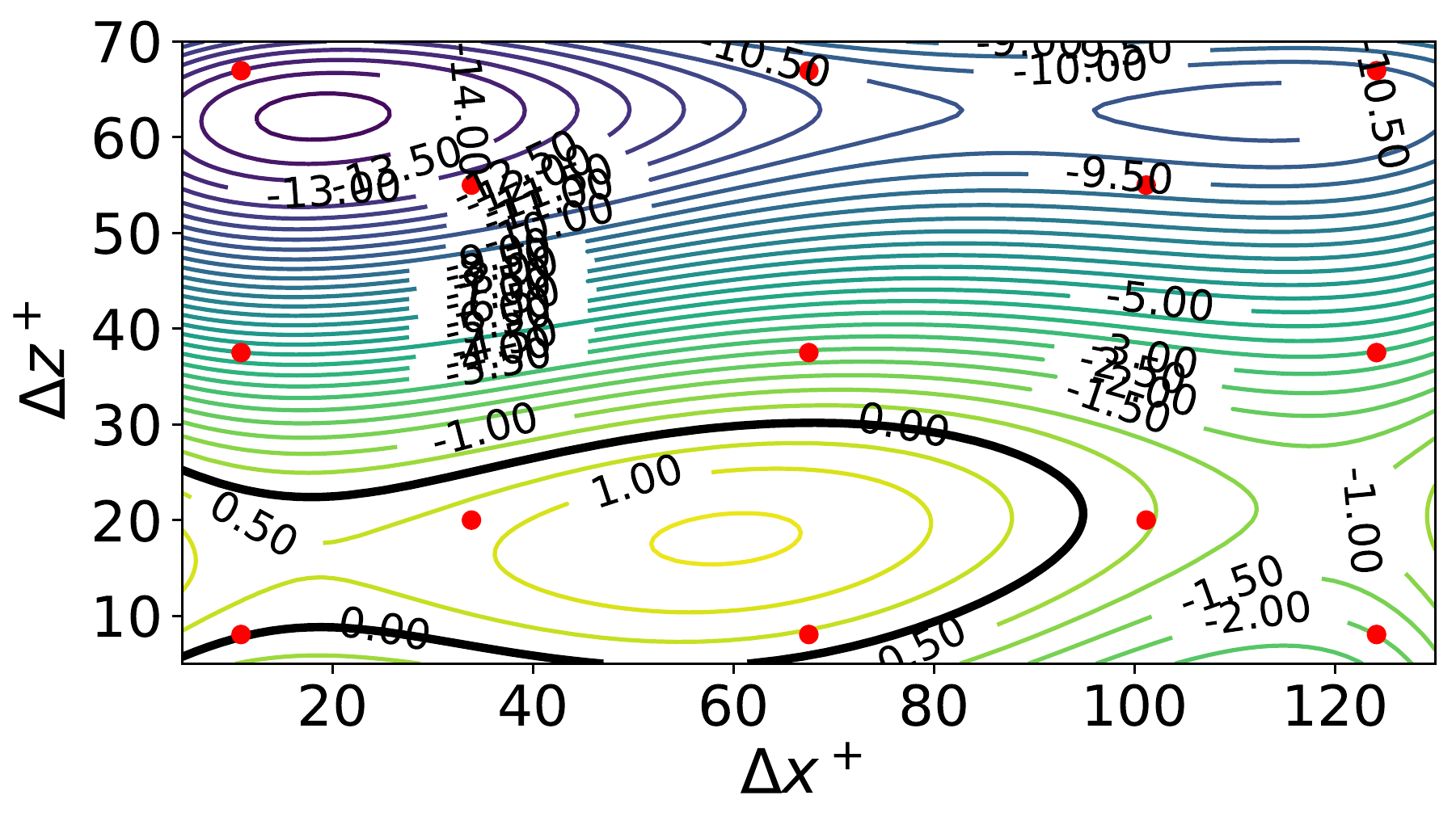} &
      \includegraphics[scale=0.4]{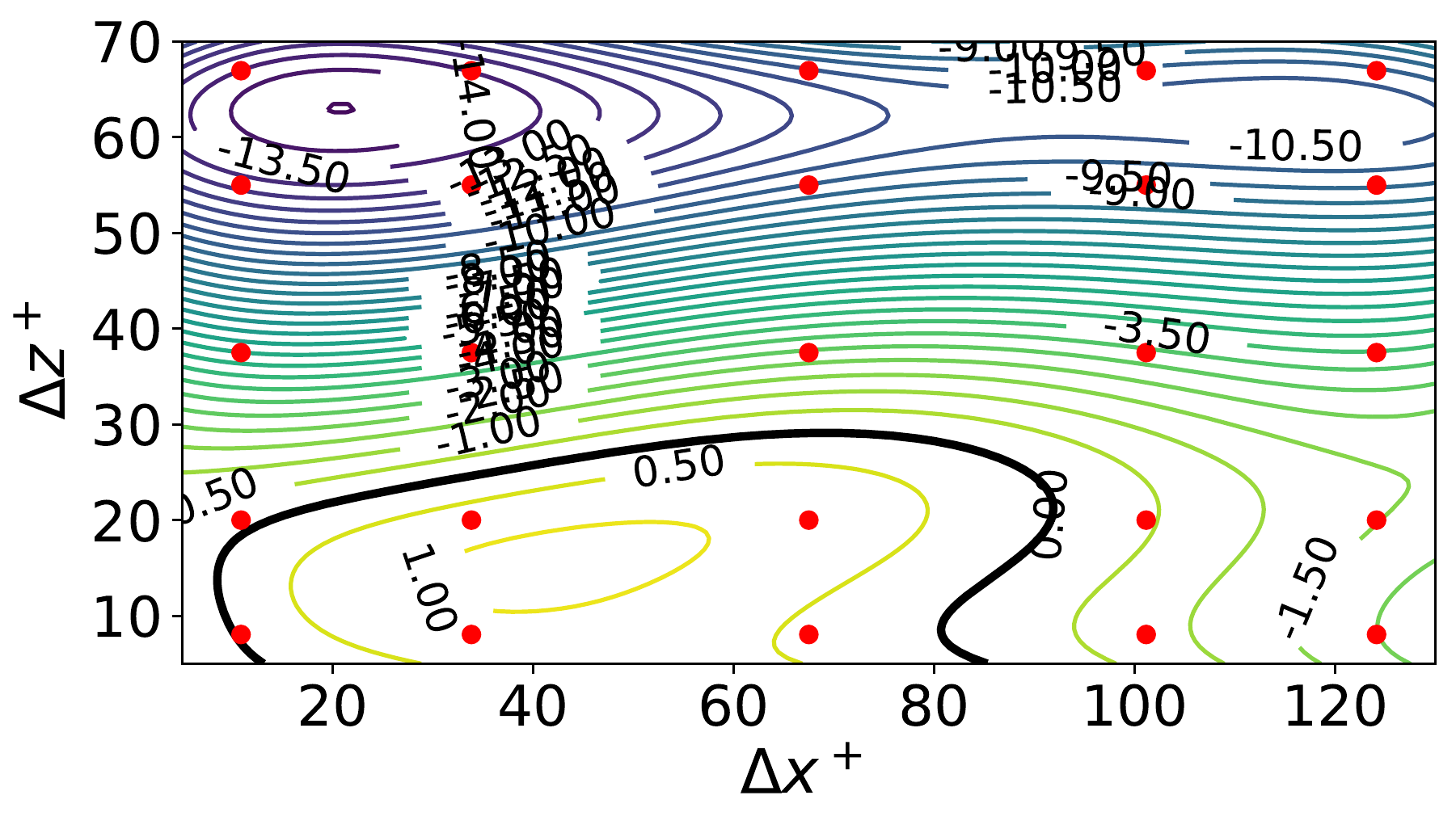} \\
      {\small{(c)}} &    {\small{(d)}} \\            
   \end{tabular}
   \caption{Isolines of $\epsilon[\lut]\%$ in $\BQ_1=[5,130]\bigotimes[5,70]$, the admissible space of $\dxp\bigotimes\dzp$. The surfaces are constructed via~(\ref{eq:pce}) using $n=4$~(a), $9$~(b), $13$~(c), and $25$~(d) training samples specified by red dots. \revs{The uncertainty in the training data due to time-averaging is assumed to be zero. The thick line represents zero error in averaged friction velocity~$\lut$.}}      
   \label{fig:duTauSurface}
\end{figure}

\fig~\ref{fig:duTauSurface} represents \revs{the isolines of} $\epsilon[\lut]\%$ in the $\dxp\dash\dzp$ plane (\ie~the parameter space). In each plot, a different number of parameter samples is used to construct the response surface using standard PCE. 
The total-order truncation scheme~(\ref{eq:pceTO}) with~$L=12$ is considered to ensure that all significant contributions in PCE~(\ref{eq:pce}) are captured. 
Since the number of samples are less than the chosen~$K=90$, the coefficients in~(\ref{eq:pce}) are computed using the compressed sensing method~(\ref{eq:cmprsdSens}).
For all cases except \fig~\ref{fig:duTauSurface}(c), tensor-product truncation~(\ref{eq:pceTP}) can alternatively be used. 
It is also emphasized that the same response surfaces could be obtained using surrogates constructed by GPR or Lagrange interpolation~\cite{smith}.
The latter is restricted to the cases with tensor-product samples in the parameter space.

\begin{figure}
\centering
   \begin{tabular}{cc}
      \includegraphics[scale=0.41]{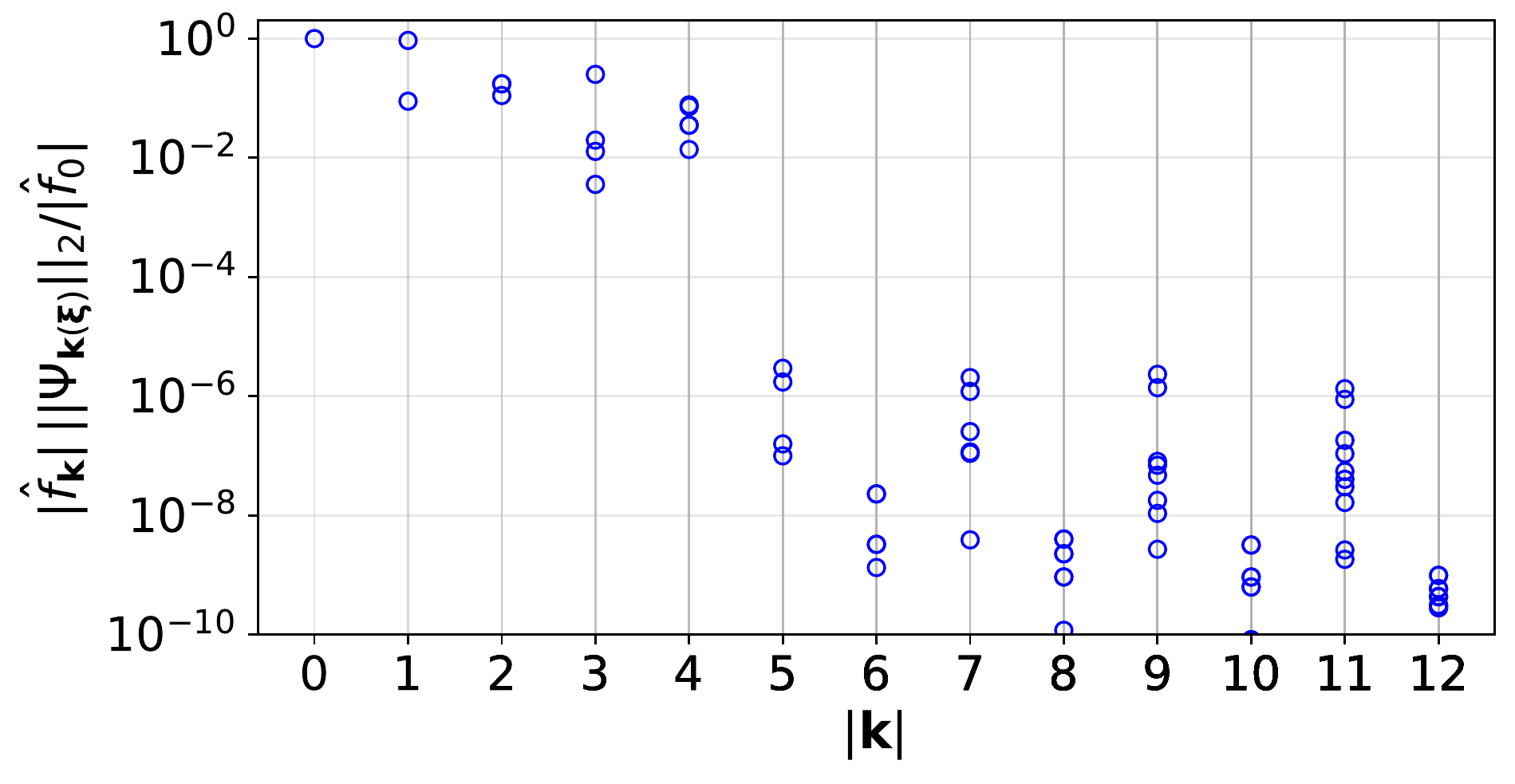} &
      \includegraphics[scale=0.41]{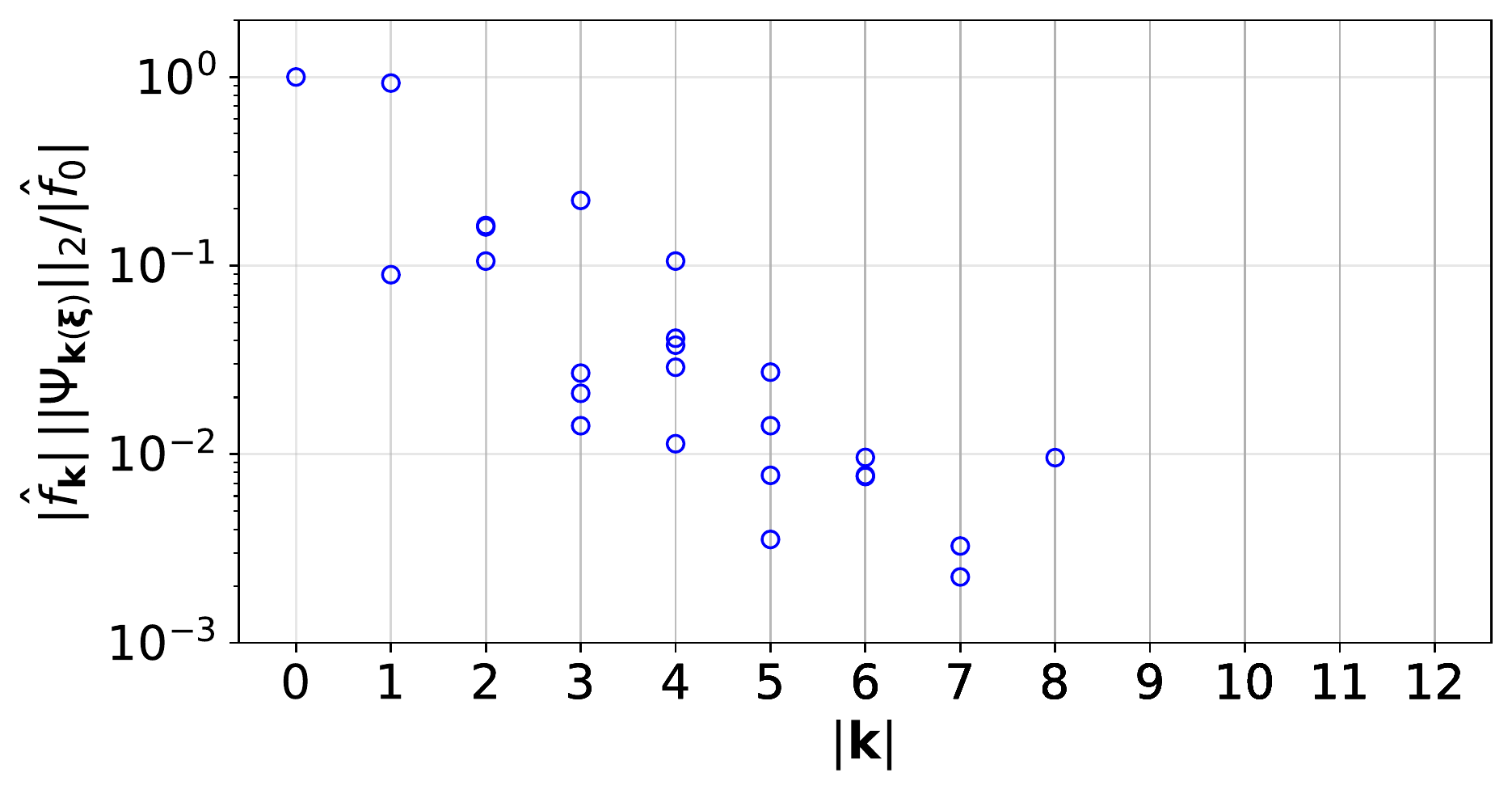} \\
      {\small{(a)}} &    {\small{(b)}} \\                  
   \end{tabular}
   \caption{Convergence of $\vartheta_\fk$ defined by~(\ref{eq:pceConvNorm}) for the PCE constructed for $\epsilon[\lut]$ using total-order truncation with $L=12$ and $n=13$~(a) and $n=25$~(b) training samples. Note that the plots~\revs{(a) and (b)} correspond to \fig~\ref{fig:duTauSurface}(c) and (d), respectively. The horizontal axis is $|\fk|=k_1+k_2$. In (a), the data with values smaller than $10^{-10}$ are not shown.} \label{fig:pceConv_uTau}
\end{figure}

As pointed out in \sect~\ref{sec:pce}, it is important to examine the quality of a surrogate in a computer experiment. 
\rev{In this regard,} \fig~\ref{fig:pceConv_uTau} shows the convergence of the PCE associated to the plots in \figs~\ref{fig:duTauSurface}(c) and~(d). 
The contribution of the terms in expansion~(\ref{eq:pce}) has been quantified by evaluating the following diagnostic,
\begin{equation}\label{eq:pceConvNorm}
\vartheta_\fk:=|\hf_\fk |\|\bm{\Psi}_\fk(\bm{\xi})\|_2/|\hf_0| \,.
\end{equation}
As a \rev{prerequisite for} ensuring that the predictions by non-intrusive PCE are valid, the values of~$\vartheta_\fk$ should exhibit a decreasing trend with~$|\fk|$.
\rev{This holds for both~$n=13$ and~$25$ cases, as shown in \fig~\ref{fig:pceConv_uTau}}.
\rev{However,} when using~$n=25$ samples, there are \rev{relatively significant} contributions from~$|\fk|\geq 5$ \rev{which are} absent from the case of~$n=13$ samples.
\rev{Further investigations (not shown here) indicate that adding samples beyond~$n=25$ does not significantly modify the contours in \fig~\ref{fig:duTauSurface}(d).}
\rev{In the case of~$n=25$ samples, there is a distinct division between the values of $\vartheta_\fk$: those with the values higher than $10^{-3}$ which are shown in \fig~\ref{fig:pceConv_uTau}(b), and the rest which are less than $10^{-20}$ (not shown here).
The $\vartheta_\fk$ illustrated in \fig~\ref{fig:pceConv_uTau}(b) are the same if the tensor-product truncation scheme~(\ref{eq:pceTP}) with the projection method~(\ref{eq:pceProj}) is alternatively used. 
Therefore, for the $n=25$ case, the use of higher number of terms in expansion (\ref{eq:pce}) does  not have a substantial effect.  
}

Returning to \fig~\ref{fig:duTauSurface}, the use of higher number of training samples clearly leads to revelation of more details of the response surfaces, although in practice there should be a balance between the number of samples and the cost of running the simulator. 
In particular, observe the variation of the black curves which specify infinite combinations of resolutions resulting in~$\epsilon[\lut]=0$. The existence of these curves, which can be due to error cancellations in the CFD simulator, besides the complexity of the error isolines which are solver-dependent, see \eg~\cite{meyers:07,salehPoF:18}, prove the challenges involved in evaluation of the accuracy of CFD~simulations. 
These also cast doubt on the legitimacy of using techniques such as classical Richardson extrapolation for the purpose of error estimation in turbulent flows, as it has been done for instance in~\cite{celik:05,klein:05} in case of LES.  
Seeking for the alternatives has, for instance, led to the development of a Bayesian extension of the Richardson extrapolation technique by Oliver~\et~\cite{oliver:14}. 
\rev{It is also important to note that the pattern and values of the error isolines can be dependent on the reference data with respect to which the accuracy in the QoIs is evaluated.}

\begin{figure}
\centering 
   \begin{tabular}{cc}
   \includegraphics[scale=0.42]{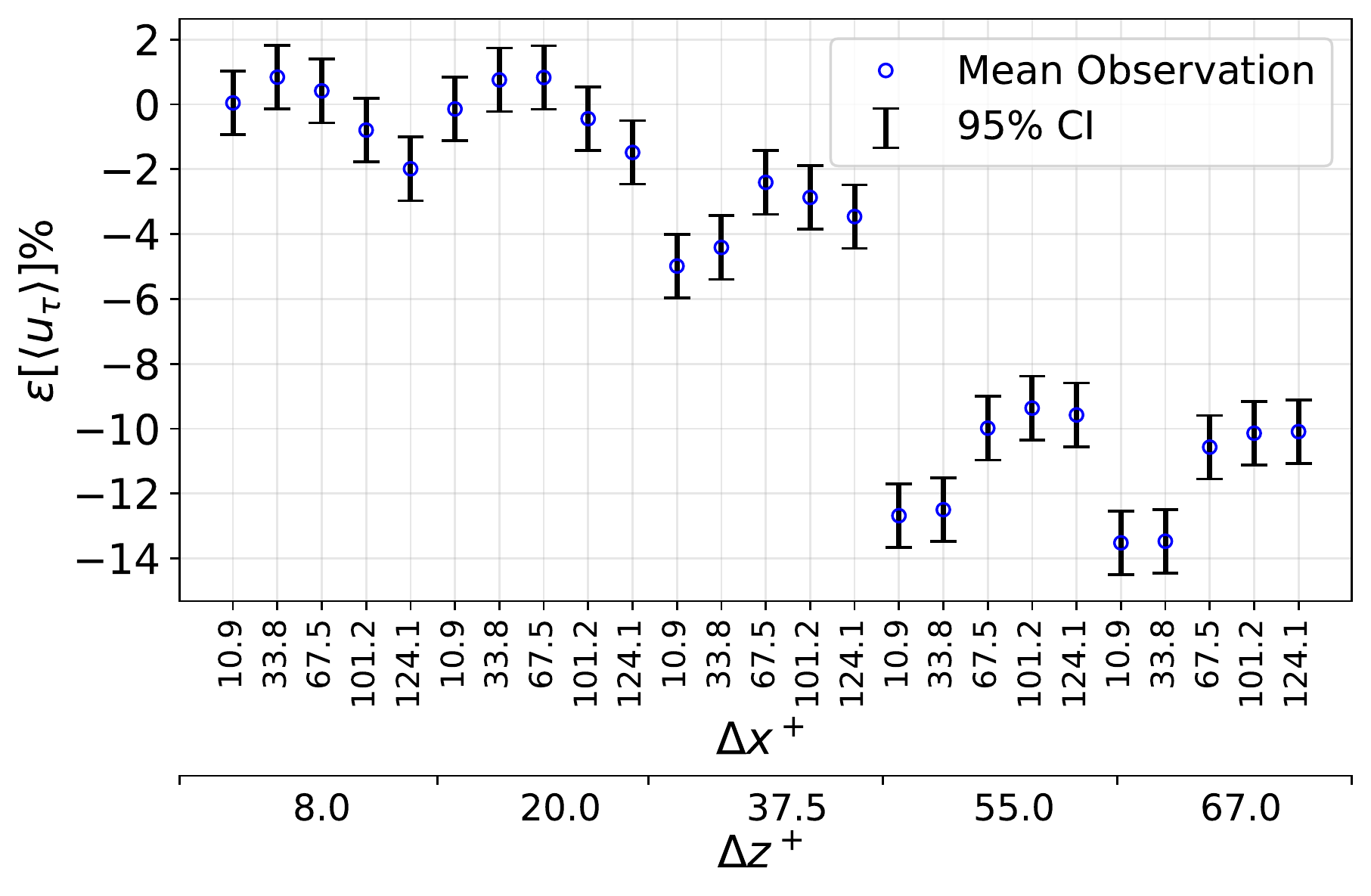} &
   \includegraphics[scale=0.42]{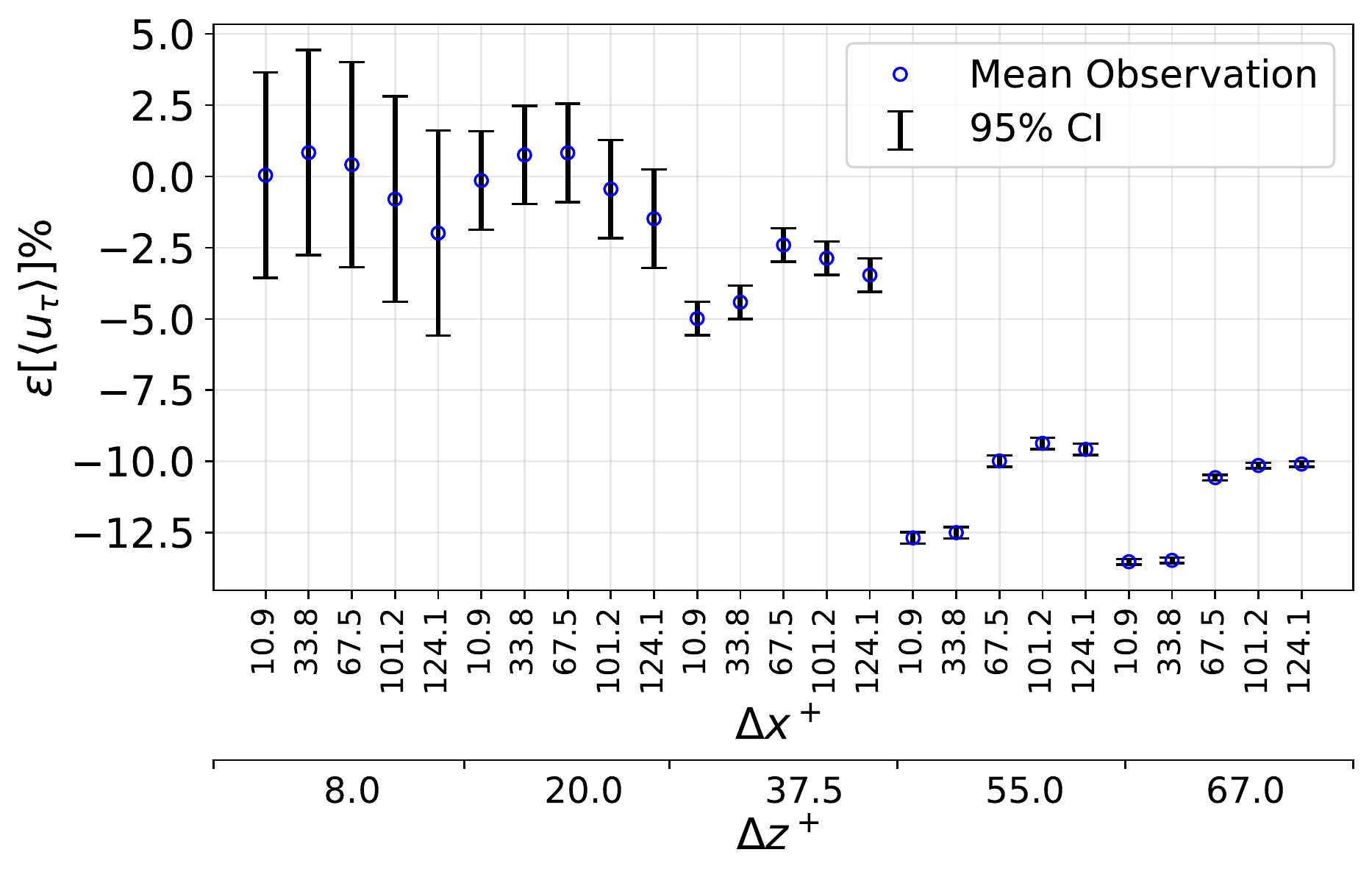} \\   
      {\small{(a)}} &    {\small{(b)}} \\        
   \end{tabular}
   \caption{\revs{Synthetic uncertainties ($95\%$ CI) added to the mean observed values of $\epsilon[\lut]\%$.}
   The mean values are taken from the 25 simulations in \fig~\ref{fig:duTauSurface}(d) \revs{(corresponding to the red dots)} in the computer experiment \revs{with $\fq=\{\dxp,\dzp\}$}. \revs{The added uncertainties are} synthetic Gaussian noise \revs{which imitate the \catII~uncertainties.} The assumed noise is $~\cN(0,\sigma^2_{d_i})$ with $i=1,2,\ldots,25$ where $\sigma_{d_i}=0.5\%$, homoscedastic noise~(a) and  $\sigma_{d_i}=0.3\exp(-2\xi_{\dzp,i})\%$, heteroscedastic noise~(b). In the latter, $\xi_{\dzp,i}\in[-1,1]$ corresponds to the $\dzp$ of the sample simulation. \revs{These synthetic data are used in \fig~\ref{fig:duTauSurfNoisy}.}}
   \label{fig:duTauGPRnoise}
\end{figure}

\begin{figure}
\centering
   \begin{tabular}{ccc}
   {\small Mean Observation} & {\small Upper $95\%$ CI} & {\small Lower $95\%$ CI}\\
   \includegraphics[scale=0.3]{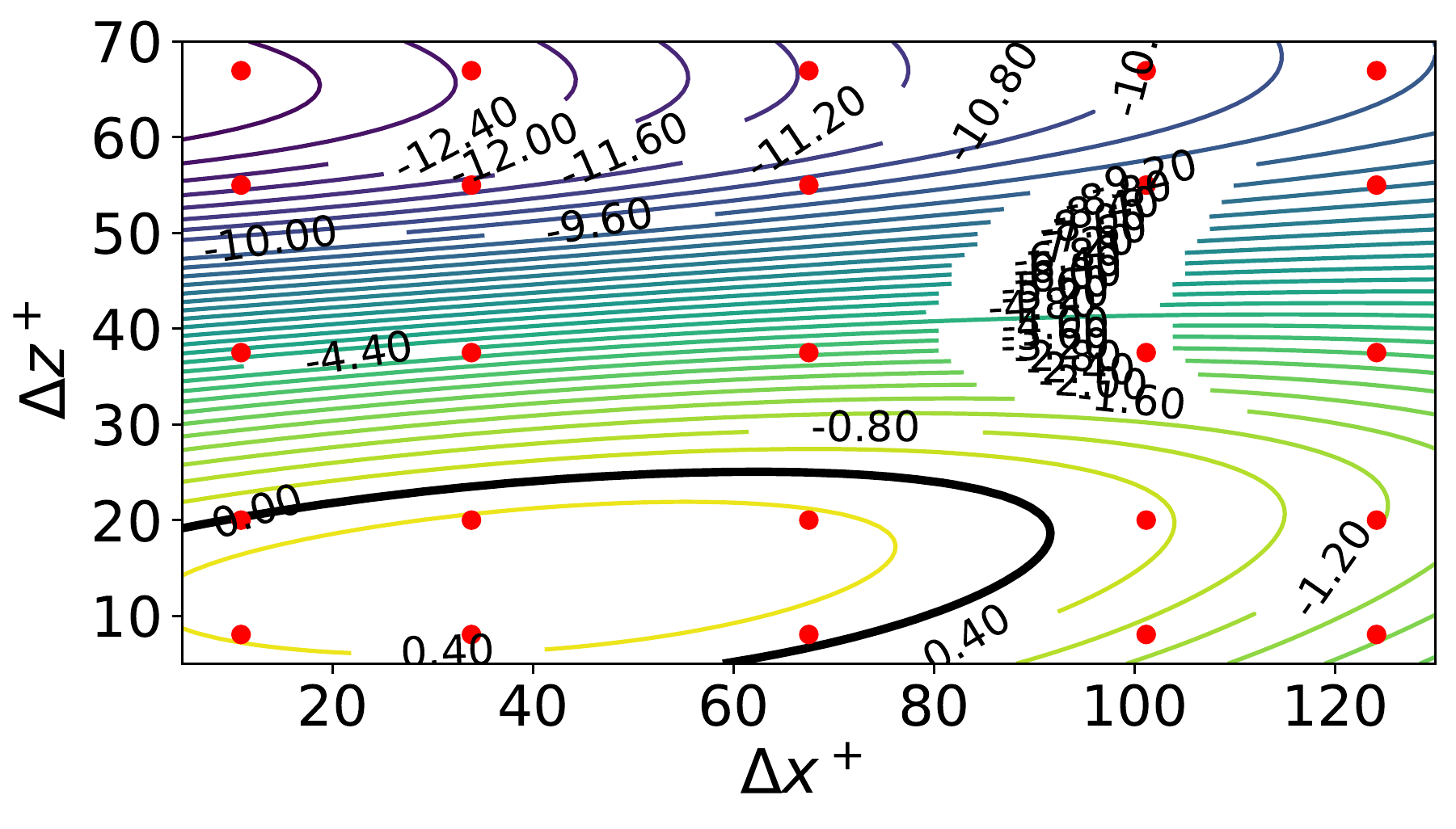} & \hspace{-0.5cm}
   \includegraphics[scale=0.3]{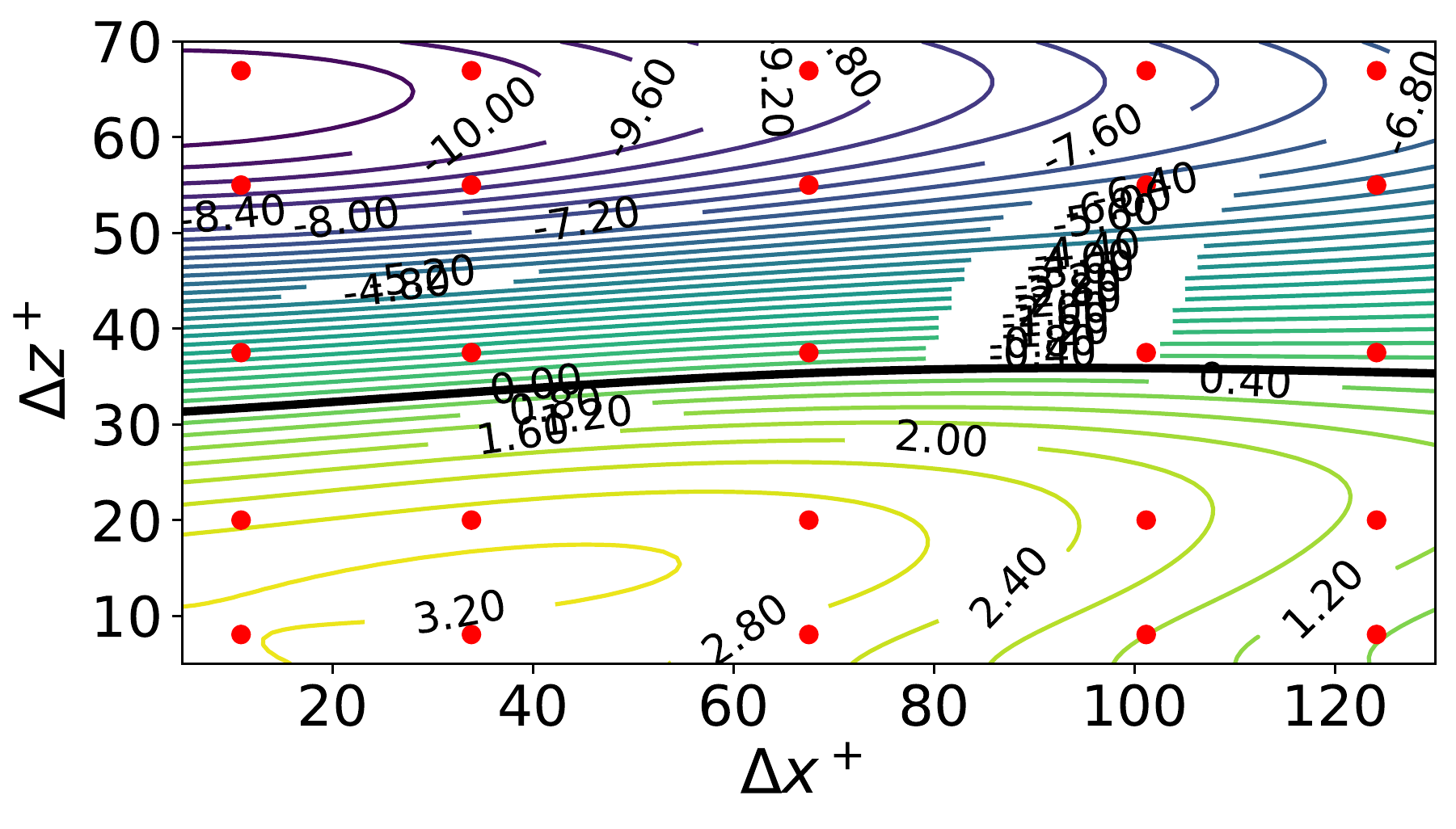} &\hspace{-0.5cm}
   \includegraphics[scale=0.3]{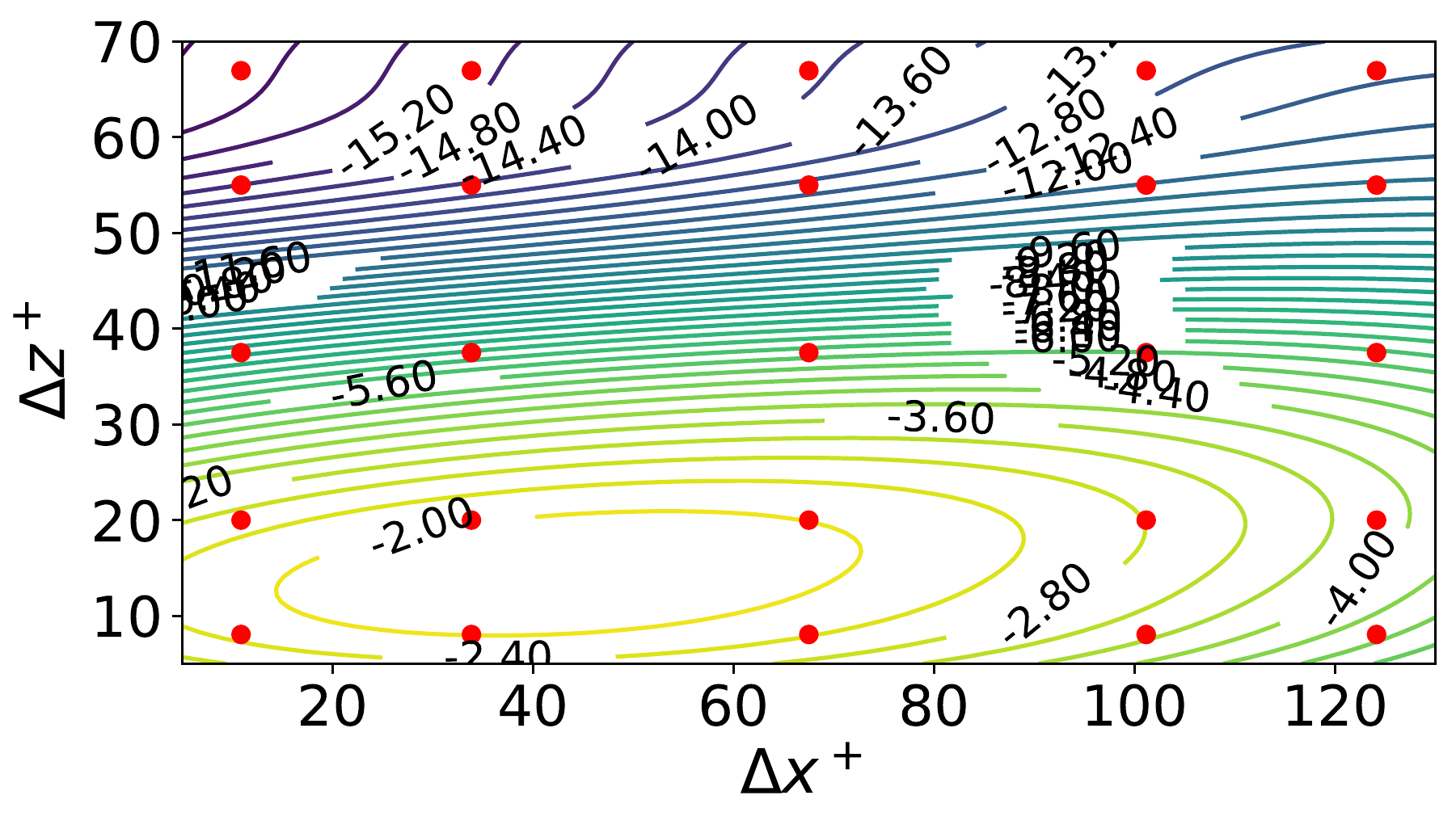} \\   
   {\small{(a)}} &    {\small{(b)}} &    {\small{(c)}}\\    
   \includegraphics[scale=0.3]{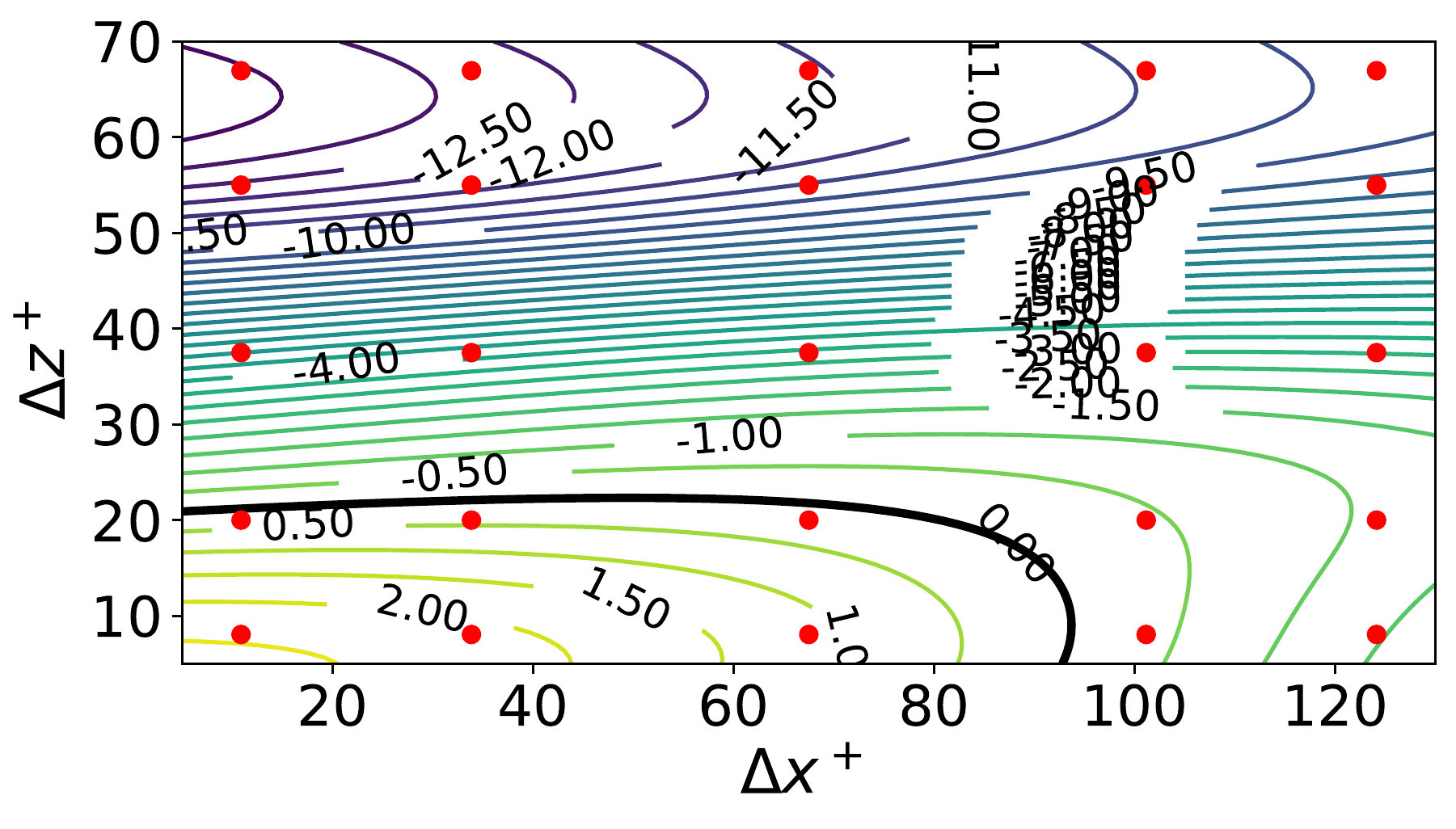} & \hspace{-0.5cm}
   \includegraphics[scale=0.3]{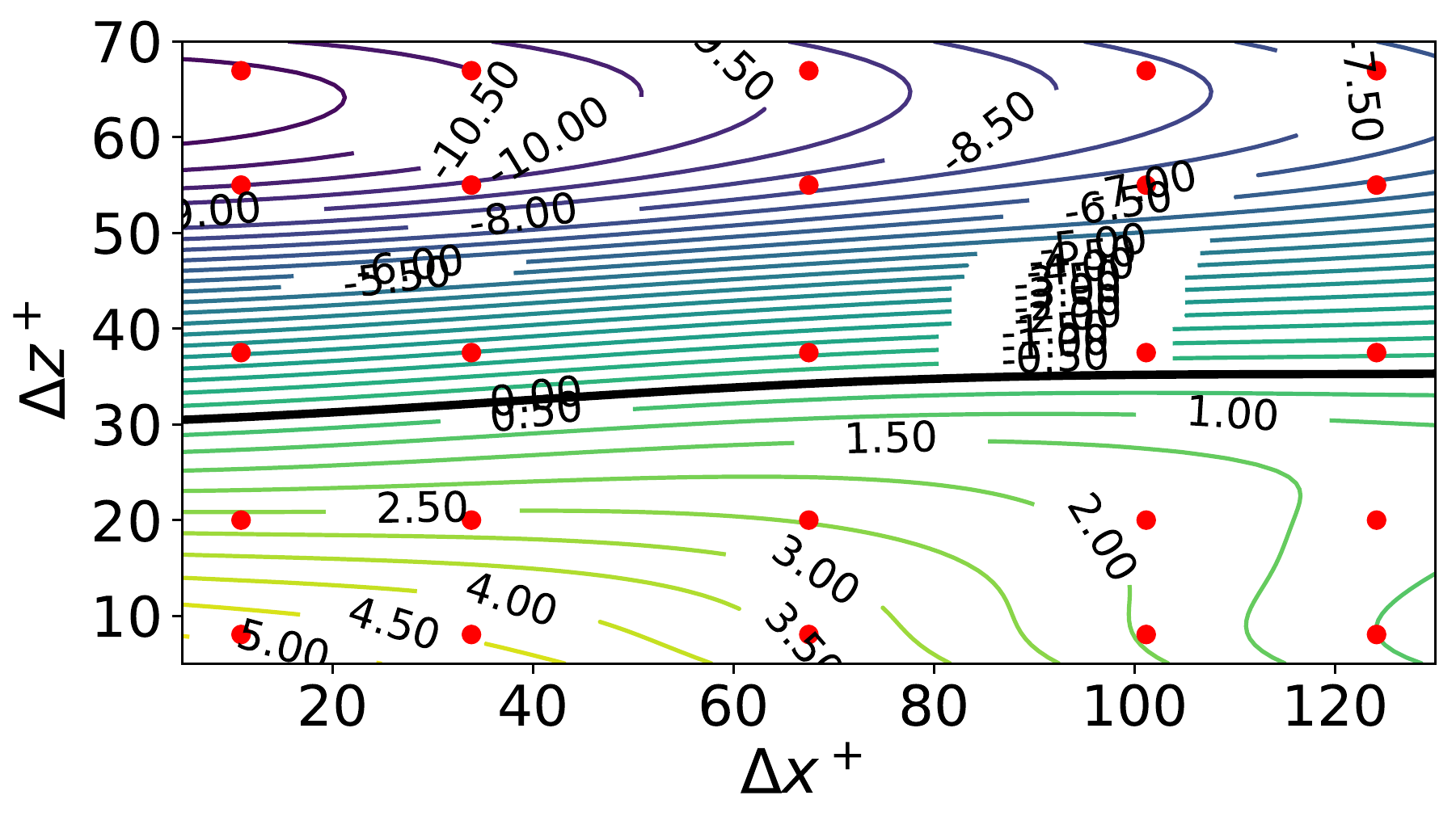} &\hspace{-0.5cm}
   \includegraphics[scale=0.3]{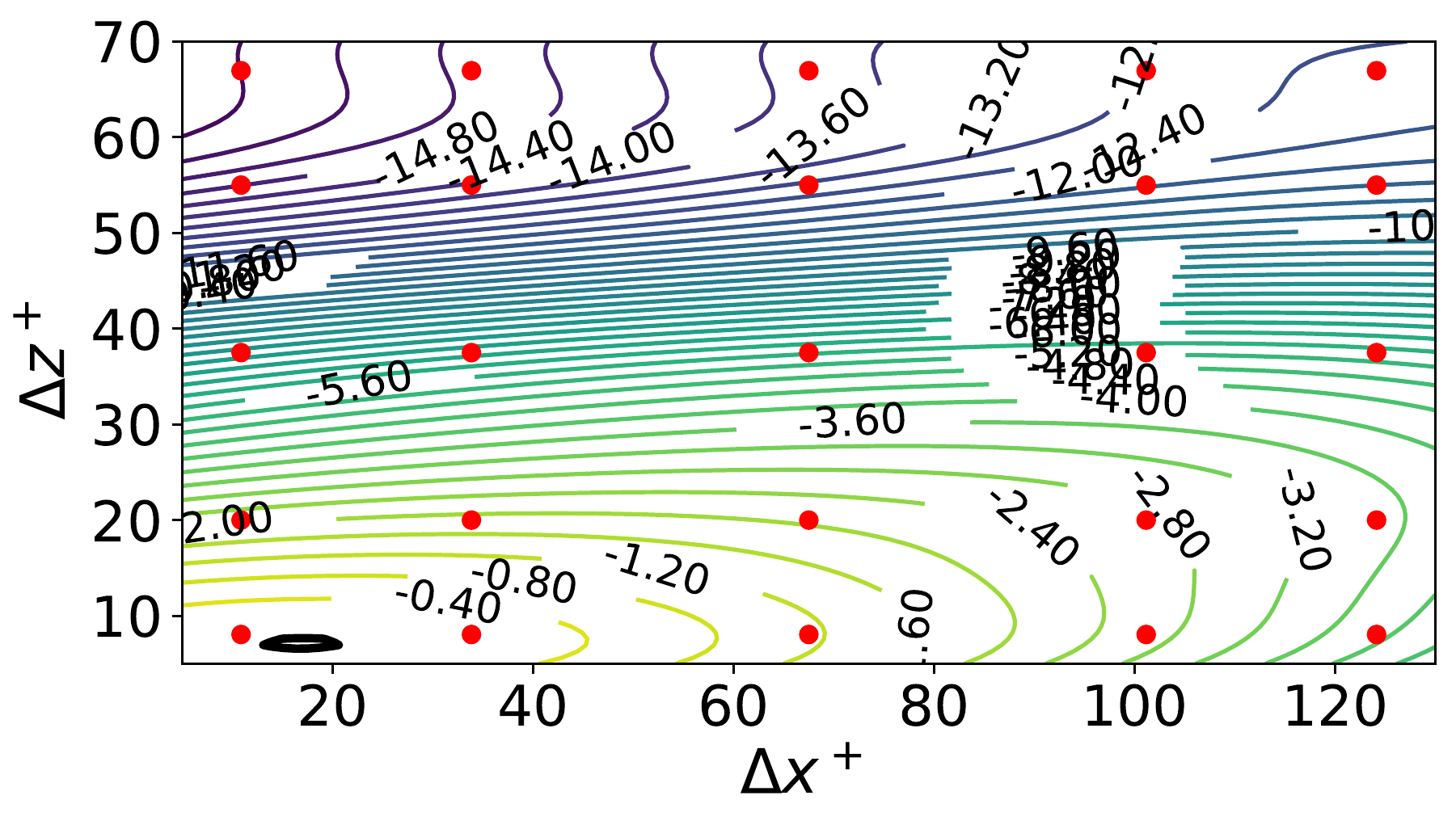} \\      
   {\small{(d)}} &    {\small{(e)}} &    {\small{(f)}}\\       
   \end{tabular}   
   \caption{Surfaces of the \revs{posterior predictive} mean (left column), and  $95\%$ CI (middle and right columns) of $\epsilon[\lut]\%$ in the $\dxp\dash\dzp$ plane. 
   These predictions are made by GPR, through \eqs~(\ref{eq:gpr_m}) and (\ref{eq:gpr_v}), using observations with synthetic homoscedastic noise shown in~\fig~\ref{fig:duTauGPRnoise}(a) (top row) and observations with synthetic heteroscedastic noise shown in~\fig~\ref{fig:duTauGPRnoise}(b) (bottom row). \rev{The 25 training samples are specified by red dots.} \revs{Using GPR, the \catII~uncertainty in the training data can be considered when constructing the surrogates in computer experiments for \catI~uncertainties.}}\label{fig:duTauSurfNoisy}   
\end{figure}

In producing \fig~\ref{fig:duTauSurface}, the \revs{\catII} uncertainty in each of the training simulations is \rev{considered} to be zero. 
\revs{Such uncertainties could be for instance due to the time-averaging.}
To account for \catII~uncertainties when studying the uncertainty due to $\fq=(\dxp,\dzp)$, we can use the GPR technique reviewed in \sect~\ref{sec:gpr}. 
For illustration purposes, two scenarios for the observation Gaussian noise $
\varepsilon_i\sim\cN(0,\sigma_{d_i}^2)$, where $i=1,2,\ldots,n$, are assumed in connection with the generic model~(\ref{eq:surrGen}). 
In the case with homoscedastic noise \rev{(same variance of all observations)}, $\sigma_d=0.5\%$ is assumed. 
In contrast, in heteroscedastic noise, where the noise level is observation-dependent, we assume $\sigma_{d_i}=0.3\exp(-2\xi_{\dzp,\,i})\%$, where $\xi_{\dzp,\,i}\in[-1,1]$ is corresponding to $\Delta z^+_i\in\BQ_\dzp$.
This construction imitates a practical situation where simulations with finer resolution in the spanwise direction are more expensive and hence may be averaged over shorter time intervals. 
Consequently their QoIs may become more uncertain. 
The plots in \fig~\ref{fig:duTauGPRnoise} represent these synthetic noise levels (shown by confidence intervals) added to the mean observations of $\epsilon[\lut]\%$ belonging to the $n=25$ case in \fig~\ref{fig:duTauSurface}(d). 
For these noisy data, surfaces of mean~$\epsilon[\lut]\%$ and associated $95\%$ confidence intervals are constructed in the~$\dxp\dash\dzp$ plane in \fig~\ref{fig:duTauSurfNoisy}.
For this purpose, posterior predictive mean and variance given by~\eqs~(\ref{eq:gpr_m}) and~(\ref{eq:gpr_v}) are used.

As a result of adding the observation noise, the isolines of the mean $\epsilon[\lut]$ may look different, compare \figs~\ref{fig:duTauSurfNoisy}(a) and (d) with \fig~\ref{fig:duTauSurface}(d).
Such difference is more clear in the case of heteroscedastic noise. 
The confidence in predicting these surfaces is reflected in \figs~\ref{fig:duTauSurfNoisy}(b) and (c) for the homoscedastic case, and in \figs~\ref{fig:duTauSurfNoisy}(e) and (f) for the heteroscedastic case. 
It is emphasized that the exaggerated noise levels are used here to show how the framework works, otherwise, based on the authors' experience, \rev{see \rf~\cite{salehPoF:18,vinuesa:16}}, the uncertainties in QoIs due to finite time-averaging intervals are smaller meaning that they may not majorly alter the plots like those in \fig~\ref{fig:duTauSurface} or other inferences due to the variation of \catI~parameters. 
Accurate quantification of the uncertainties due to finite time-averaging using methods similar to those investigated in~\cite{russo:17,oliver:14,vinuesa:16} will be considered in the feature.

\begin{figure}
   \centering
   \begin{tabular}{c}
   \includegraphics[scale=0.5]{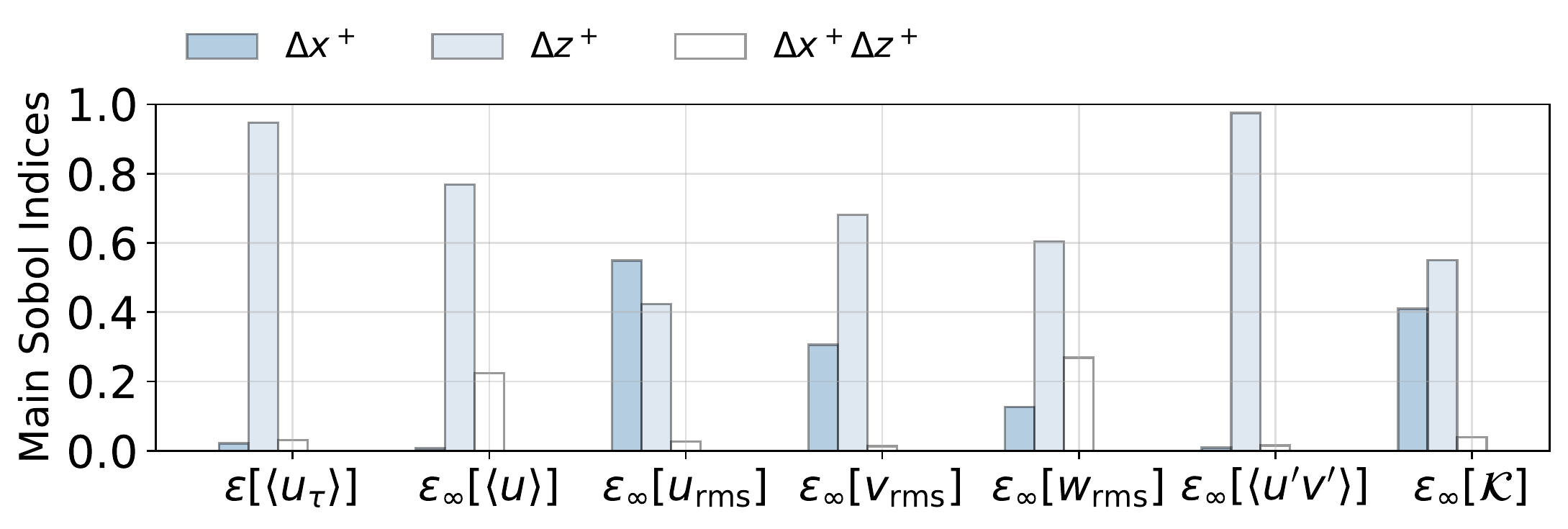} \\
   \includegraphics[scale=0.5]{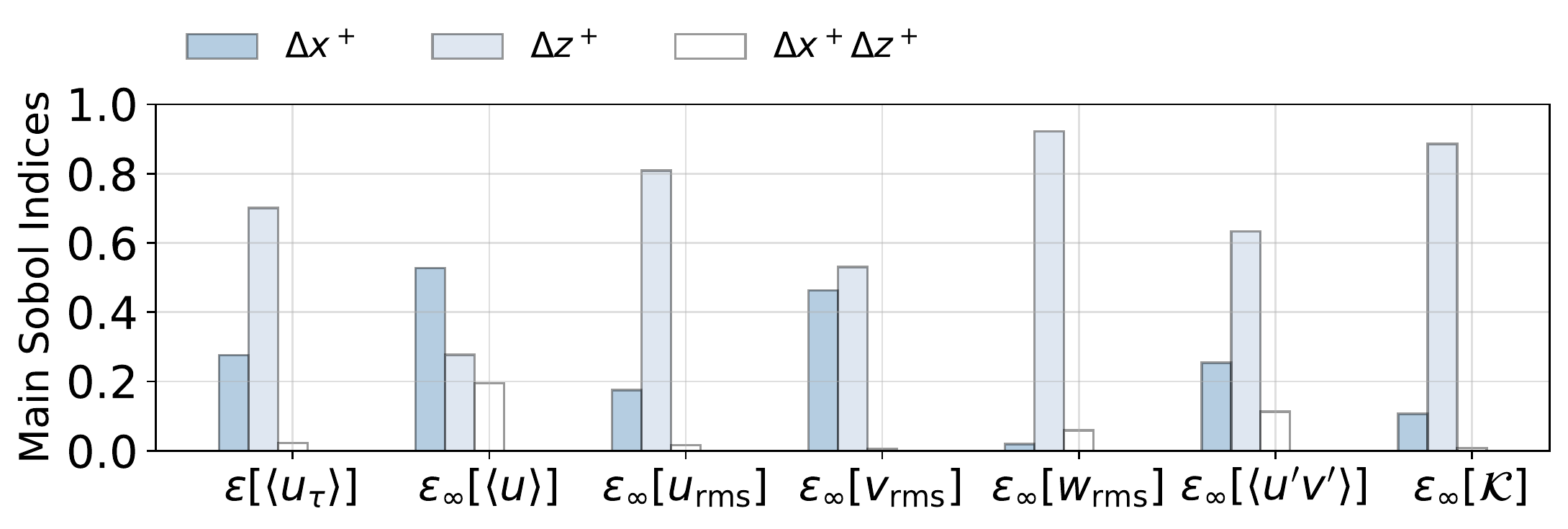}
   \end{tabular}
   \caption{Main Sobol indices~(\ref{eq:sobol}) for different error measures when the admissible space for $\dxp\bigotimes\dzp$ is $\BQ_1=[5,130]\bigotimes[5,70]$~(top), and $\BQ_2=[5,50]\bigotimes[5,30]$~(bottom). \revs{The surrogate constructed by $n=25$ training samples over $\BQ_1$, see \fig~\ref{fig:duTauSurface}(d), is employed to interpolate QoIs over $\BQ_2$.}}\label{fig:sobolError}
\end{figure}

Applying the global sensitivity analysis described in \sect~\ref{sec:sobol}, 
the Sobol indices indicating the sensitivity of the error responses with respect to variation of~$\dxp$ and~$\dzp$ are obtained, and they are shown in \fig~\ref{fig:sobolError}.
It is recalled that the Sobol indices represent the global behavior of the error responses, in contrast to the error portraits, which reflect the local behavior. 
Note that in \fig~\ref{fig:sobolError}, the uncertainty due to time-averaging is neglected. 
\rev{The QoIs whose errors are considered in the sensitivity analysis in this figure are averaged friction velocity~$\lut$, mean streamwise velocity~$\lu$, rms velocity fluctuations~$u'_\rms$, $v'_\rms$, and~$w'_\rms$ in the streamwise, wall-normal and spanwise directions, respectively, Reynolds shear stress~$\uv$, and turbulent kinetic energy $\cK=(u'^2_\rms+v'^2_\rms+w'^2_\rms)/2$.}
\rev{Clearly over a given admissible range, the Sobol indices showing sensitivity with respect to~$\dxp$ and~$\dzp$ change between different error measures. 
With the exceptions~$\einf[u'_\rms]$ on $\BQ_1=[5,130]\otimes[5,70]$ and~$\einf[\lu]$ on $\BQ_2=[5,50]\otimes[5,30]$, all the considered error measures are more sensitive to the variation of~$\dzp$ than~$\dxp$, regardless of the admissible range of these parameters. 
This behavior is not surprising, although computing Sobol indices provides a quantitative way to study it. 
In fact, the relevant near-wall structures in turbulent flows have smaller length-scales in the spanwise ($z$) direction than in the streamwise~($x$) direction, see~\eg~\cite{jimenez:13}.
Thus, employing coarse~$\dzp$ eliminates the possibility of correctly resolving such structures which in turn largely impacts the numerical modeling of the flow physics. 
Another interesting observation is that changing the admissible range,~$\BQ$, it is only for the streamwise velocity component (mean or rms fluctuation) that the sensitivity with respect to~$\dxp$ may become larger than that of~$\dzp$.
In particular,~$\dxp$ is the most influential factor on~$\einf[u'_\rms]$ over~$\BQ_1$ and on~$\einf[\lu]$ over $\BQ_2$.}

\subsubsection{Robustness, Sensitivity, and Observation Probability of QoIs}
In a computer experiment, the profiles of flow QoIs are affected by variation of $\fq=(\dxp,\dzp)$ over~$\BQ$.
In this regard, the focus of this section is on three important and insightful analyses which are complementary to each other. 
These analyses are novel and can be applicable to any CFD practice. 
First, by evaluating the robustness, an estimate of the amount of uncertainty propagated to each point of the profile of QoIs \revs{due to the variation of~$\fq$}, is obtained. 
Second, the contribution of each parameter \revs{(among~$\fq$)} in such propagated uncertainty is measured by computing the Sobol sensitivity indices.
Finally, it is explained how to compute the probability of observing a particular value for the QoIs conditioned on letting the parameters~$\fq$ vary.

\begin{figure}[!htbp]
   \centering
   \begin{tabular}{cc}
      \includegraphics[scale=0.42]{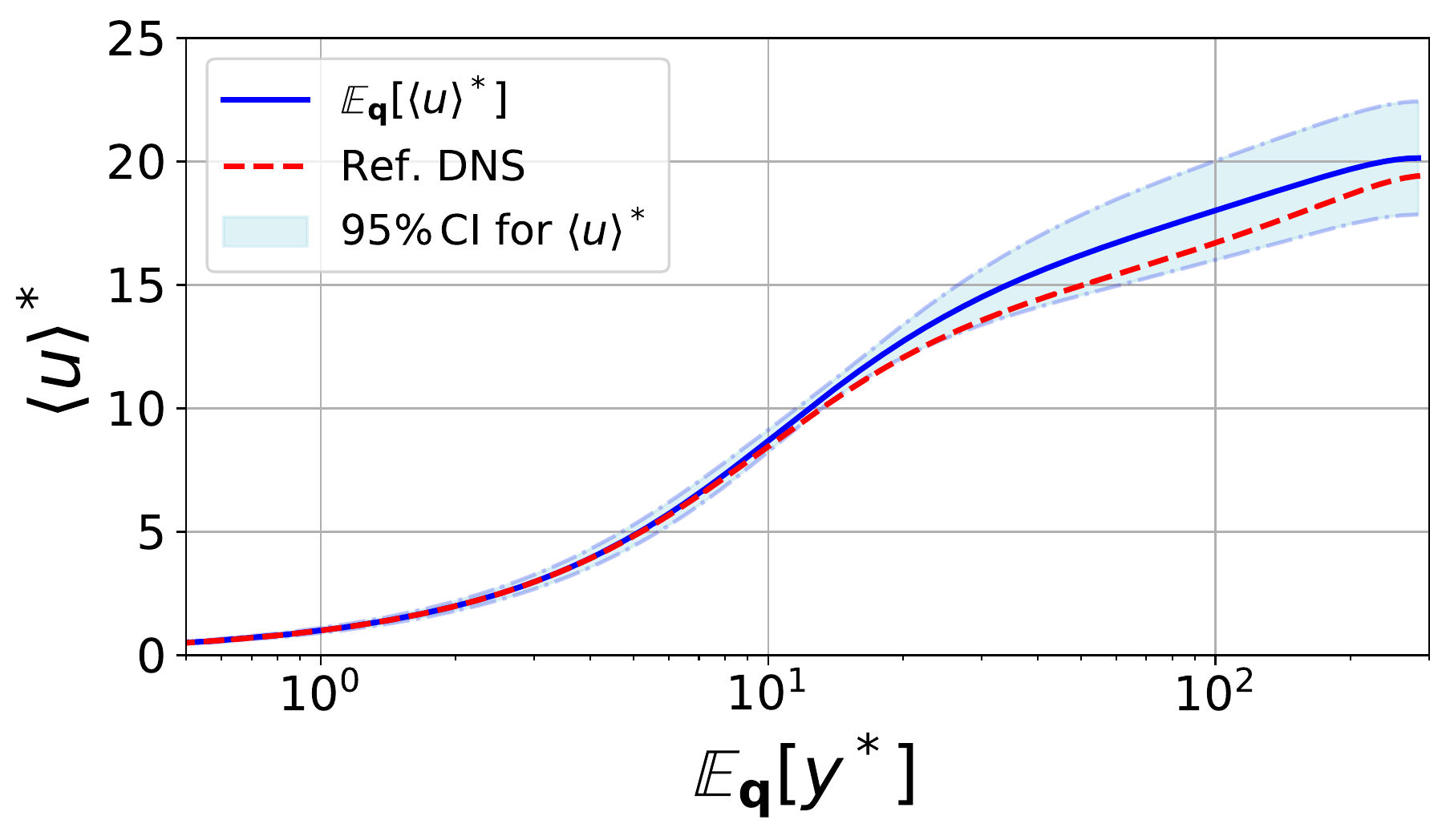} &
      \includegraphics[scale=0.42]{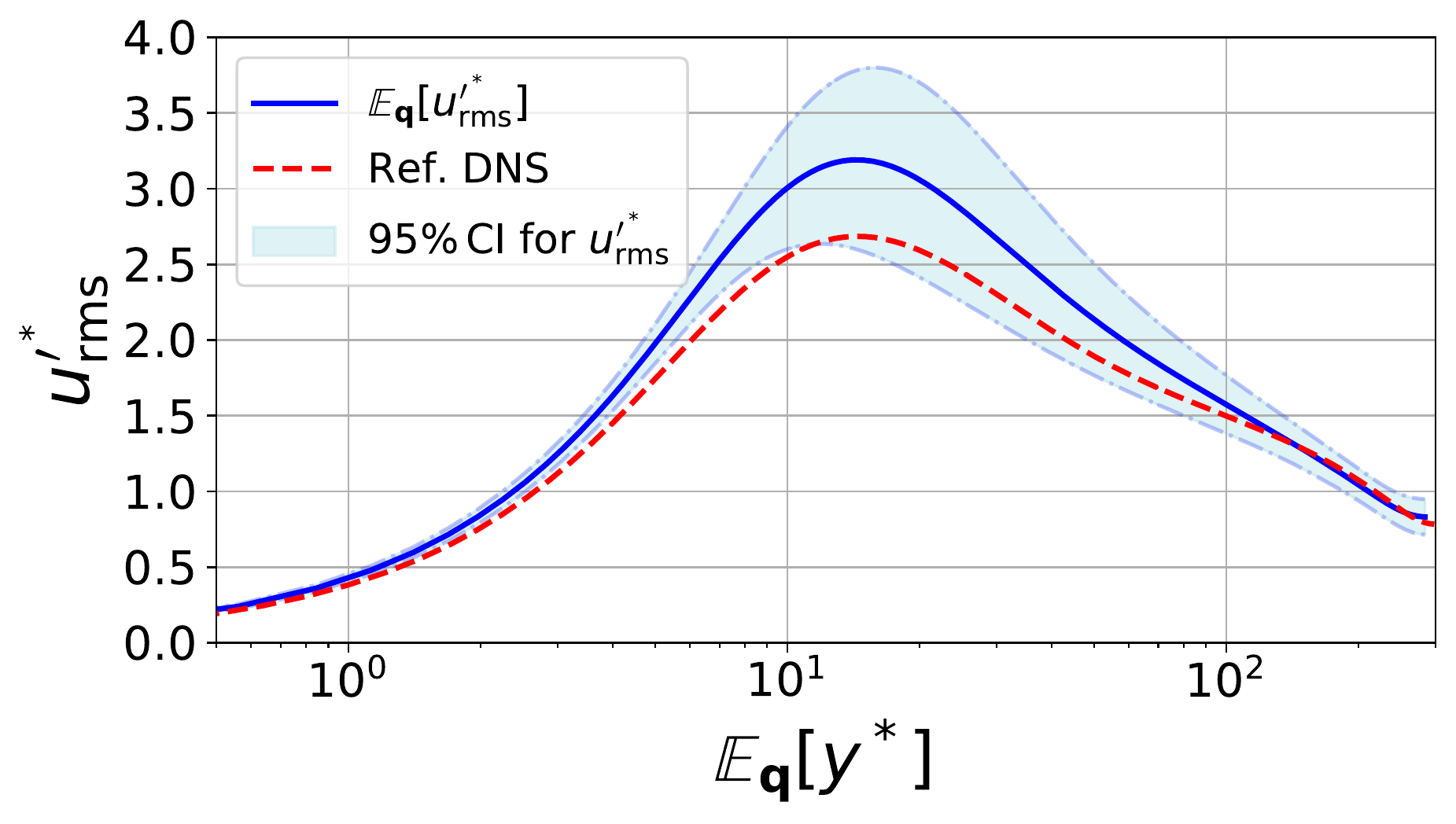}\\
      \includegraphics[scale=0.42]{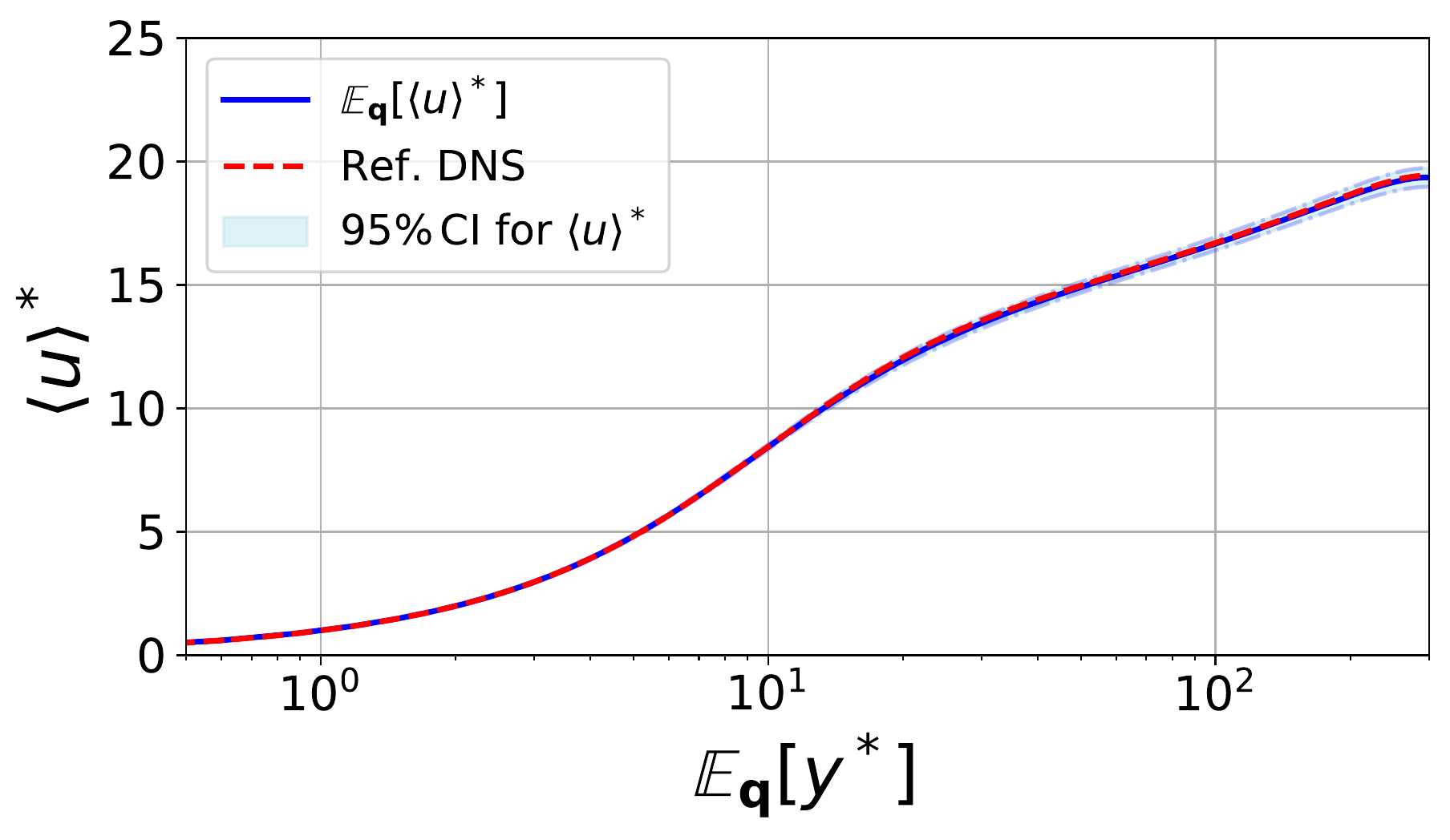}&
      \includegraphics[scale=0.42]{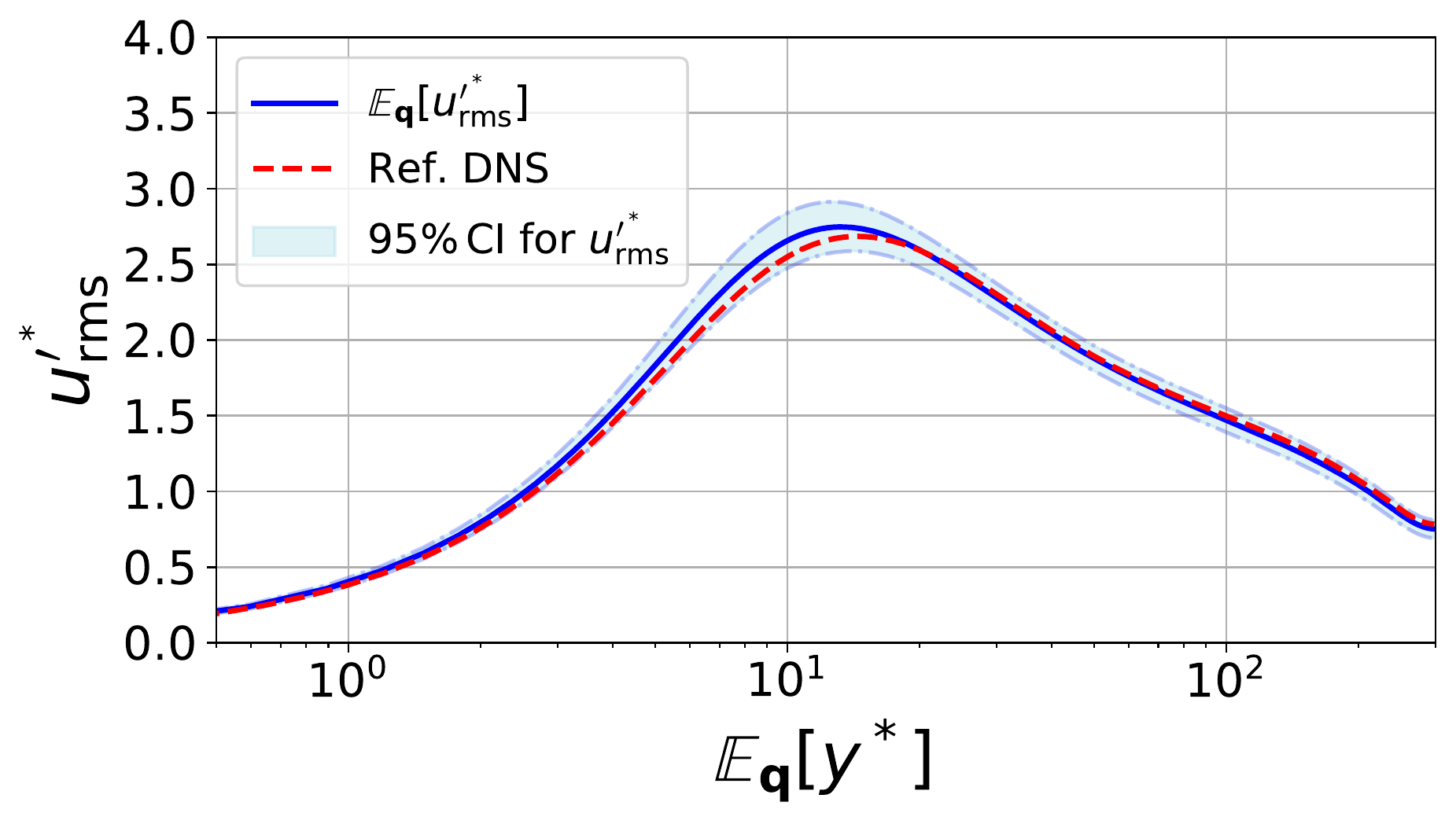}\\
   \end{tabular}
   \caption{Expected value of $\lu^*$ and $u'^*_\rms$ profiles along with the associated 95\% confidence intervals for the variation of~$\dxp$ and~$\dzp$ over $\BQ_1=[5,130]\bigotimes[5,70]$~(top), and $\BQ_2=[5,50]\bigotimes[5,30]$~(bottom). For both cases, $n=25$ training samples are used. The top plots correspond to the case shown in \fig~\ref{fig:duTauSurface}(d). \revs{Any \catII~uncertainty, e.g. due to time-averaging, in the training data is neglected.} The DNS data of Iwamoto~\et~\cite{iwamoto:02} are provided as reference. \revs{Note that the expected values $\BE_\fq[\lu^*]$ and $\BE_\fq[u'^*_\rms]$ (the blue lines) do not show the best results.}} \label{fig:exCI}
\end{figure}

The discrete profile of a time-averaged QoI is represented by $f_i(\fq)=f(y_i,\fq)$ for $i=1,2,\cdots,N_g$. 
Following~\sect~\ref{sec:uqCFD}, confidence intervals for the expected value $\BE_\fq[f_i(\fq)]$ are constructed for~$\fq\in \BQ$.
As a result, we obtain the plots in \fig~\ref{fig:exCI} showing the uncertainty in the inner-scaled profiles of mean and rms fluctuations of the streamwise velocity component, $\langle u \rangle^*=\lu/\lut$ and $u'^*_{\rms}=u'_{\rms}/\lut$, as a result of variation of~$\dxp$ and~$\dzp$ over~$\BQ_1$ and~$\BQ_2$. 
\rev{For the sake of brevity, we do not present the results for profiles of Reynolds shear stress $\uv$ and other components of the rms fluctuation velocity.}
Note that since the computed~$\lut$ (not the reference value) are used for scaling the quantities \rev{of the corresponding simulation in the computer experiment}, we have used the superscript~$^*$ instead of~$^+$.
Moreover, the horizontal axis in the plots represents~$\BE_\fq[y^*]$ \rev{which is the expected value of~$y^*$ over the simulations in the experiment.}
\rev{Therefore,} the illustrated confidence intervals only reflect the variation in the QoI which is on the vertical axis. 
Several interesting insights can be gained from~\fig~\ref{fig:exCI}.
First, as expected, the way in which the uncertainty \revs{due to the variation of~$\fq$} is propagated depends on the QoI. 
However, for any QoI the robustness of the profile varies with the distance from the wall. 
In the near-wall region, profiles of first- and second-order velocity moments are found to be more robust compared to the outer part of the boundary layer. 
Maximum uncertainty in~$\lu^*$ is observed close to the channel center line, while for~$u'^*_{\rms}$, it is around the peak where the highest influence of the variation of~$\dxp$ and~$\dzp$ is observed.
Moreover, comparison between the plots in the top and bottom rows of \fig~\ref{fig:exCI} reveals that reducing the range of variation of~$\fq$ may result in smaller propagation of uncertainty in QoIs.
More notably, the~$\lu^*$ is found to be very robust when~$\dxp$ and~$\dzp$ change over $[5,50]$ and $[5,30]$,~respectively. 
\revs{As explained in \sects~\ref{sec:uq} and~\ref{sec:uqCFD}, $\BE_\fq[\cdot]$, the expected value of the QoIs in a computer experiment based on variation of~$\fq$ which is specified by the blue line in \fig~\ref{fig:exCI}, does not reflect the most accurate value of the QoIs.
It however, provides a baseline around which the confidence interval based on associated $\BV_\fq[\cdot]$ is built.}

\begin{figure}
   \centering
   \begin{tabular}{cc}
   \includegraphics[scale=0.42]{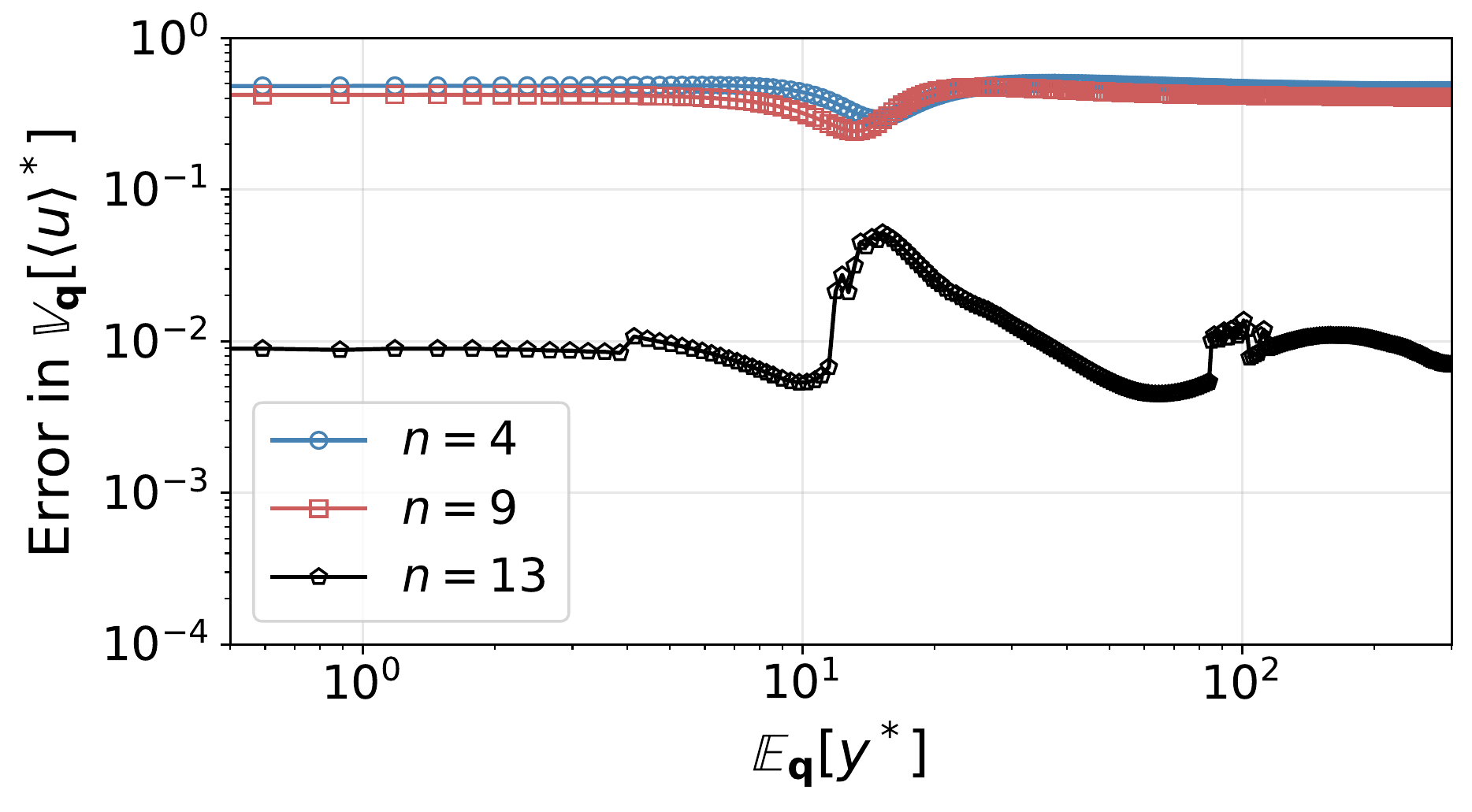} &
   \includegraphics[scale=0.42]{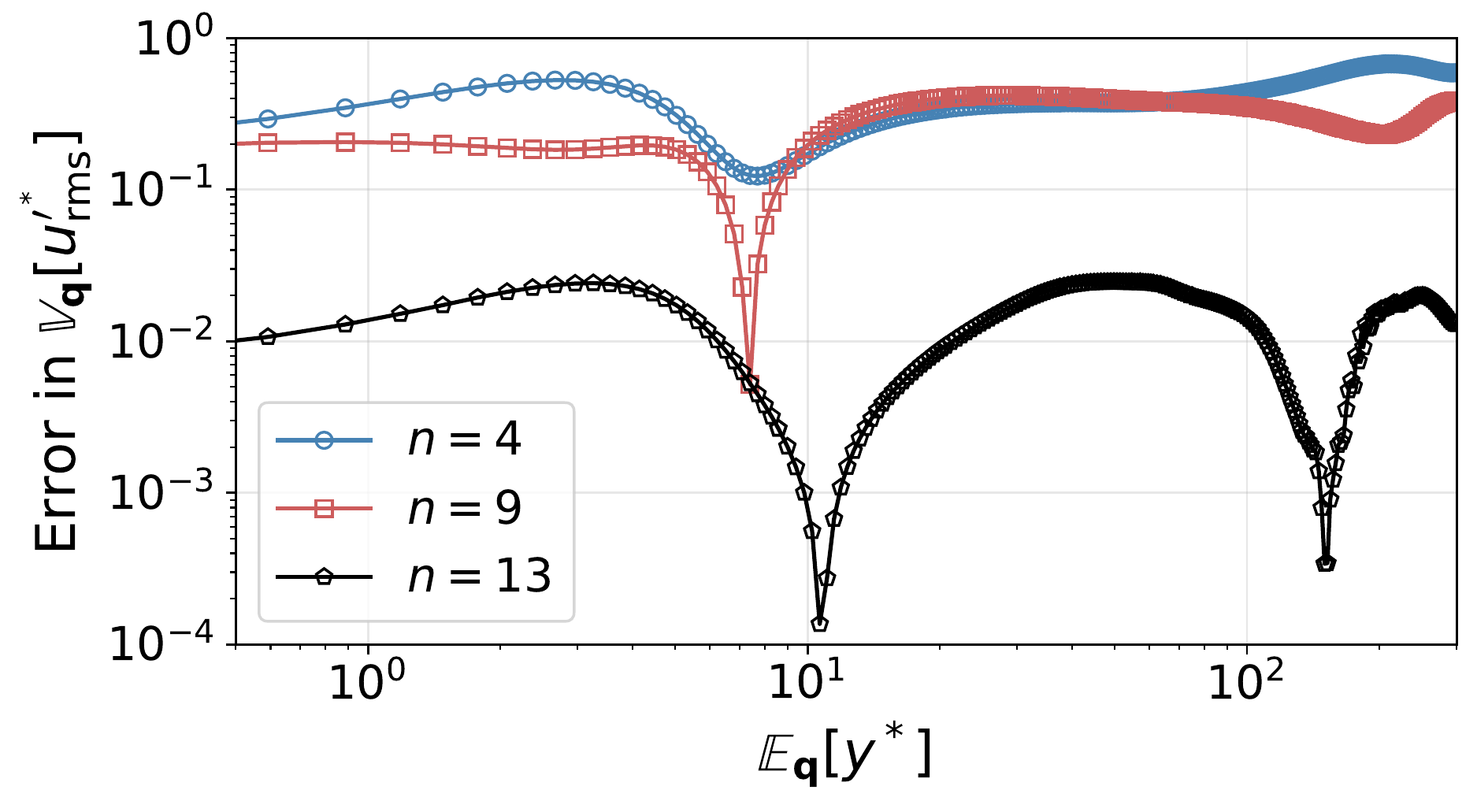} \\   
   \small{(a)} & \small{(b)} \\
   \end{tabular}
   \caption{Normalized relative error in the estimated $\BV_\fq[\lu^*]$~(a) and $\BV_\fq[u'^*_\rms]$~(b) using $n=4,\,13,\,25$ training samples. The reference values \rev{for computing the error and normalizing it} are provided by the PCE with $n=25$ samples. In all cases the TOM scheme~(\ref{eq:pceTO}) with $L=12$ is used for truncation of the PCE. }\label{fig:convPCE_subC8}
\end{figure}

The top plots in \fig~\ref{fig:exCI} correspond to the computer experiment with the~$25$ training samples shown by dots in \fig~\ref{fig:duTauSurface}(d). 
We can investigate how much error would be involved in the confidence intervals if lower number of simulations were considered \revs{in the computer experiment}. 
To this end, we compute the relative error between~$\BV_\fq[f(y,\fq)]$ estimated by~$n=4\,,9\,,13$ samples and the reference value given by~$n=25$ case.
\revs{These samples are represented by the red dots over~$\BQ_1$ in \fig~\ref{fig:duTauSurface}.}
The resulting errors for~$\lu^*$ and~$u'^*_\rms$ are plotted in \fig~\ref{fig:convPCE_subC8}. 
Clearly, using~$4$ and~$9$ simulations would be \revs{considerably} erroneous. 
But, by including~$n=13$ simulations, the error in the~variance of~$\lu^*$ and~$u'^*_\rms$ is less than approximately~$6\%$ and~$4\%$, respectively.
Therefore, in a computer experiment fewer samples may be required to assess robustness of the profiles compared to what is needed to get an accurate error portrait, see~\fig~\ref{fig:duTauSurface}. 
It is noted that besides the above error analysis, the non-intrusive PCEs in each experiment can be examined by evaluating the norm~(\ref{eq:pceConvNorm}) at each point on the profile of a QoI.

\begin{figure}
\centering
   \begin{tabular}{cc}
   \includegraphics[scale=0.42]{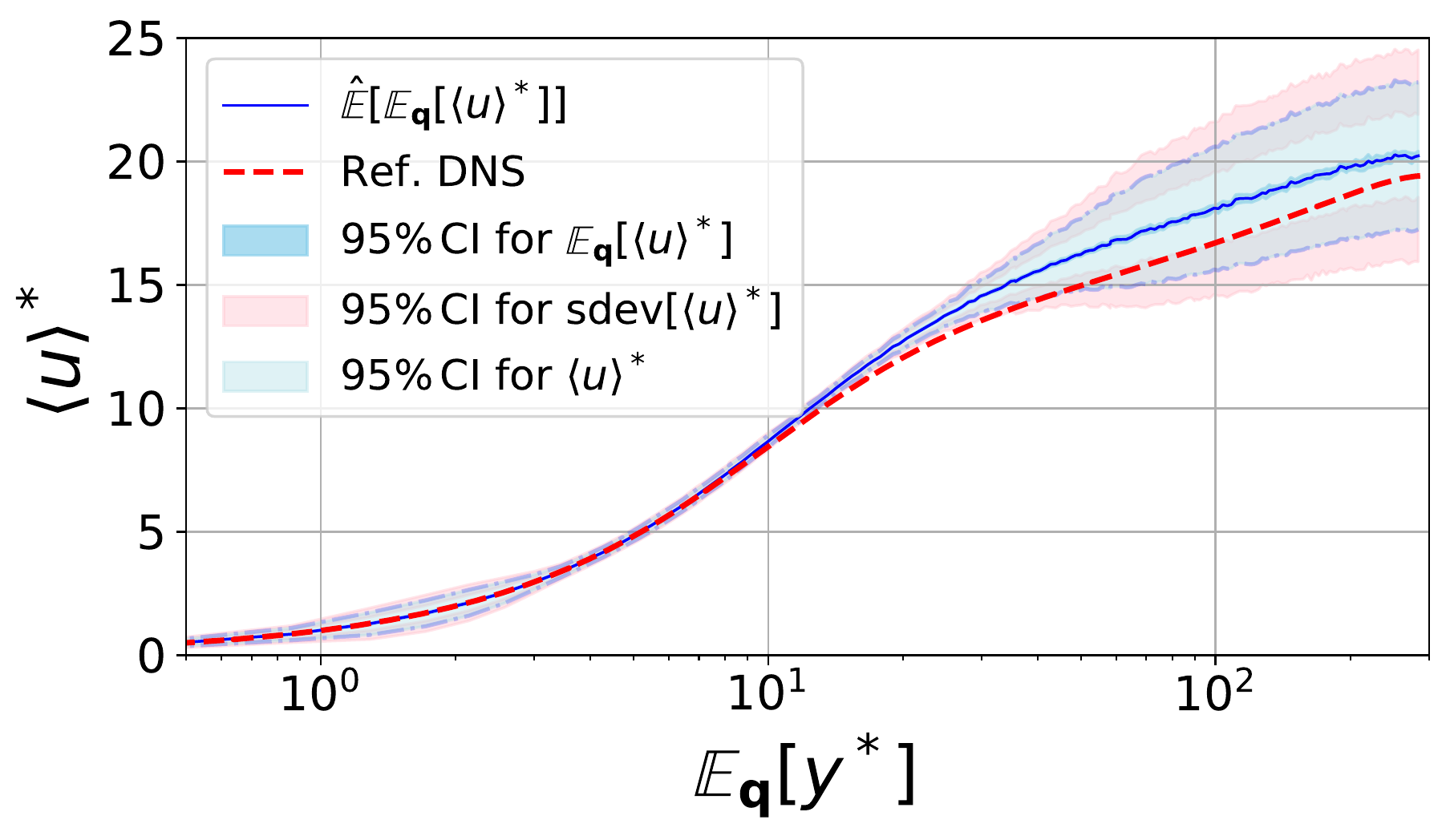} &
   \includegraphics[scale=0.42]{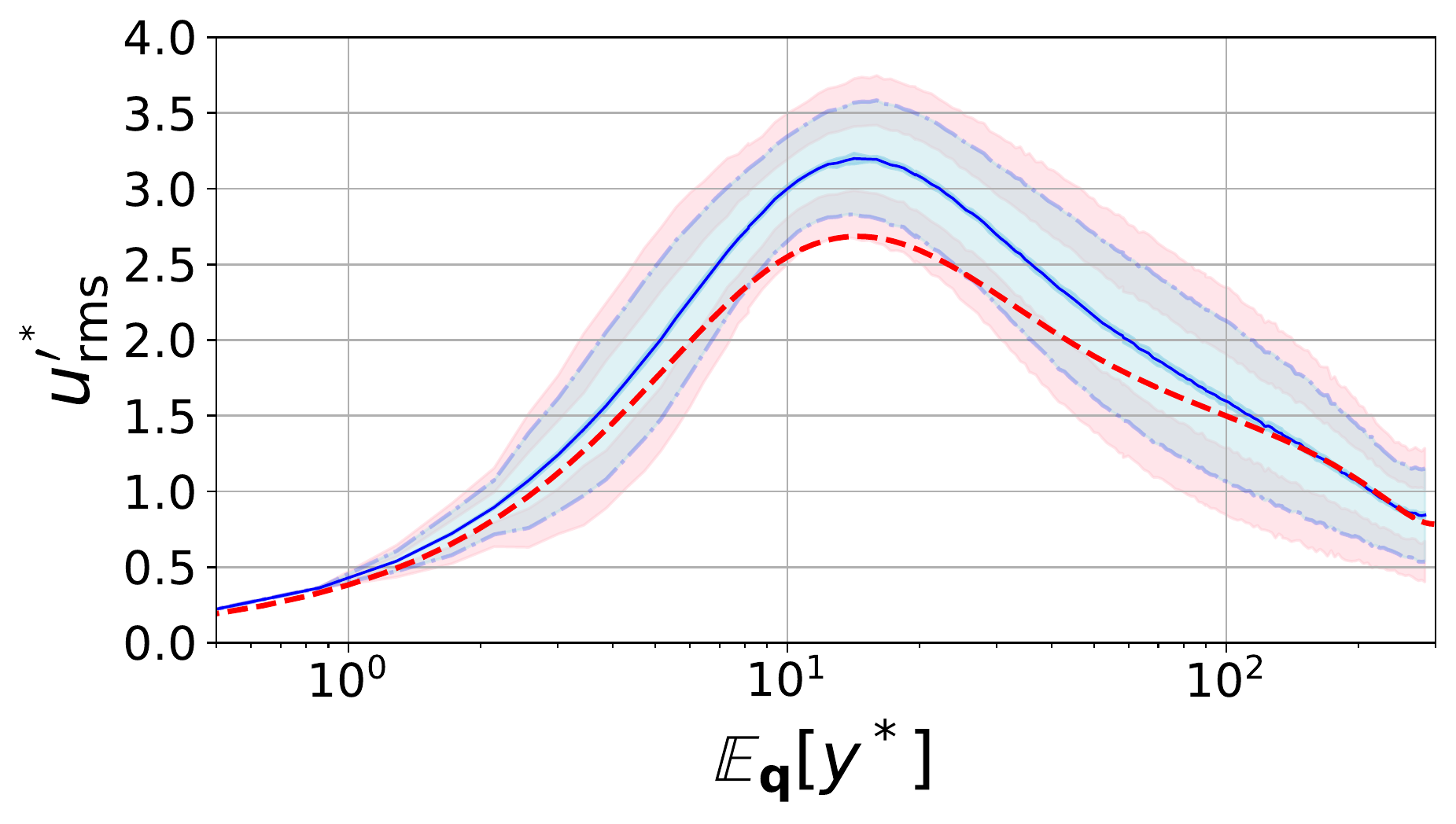} \\
   \small{(a)} & \small{(b)} \\   
   \end{tabular}
   \caption{Propagation of uncertainty into $\lu^*$~(a) and $u'^*_\rms$~(b) profiles due to the variation of~$\dxp$ and~$\dzp$ over $\BQ_1$ when the simulator's data are uncertain with a \revs{Gaussian noise} with synthetic standard deviations~(\ref{eq:chanTimeNoise}).
   The experiment with $25$ simulations as in \fig~\ref{fig:duTauSurface}(d) and \fig~\ref{fig:exCI} is considered \revs{for the expected value of the simulated QoIs}. \revs{The shaded regions show the followings: dark blue: $\hBE[\BE_\fq[f(y^*,\fq)]]\pm t_{95\%} \sqrt{\hBV[\BE_\fq[f(y^*,\fq)]]}$, light blue: $\hBE[\BE_\fq[f(y^*,\fq)]]\pm t_{95\%} \sqrt{\hBE[\BV_\fq[f(y^*,\fq)]]}$, pink: $\hBE[\BV_\fq[f(\chi,\fq)]]\pm t_{95\%} \sqrt{\hBV[\BV_\fq[f(\chi,\fq)]]}$, see \sect~\ref{sec:uqCFD}.} }
   \label{fig:C8_ppce}
\end{figure}

In the above analysis of robustness against variation of~$\fq=(\dxp,\dzp)$, \fig~\ref{fig:exCI}, the training values of QoIs (\ie~$\lu^*$ and~$u'^*_\rms$) were assumed to have negligible uncertainties. 
Such uncertainty could, for instance, have been due to insufficient averaging over time.
For the purpose of illustrating the novel functionality of the PPCE method of \sect~\ref{sec:ppce}, synthetic yet reasonable uncertainties are considered to be involved in the training QoIs. 
According to a preliminary analysis (not shown here), the time-averaging uncertainty in the profile of a channel flow QoI can be a function of wall distance. 
Analyzing the time-series of a set of channel flow simulations at different~$\dxp$ and $\dzp$ values, approximate curves for~$g(y^*)$ are obtained, where $\sqrt{\BV_t[r(y^*)]}/\BE_t[r(y^*)] = c_0\, g(y^*)$. 
Note that $g(y^*)$ has the the value one at the first off-wall grid point and takes different forms for different quantities~$r(y^*)$. 
In this expression,~$c_0$ is a constant value, and $\BV_t[\cdot]$ and $\BE_t[\cdot]$ are estimated variance and expectation of a time-averaged quantity.
We repeat the computer experiment designed for investigating the impact of $\fq=(\dxp,\dzp)$, with the results shown in \fig ~\ref{fig:exCI}(top), considering Gaussian noise $\varepsilon_i\sim \cN(0,\sigma^2_{d_i})$ in~(\ref{eq:surrGen})  where $i=1,2,\cdots,25$. 
The observation-dependent noise levels are constructed as,
\begin{equation}\label{eq:chanTimeNoise}
\sigma_{d_i}(y^*) = c_0 \, g(y^*) \BE_t[r^{(i)}(y^*)] \,,
\end{equation}
where, $\BE_t[r^{(i)}(y^*)]$ is substituted by~$\lu^{*^{(i)}}$ and~$u^{*^{(i)}}_\rms$ obtained by averaging over long enough time \revs{(what used in \figs~\ref{fig:duTauSurface}, \ref{fig:exCI}, for instance)} and the value of~$c_0$ is assumed to be~$0.01$.
\rev{This means that if for instance~$g(y^*)=1$, then $1\%$ of the value of the QoIs at each wall-normal location in the channel is uncertain due to time-averaging.}
Using the PPCE method of \sect~\ref{sec:ppce}, the plots in \fig~\ref{fig:C8_ppce} are obtained. 
Three shaded areas representing different confidence intervals can be observed which are constructed following the discussion in \sect~\ref{sec:uqCFD}.
The dark-blue area represents  
$\hBE[\BE_\fq[f(y^*,\fq)]]\pm t_{95\%} \sqrt{\hBV[\BE_\fq[f(y^*,\fq)]]}$. 
This region is very small \revs{(almost invisible)} meaning a high confidence in estimating the $\hBE[\BE_\fq[f(y^*,\fq)]]$ profile (solid blue line).
Also, note that this profile remains unaffected in the plots in \fig~\ref{fig:C8_ppce} and \fig~\ref{fig:exCI}(top). 
The light blue shaded region with dash-dotted boundaries shows $\hBE[\BE_\fq[f(y^*,\fq)]]\pm t_{95\%} \sqrt{\hBE[\BV_\fq[f(y^*,\fq)]]}$. 
This region can be compared to the corresponding CI \revs{for the variation of $\fq$} in the plots in \fig~\ref{fig:exCI}(top). 
As a result of including noise~(\ref{eq:chanTimeNoise}), the robustness of~$\lu^*$ with respect to variation of $\fq$ remains almost unchanged, except at  small~$y^*$ (near the wall). 
On the other hand, for $u'^*_\rms$, the changes are dramatic.  
This clearly indicates the importance of accounting for uncertainties in the training data, upon their existence, in a computer experiment. 
Finally, the pink area in \fig~\ref{fig:C8_ppce} reflects the uncertainty involved in the estimated light-blue confidence regions.

\begin{figure}[!htbp]
   \centering
   \begin{tabular}{cc}
      \includegraphics[scale=0.4]{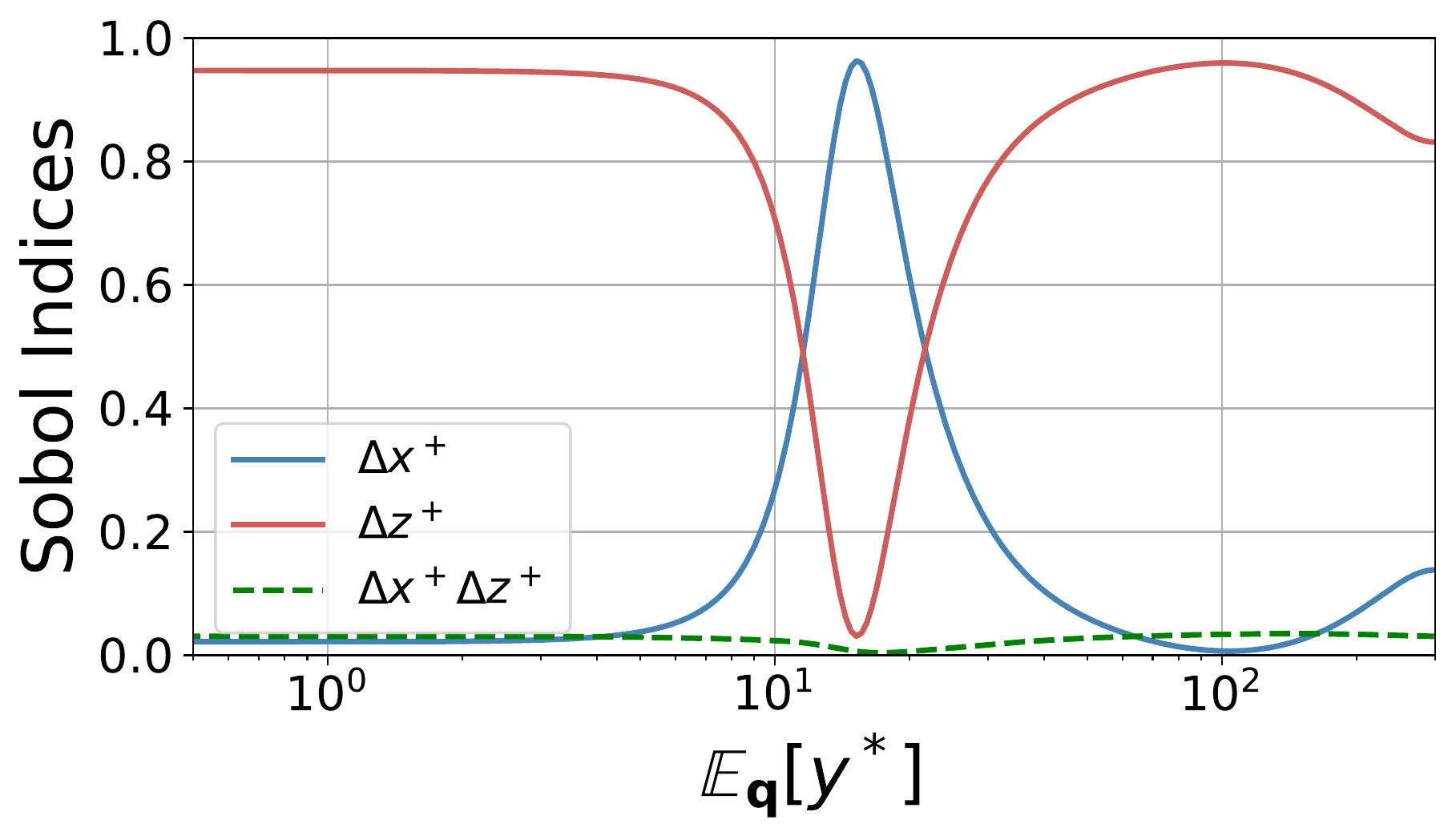} &
      \includegraphics[scale=0.4]{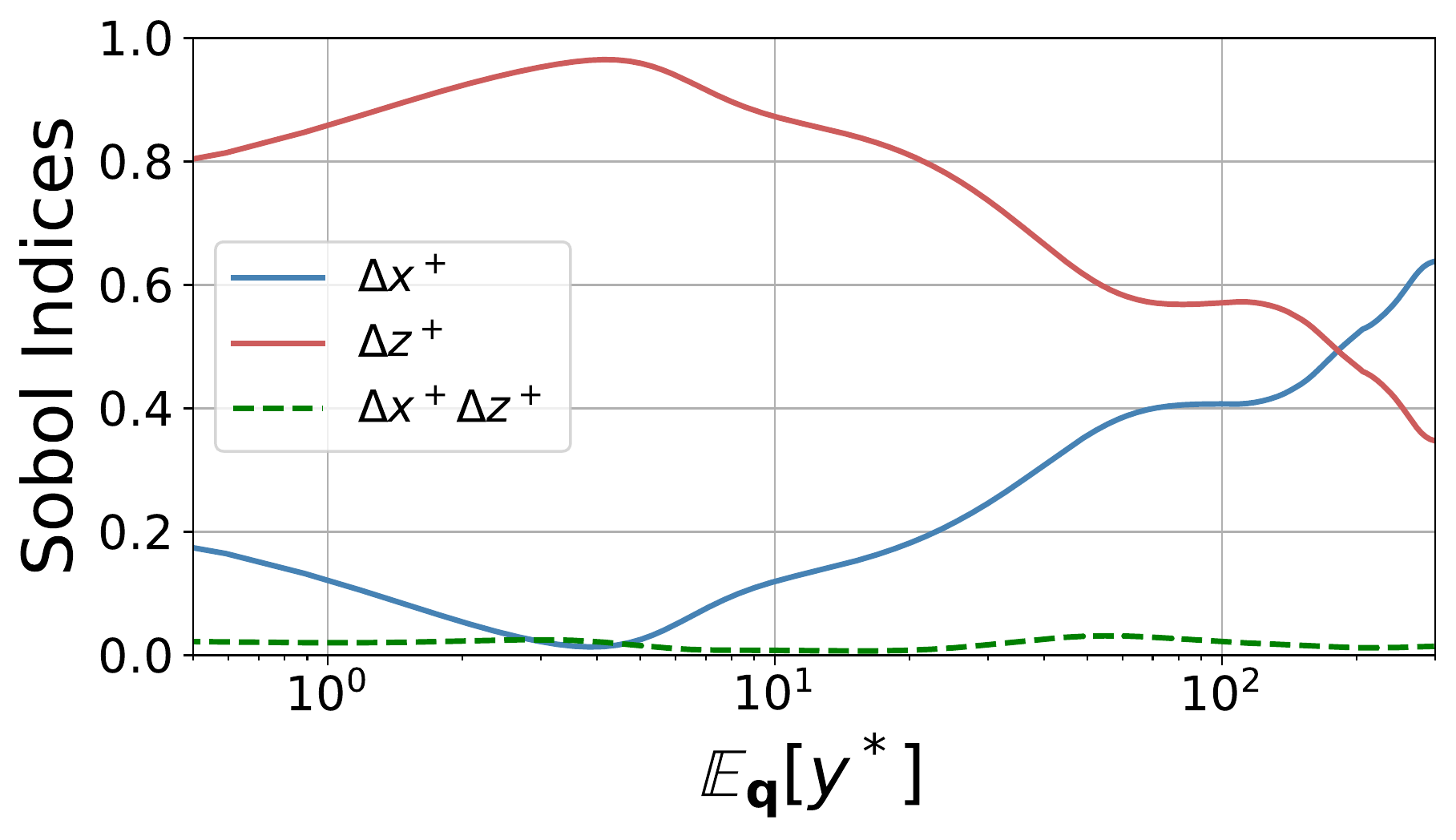}\\
      \includegraphics[scale=0.4]{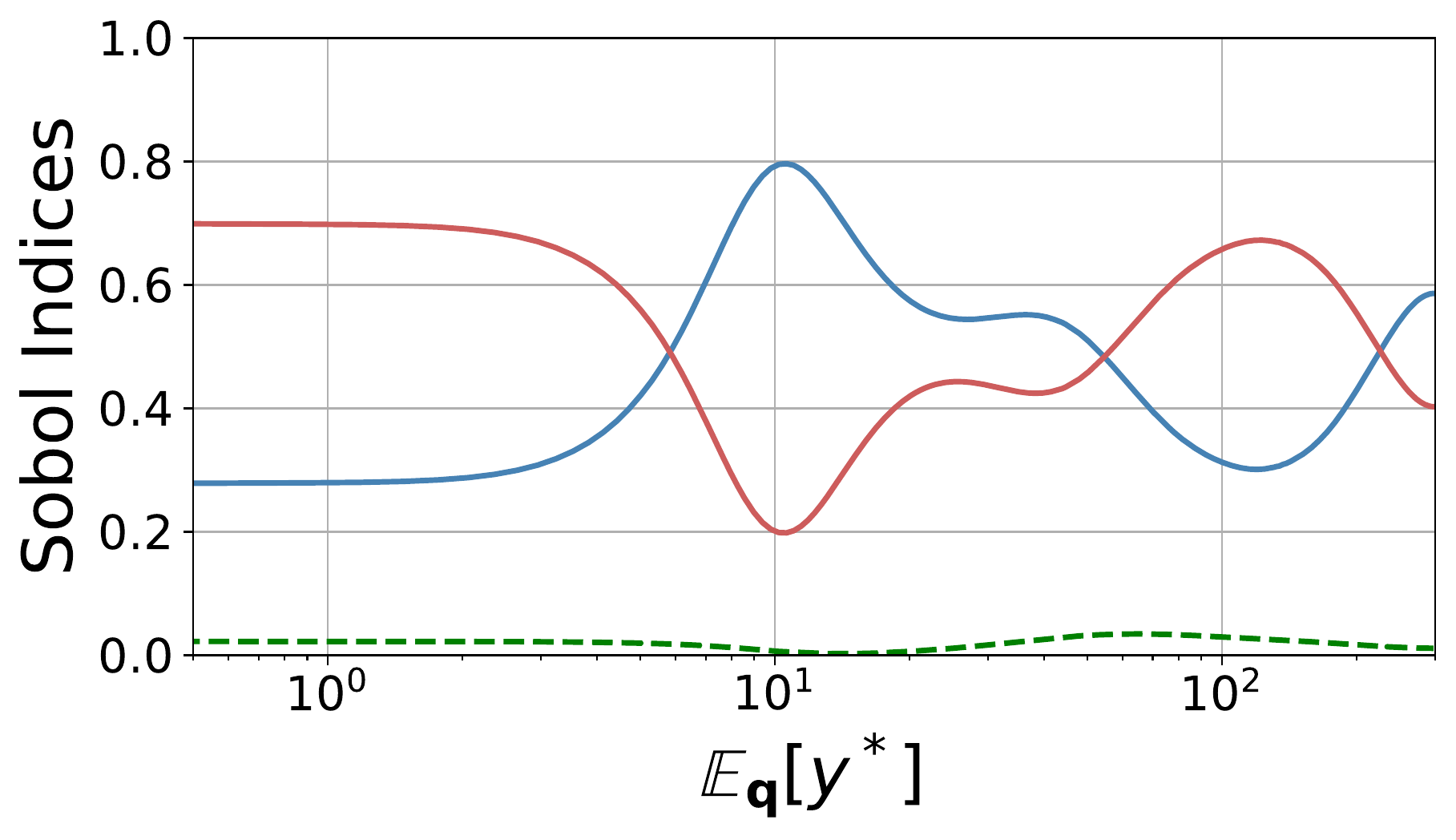}&
      \includegraphics[scale=0.4]{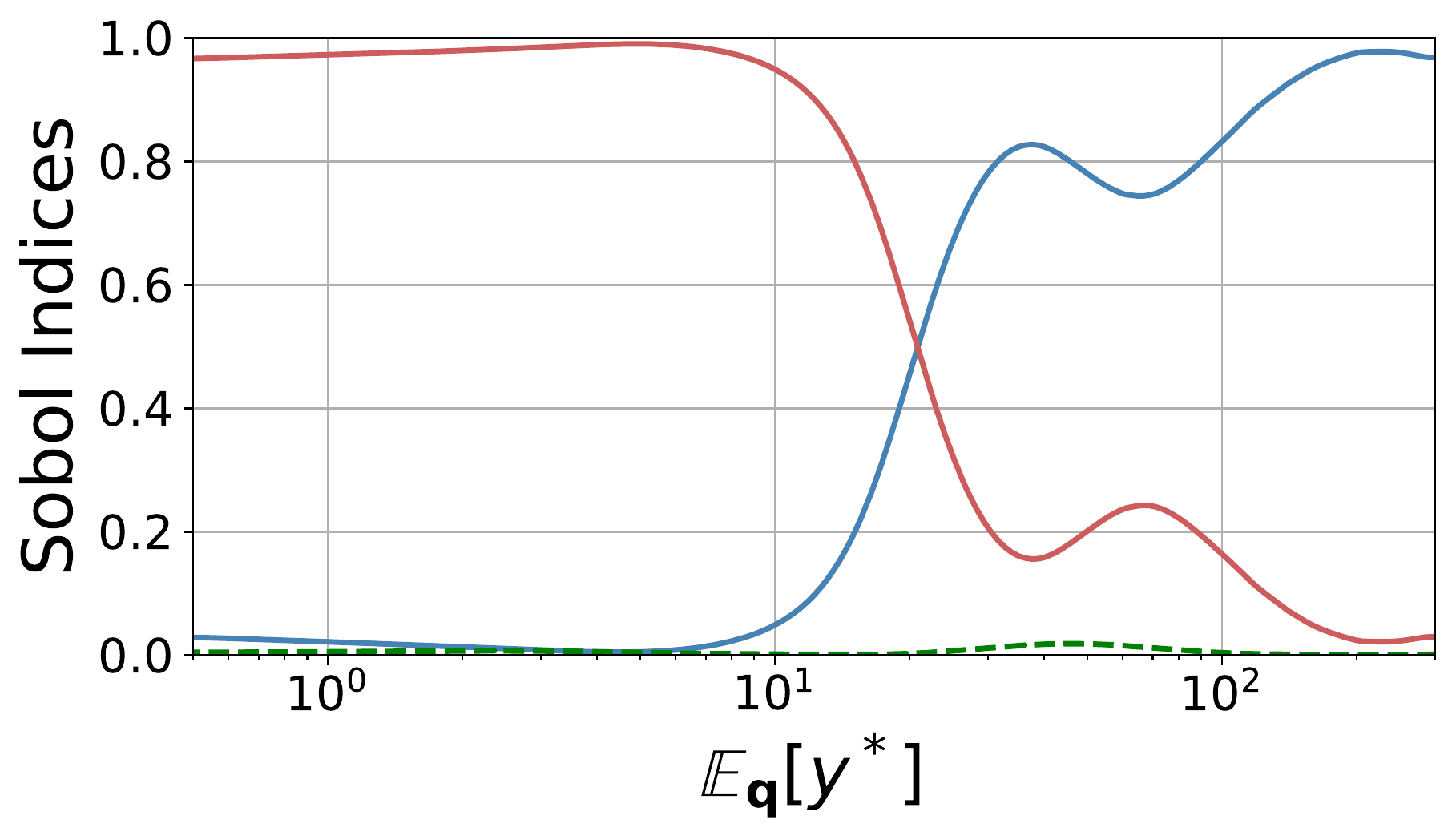}\\
   \end{tabular}
   \caption{Main Sobol indices for $\lu^*$ (left column) and $u'^*_\rms$ (right column) profiles showing the sensitivity with respect to~$\dxp$ and~$\dzp$ as they vary over $[5,130]$ and $[5,70]$~(top row) and $[5,50]$ and $[5,30]$~(bottom row), respectively.}\label{fig:exProfSobol}
\end{figure}

Henceforth, we do not consider the synthetic uncertainties in the training data in the computer experiments for~$\fq$. 
In order to specify the contribution of each component of~$\fq$ compared to the others in driving the uncertainty in the profiles of QoIs, the Sobol sensitivity indices are computed. 
\fig~\ref{fig:exProfSobol} shows the profiles of these indices for~$\lu^*$ and~$u'^*_\rms$ belonging to the cases represented in~\fig~\ref{fig:exCI}. 
The information deduced from the plots in these two figures are complementary to each other. 
For instance, the highest uncertainty in the~$u'^*_\rms$ profile that is near the peak turns out to be mostly driven by the variation in~$\dzp$. 
Further, at different wall distances, the relative influence of a parameter on the uncertainty of the QoI can change. 
However, near the wall, \ie~for $\BE_\fq[y^*]\lesssim 7$, it is~$\dzp$ which dominates. 
It is emphasized that this type of interpretation and conclusions are neither trivial nor achievable without using the current UQ framework. 
In particular, note the fact that the admissible space of~$\dxp$ is taken to be always larger than the space of~$\dzp$.

\begin{figure}
\centering
   \begin{tabular}{c}
   \includegraphics[scale=0.5]{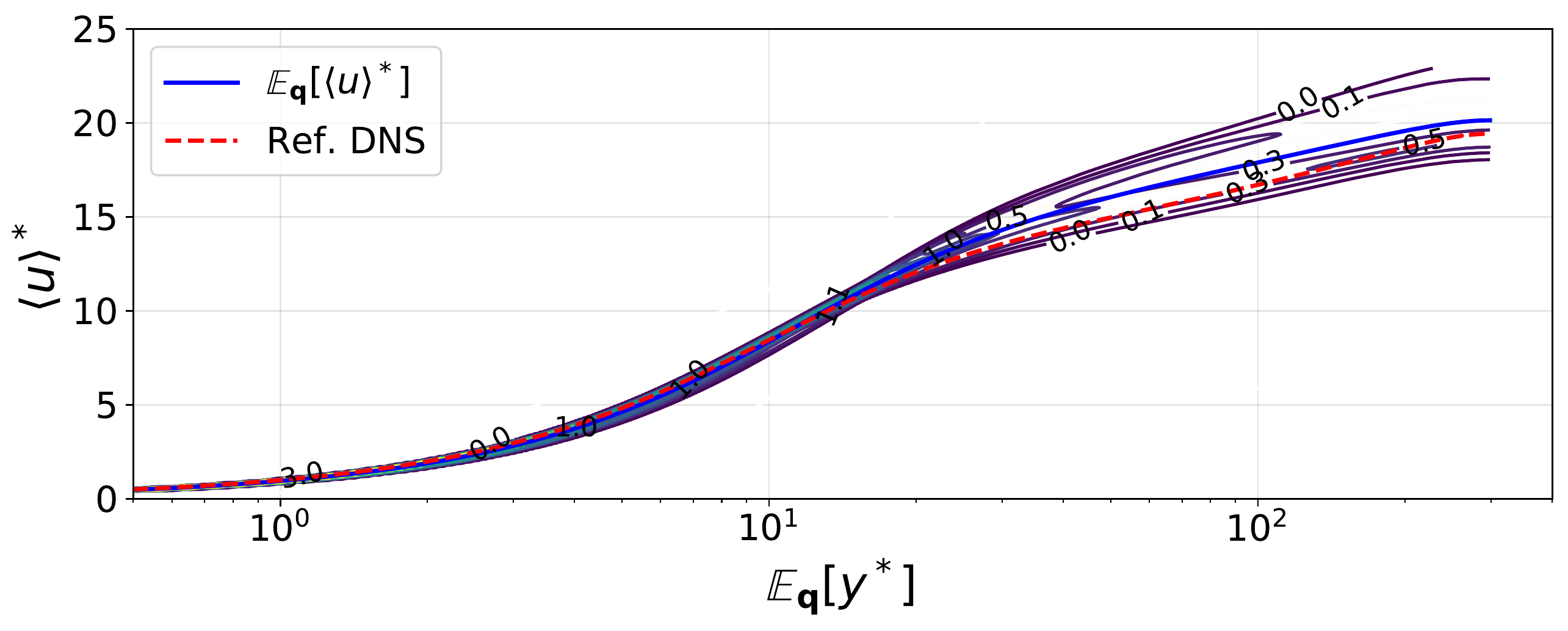} \\
   \includegraphics[scale=0.46]{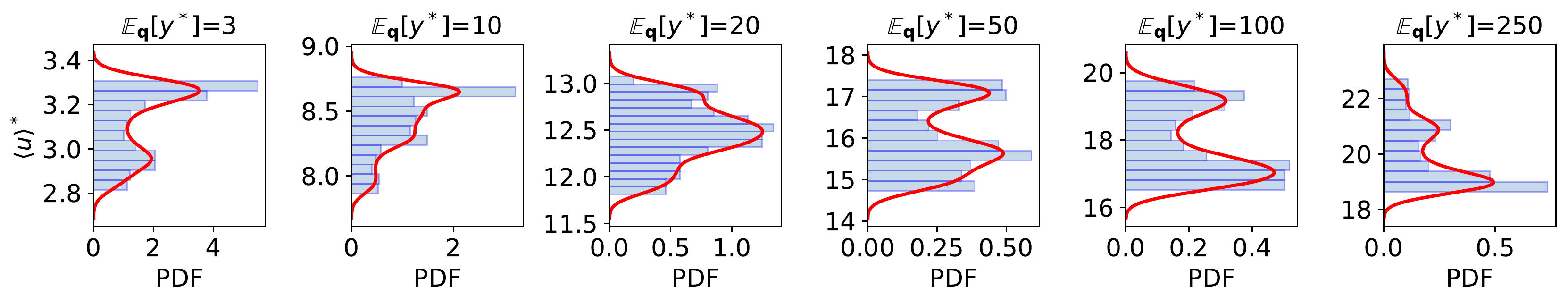} \\
   \end{tabular}
   \caption{PDF contours (top) and histograms at specific wall-normal distances (bottom) of $\lu^*$ due to the variation of $\dxp$ and $\dzp$ over $[5,130]$ and $[5,70]$, respectively. \rev{The numbers on the isolines in the top plot show the PDF values.} In the bottom plots, the blue bars and solid line respectively show histogram and kernel density estimation of the PDF.}\label{fig:uqPDF_up}
\end{figure}

\begin{figure}
\centering
   \begin{tabular}{c}
   \includegraphics[scale=0.5]{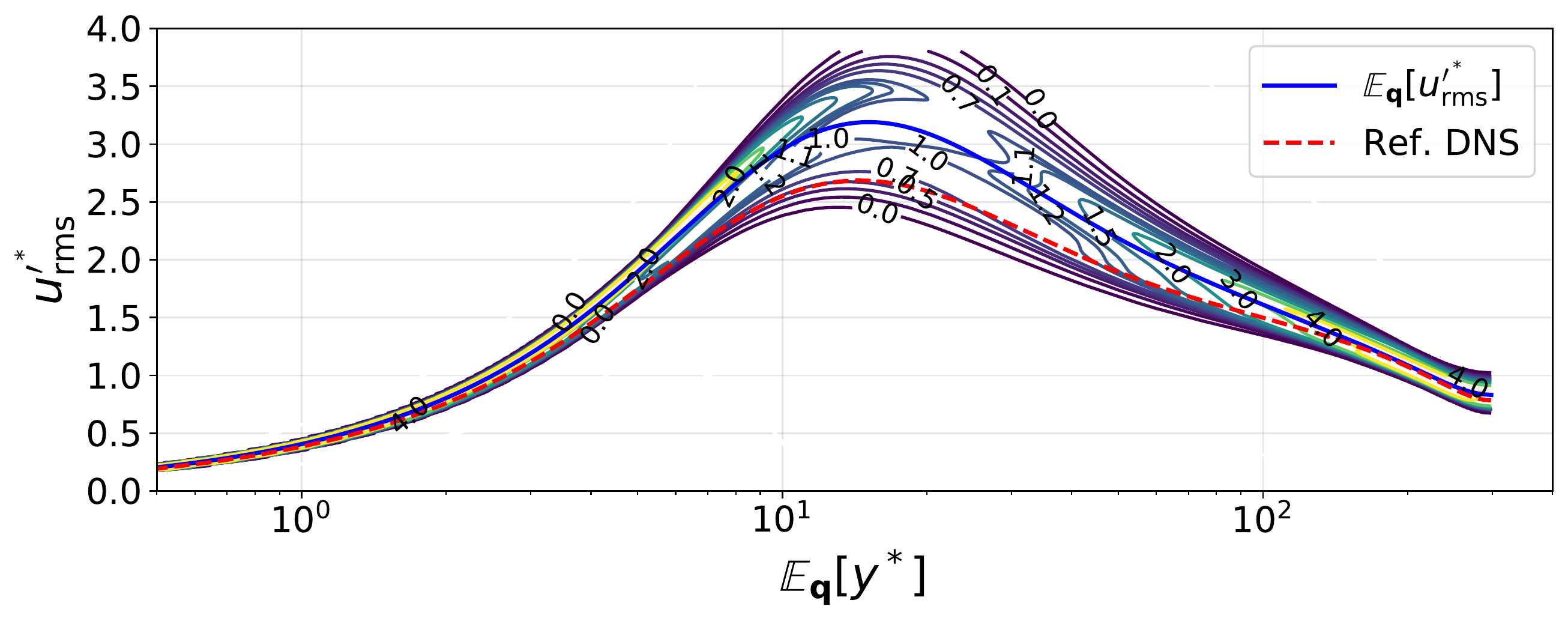} \\
   \includegraphics[scale=0.46]{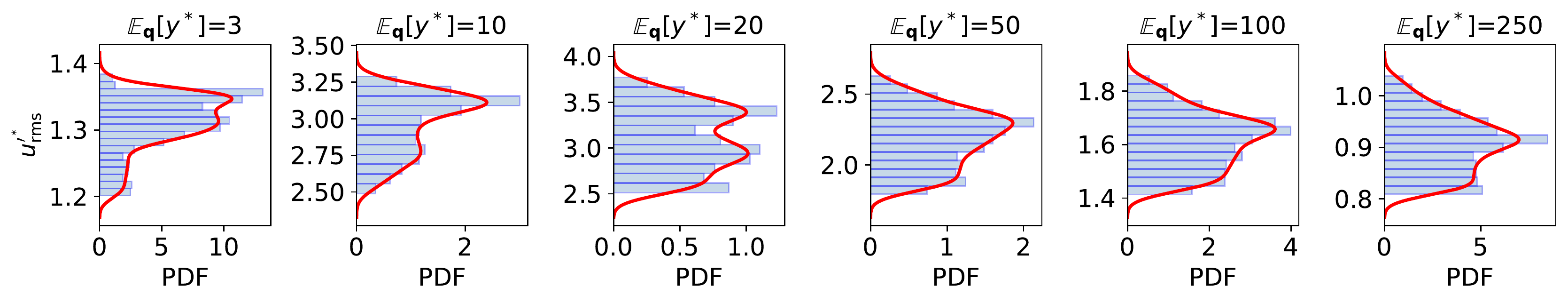} \\
   \end{tabular}
   \caption{PDF contours~(top) and histograms at specific wall-normal distances (bottom) of $u'^*_\rms$ due to the variation of $\dxp$ and $\dzp$ over $[5,130]$ and $[5,70]$, respectively. \rev{The numbers on the isolines in the top plot show the PDF values.} In the bottom plots, the blue bars and solid line respectively show histogram and kernel density estimation of the PDF.}\label{fig:uqPDF_upp}
\end{figure}

We move on to the third type of analysis mentioned at the beginning of this section. 
The idea is to estimate how probable is to observe a particular value of the QoIs as a result of variation of~$\fq$ over~$\BQ$. 
To this end, probability density functions (PDF) of the QoIs are constructed through evaluating the surrogate~$\tf(\fq)$ at enough number of samples of~$\fq$.
The surrogate can be constructed for instance by PCE~(\ref{eq:pce}) or GPR~(\ref{eq:gpr}) or any other method, see~\eg~\cite{uqHandbook,smith,gramacy:20}.

Shown in~the top plots of \figs~\ref{fig:uqPDF_up} and~\ref{fig:uqPDF_upp} are the isolines of PDFs of~$\lu^*$ and~$u'^*_\rms$ constructed along the associated profiles.
The bottom plots represent the PDFs at a few off-the-wall locations extracted from the top plots.
Clearly, the PDF of the QoI varies with the wall distance and is not the same as any \rev{standard} distributions. 
Moreover, at a fixed $\BE_\fq[y^*]$, there might be more than one value for the QoIs which have high probability to be observed. 
In other words, the PDF of the QoI can be multi-modal. 
As an extension of the current analysis, one can estimate the distribution of the combinations of the parameters $\fq$ which lead to a specific value for the QoIs. 
This can be achieved through conducting a UQ inverse problem in a Bayesian framework, see \eg~\cite{uqHandbook,smith}, where the surrogate in model~(\ref{eq:surrGen}) specifies the relationship between the input parameters and outputs.

\section{Summary and Conclusions}\label{sec:conclusions}
Validation and verification (V\&V) aim at assessing the accuracy and reliability of the outputs of computational simulations of different physics problems. 
A computational model depends on a set of inputs and parameters (together called factors) which can influence the simulator's outputs. 
Tracing such impacts through closed mathematical relations is not usually feasible, considering the complexities in computational, numerical and mathematical models. 
\rev{This is particularly challenging in the case of the Navier--Stokes equations being numerically solved for studying flow turbulence.}
Therefore, designing appropriate computer experiments is inevitable for the purpose of V\&V. 
The present study develops a unified framework by combining different UQ techniques and computer experiments in order to quantify accuracy, robustness, and sensitivity in computational physics. 
Although some of the UQ techniques may be well-established, utilizing them for the purpose of this study is novel. 
As a UQ way of thinking, all the simulator's factors and outputs are considered to be random, and thus probabilistic approaches can be applied. 
According to this view, the concepts of accuracy, robustness and sensitivity of the outputs with respect to variation of the factors are defined. 
Depending on the reference true data used for assessment of accuracy, both validation and verification can be addressed.

The framework is general and flexible so it can be applied to any computational physics problem. 
However, the examples and discussions in this manuscript are provided with focus on CFD, in general, and high-fidelity simulations of wall turbulence, in particular.  
In this regard, two sets of factors for the computer simulators, e.g. CFD solvers, are considered. 
\catI~parameters are either uncertain by definition or assumed to be so.
Hence they are allowed to vary according to a given probability distribution over an admissible space. 
Based on a limited number of (training) samples for these factors, a surrogate for the actual simulator is constructed in a computer experiment.
To estimate variation in the outputs due to the \catI~parameters, different constructions for non-intrusive PCE are discussed. 
To assess sensitivity of QoIs with respect to the uncertain factors, Sobol indices based on analysis of variance (ANOVA) are computed. 
An overview on Gaussian process regression is given, which is a powerful method to include uncertainties due to \catII~parameters which affect the simulator's outputs in the computer experiment designed based on \catI~parameters. 
As a novel contribution, it is shown how to use GPR with heteroscedastic noise structure which enables us to incorporate observation-dependent uncertainties. 
It is recalled that usually GPR is used with homoscedastic noise levels. 
Another novelty of this study is to combine standard PCE with uncertain predictions by GPR to derive probabilistic PCE for the purpose of estimating ``uncertain propagation of uncertainties" in the simulator's outputs due to the combination of \catI~and \catII~parameters.

Different capabilities of the developed framework are illustrated by applying it to a computer experiment based on scale-resolving simulations of turbulent channel flow. 
The CFD simulations are carried out using the open-source spectral-element solver \nek~\cite{nek}. 
The inner-scaled grid resolutions $\dxp$ and $\dzp$ in the wall-parallel directions of the channel are the \catI~parameters. 
The quantities of interest are the averaged friction velocity $\lut$ and the profiles of mean and rms fluctuation of velocity, \ie~$\langle u_i\rangle$ and $\langle u'_i u'_j\rangle$ for $i,j=1,2,3$, respectively. The \catII~factors may, for instance, be the insufficiency of samples when computing time-averaged QoIs. 
Although, the values of such an uncertainty are negligible in the data used in the present study, we employ synthetic values for this uncertainty to illustrate the features of the framework.

Through the illustrative example, it is shown how the UQ framework facilitates finding quantitative estimates to the following questions which are relevant to any CFD simulation: 
\emph{i)} How much do the QoIs vary as a result of variation of different factors?
Through this uncertainty-propagation problem, robustness of the QoIs with respect to the input variations is measured. 
\emph{ii)} In a given variation in a QoI, what is the contribution of each factor? This is answered by global sensitivity analysis. 
\emph{iii)} What is the probability of observing a particular value of the QoIs when inputs vary?
Any of these analyses can also be applied to the accuracy of a QoI using experimental or high-fidelity reference data.

For the case of turbulent channel flow, the portrait of the error in~$\lut$ in the $\dxp\dash\dzp$ space, \fig~ \ref{fig:duTauSurface}, gives an understanding of the complexity in the variation of error with grid resolution and hence clarifies the difficulty that analytical error estimators have to deal with.  \rev{It should be highlighted that our approach is not intended to find the \emph{best} combination of the factors to be closest to the reference data, but rather quantifies the sensitivity, robustness and interdependence of the various factors.}
Considering \catII~uncertainties, the isolines are slightly affected, see \fig~\ref{fig:duTauSurfNoisy}.
The largest variation is observed for the isoline of zero~$\epsilon[\lut]$. 
The Sobol indices computed for error in different QoIs (see \fig~\ref{fig:sobolError}) are informative, since they specify which factor has to be altered in order to have the highest impact on a specific error measure. 
A discussion is made on the number of samples from the uncertain factors which have to be included in a computer experiment. 
With regard to the non-intrusive PCE, the use of the diagnosis tool~(\ref{eq:pceConvNorm}) is represented in \fig~\ref{fig:pceConv_uTau}. 
When the purpose is estimating the propagated uncertainty into the inner-scaled profiles of mean and rms fluctuation of streamwise velocity,~$\lu^*=\lu/\lut$ and~$u'^*_\rms = u'_\rms/\lut$ (see \fig~\ref{fig:convPCE_subC8}), the inclusion of~$13$ samples is found to be insignificantly erroneous compared to the case with almost double number of samples. 
However, in any case it is important to use a space-filling sampling method.

The propagated uncertainties in~$\lu^*$ and $u'^*_\rms$ profiles due to the variation of~$\dxp$ and~$\dzp$ are represented in \fig~\ref{fig:exCI}. 
Interestingly, in the \rev{immediate} near-wall region \revs{with $\BE_\fq[y^*]\lesssim 5$},  the profiles are very robust against changing the resolution. 
The most influenced region is near the channel center for $\lu^*$ and near the peak for~$u'^*_\rms$. 
Moreover, the admissible space of the uncertain parameters is shown to influence the propagated uncertainty. 
Such influence varies between the QoIs. 
The information from these UQ forward problem should be combined with the results of global sensitivity analysis. 
The plots in~\fig~\ref{fig:exProfSobol} show how much influence each factor has on the QoIs at different distances from the wall. 
The profiles of Sobol indices vary between the QoIs and also with the admissible space. 
However, the most influential factor at very close distances to the wall is always observed to be~$\dzp$, \ie~the grid resolution in the spanwise direction. 
As the third type of information which could be extracted from the framework, the PDF of profiles of $\lu^*$ and $u'^*_\rms$ are plotted in \fig~\ref{fig:uqPDF_up} and \fig~\ref{fig:uqPDF_upp}, respectively. 
At any wall-distance, the PDFs of the QoIs are multimodal, even though the factors $\dxp$ and $\dzp$ were assumed to be uniform random variables. 

Finally, we have shown how to combine the variation of grid resolutions with uncertainty from other sources which affect the simulator's outputs, for instance, insufficient averaging in time. 
The novel plots of \fig~\ref{fig:C8_ppce} are obtained as a result of applying the probabilistic PCE of \sect~\ref{sec:ppce}. 
No need to emphasize that many of the resulting interpretations and conclusions could not be achieved without using the relevant techniques of the introduced framework.

The present study can be applied and extended in various ways, including the followings.
The framework can be applied to compare performance of different simulators for benchmark problems. 
The uncertainties due to insufficient time-averaging in turbulent simulations are needed to be accurately quantified and then be used in the present framework.
A UQ inverse problem can be designed and solved to estimate with confidence the grid resolutions which could result in specific PDFs of QoIs similar to what is shown in \fig~\ref{fig:uqPDF_up} and \fig~\ref{fig:uqPDF_upp}. 
\rev{Another relevant application area of the present framework is the study of sensitivity within turbulence modelling (LES and RANS), or the consideration of physical factors such as surface roughness and/or geometrical uncertainties. }

\section*{Acknowledgments}
This work has been supported by the EXCELLERAT project which has received funding from the European Union's Horizon 2020 research and innovation programme under grant agreement No 823691.
Financial support by the Linn{\'e} FLOW Centre at KTH for SR is gratefully acknowledged.
RV acknowledges the financial support from the Swedish Research Council (VR), and PS funding by the Knut and Alice Wallenberg (KAW) foundation as part of the Wallenberg Academy Fellow programme.
The channel flow simulations in \sect~\ref{sec:example} were performed on the resources provided by the Swedish National Infrastructure for Computing
(SNIC) at PDC (KTH Royal Institute of Technology), HPC2N (Ume{\r a} University), and NSC (Link{\"o}ping University),~Sweden.

\bibliographystyle{abbrvnat}
\bibliography{bib_uqRobustWRChan}

\end{document}

%% file: commands.tex
\newcommand{\et}{et al.}
\newcommand{\eg}{e.g.}
\newcommand{\ie}{i.e.}
\newcommand{\fig}{Figure}
\newcommand{\figs}{Figures}
\newcommand{\eq}{Eq.}
\newcommand{\eqs}{Eqs.}
\newcommand{\sect}{Section}
\newcommand{\sects}{Sections}
\newcommand{\rf}{Ref.}
\newcommand{\rfs}{Refs.}

\newcommand{\nek}{Nek5000}
\newcommand{\catI}{Category-I}
\newcommand{\catII}{Category-II}

\newcommand{\BE}{\mathbb{E}}
\newcommand{\BR}{\mathbb{R}}
\newcommand{\BQ}{\mathbb{Q}}
\newcommand{\BN}{\mathbb{N}}
\newcommand{\BZ}{{\mathbb{Z}}}
\newcommand{\BV}{\mathbb{V}}
\newcommand{\hBE}{\hat{\mathbb{E}}}
\newcommand{\hBV}{\hat{\mathbb{V}}}
\newcommand{\cR}{r}

\newcommand{\cN}{\mathcal{N}}
\newcommand{\cD}{\mathcal{D}}
\newcommand{\cK}{\mathcal{K}}
\newcommand{\GP}{\mathcal{GP}}

\newcommand{\dd}{\mbox{d}}
\newcommand{\dash}{\mbox{-}}

\newcommand{\fq}{\mathbf{q}}
\newcommand{\fQ}{\mathbf{Q}}
\newcommand{\fQs}{\mathbf{Q}^*}
\newcommand{\fxi}{\bm{\xi}}
\newcommand{\fk}{\mathbf{k}}
\newcommand{\feps}{\bm{\varepsilon}}
\newcommand{\fC}{\mathbf{C}}
\newcommand{\fR}{\mathbf{R}}

\newcommand{\hf}{\hat{f}}
\newcommand{\hbf}{\hat{\mathbf{f}}}
\newcommand{\tf}{\tilde{f}}
\newcommand{\fTheta}{{\bm{\Theta}}}
\newcommand{\fThetaEps}{{\bm{\Theta}}_\varepsilon}

\newcommand{\rey}{\mbox{Re}}
\newcommand{\reyt}{\mbox{Re}_\tau}
\newcommand{\reyb}{\mbox{Re}_b}

\newcommand{\dxp}{{\Delta x^+}}
\newcommand{\dyp}{\Delta y^+}
\newcommand{\dzp}{{\Delta z^+}}

\newcommand{\lut}{\langle u_\tau \rangle}
\newcommand{\lu}{\langle u \rangle}
\newcommand{\uv}{\langle u'v'\rangle}
\newcommand{\rms}{{\rm rms}}

\newcommand{\einf}{\epsilon_{\infty}}
\newcommand{\rk}{{\rm k}}
\newcommand{\rw}{{\rm w}}

